\documentclass[12pt,a4paper,final]{iopart}

\usepackage{iopams}
\usepackage{graphicx}
\usepackage{epstopdf}
\usepackage[breaklinks=true,colorlinks=true,linkcolor=blue,urlcolor=blue,citecolor=blue]{hyperref}

\begin{document}

\title[Correlations and synchronization in a Bose-Fermi mixture]{Correlations and synchronization in a Bose-Fermi mixture}

\author{Pablo~D\'{i}az}
\address{Departamento de Ciencias F\'{i}sicas, Universidad de La
Frontera, Casilla 54-D, Temuco, Chile.}
\author{David~Laroze}
\address{SUPA School of Physics and Astronomy, University of Glasgow, Glasgow G12 8QQ, United Kingdom}
\address{Instituto de Alta Investigaci\'{o}n, Universidad de
Tarapac\'{a}, Casilla 7D, Arica, Chile}
\ead{dlarozen@uta.cl}

\author{Boris A. Malomed}
\address{Department of Physical Electronics, School of Electrical Engineering,
Faculty of Engineering, Tel Aviv University, Tel Aviv 69978, Israel}

\begin{abstract}
We study a Bose-Fermi mixture within the framework of the mean-field theory,
including three possible regimes for the fermionic species: fully polarized,
BCS, and unitarity. Starting from the 3D description and using the
variational approximation (VA), we derive 1D and 2D systems of equations, under the
corresponding confining potentials. This method produces a pair of nonlinear
Schr\"{o}dinger (NLS) equations coupled to algebraic equations for the transverse
widths of the confined state. The equations incorporate interactions between
atoms of the same species and between the species, assuming that the latter
can be manipulated by means of the Feshbach resonance (FR). As an
application, we explore spatial density correlations in the ground state
(GS) between the species, concluding that they strongly depend on the sign
and strength of the inter-species interaction. Also studied are the dynamics
of the mixture in a vicinity of the GS and the corresponding spatiotemporal
inter-species correlation. The correlations are strongly affected by the
fermionic component, featuring the greatest variation in the unitary regime.
Results produced by the VA are verified by comparison with full numerical solutions.
\end{abstract}

\vspace{2pc}
\submitto{\JPB}

\section{Introduction}

\label{sec-1}

Cold atom gases trapped in magneto-optical potentials have tremendously
extended the range of experimental possibilities in condensed-matter
physics. The possibility of controlling the intensity of the interactions,
the depth of the trapping potential lattices, and their geometry enables
emulating a broad variety of solid-state settings and the creation of novel
states of the quantum matter \cite{Bloch08}-\cite{spin-orbit2}.

In this context, a great variety of experimental and theoretical studies
were stimulated by the possibility of mixing degenerate bosonic and
fermionic gases. The first experimental works on Bose-Fermi mixtures
(BFMs) were conducted with lithium isotopes \cite{Partridge01,Schreck01},
which was a precursor to the study of new combinations, such as $^{174}$Yb-$%
^{6}$Li \cite{Hansen11} and $^{87}$Rb-$^{40}$K \cite{Heinze11}. Particular
attention has been paid to heavy-atom mixtures, e.g., $^{87}$Sr-$^{84}$Sr
\cite{Tey10} isotopes with a large nuclear spin, that have been proposed as
prototypes for handling the quantum information. One of important points of
the work in this direction is the use of Feshbach resonances (FRs) in the
mixtures, as they make it possible to control the two-body interactions
between the species. The FR has been observed for the $^{87}$Rb-$^{40}$K
mixture \cite{Best09,Cumby13}, and a giant FR has been reported for $^{85}$%
Rb-$^{6}$Li \cite{Deh10}. Further, five FRs were observed in $^{6}$Li-$%
^{133} $Cs \cite{Tung13}, and over 30 resonances are known to occur in the $%
^{23}$Na-$^{40}$K mixture \cite{Park12}. In Ref. \cite{Wu11}, a multiple
heteronuclear FRs was reported in a triply quantum-degenerate mixture of
bosonic $^{41}$K and two fermionic species, $^{40}$K and $^{6}$Li. In the
latter context, a wide $s$-wave resonance for combination $^{41}$K-$^{40}$K
transforming the system into a strongly interacting isotopic BFM immersed
into a Fermi sea.

Parallel to the experiment, several theoretical works addressed the dynamics
of BFMs \cite{Lelas09}-\cite{Bertaina13}. An approach that has proven to be
very useful for describing the ground state (GS) of the mixtures is provided
by the mean field theory \cite{Adhikari08}-\cite{Nishida06}. In this
context, the FRs for mixtures of $^{23}$Na-$^{6}$Li, $^{87}$Rb-$^{40}$K, $%
^{87}$Rb-$^{6}$Li, $^{3}$He-$^{4}$He, $^{173}$Yb-$^{174}$Yb, and $^{87}$Sr-$%
^{84}$Sr have been studied \cite{Salasnich07a,Gautam11}.

The possibility of holding ultracold gases in magneto-optical traps makes it
possible to create effectively one- or two-dimensional (1D or 2D)
situations. A great number of works addressed such 1D settings \cite%
{Salasnich07b}-\cite{Abduallev13}. In particular, the variational
approximation makes it possible to reduce the 3D description to 1D or 2D
\cite{Adhikari10,Cheng11}. In many cases, these approximations do not take
into regard the fact that the transverse confinement width(s) may vary with
the particle density in the unconfined direction. The variational
approximation that does take this fact into account was developed for bosons
in Ref. \cite{Salasnich02} and for fermions in Ref. \cite{Diaz12},
demonstrating the high accuracy of the 1D and 2D approximations in
comparison with the full the 3D description.

In this paper we analyze the BFM at zero temperature confined to 1D
or 2D by external potentials. First, we consider the cigar-shaped
configuration produced by the strong confinement with the
cylindrical symmetry acting in the transverse plane, while a generic
weak potential may act in the axial direction. Second, we consider
the disc-shaped BFM corresponding to the strong confinement applied
in one direction, and a weak 2D potential acting in the
perpendicular plane. We use the variational method developed in
Refs. \cite{Salasnich02,Diaz12} to derive effective 1D and 2D
nonlinear Schr\"{o}dinger (NLS) equations for the cigar-shaped and
disc-shaped configurations, respectively. In particular, our results
apply to the $^{7}$Li-$^{6}$Li BFM. In both cases (1D and 2D), we
analyze the influence of the density of both species on the width of
the mixture in the confined dimensions, the main control parameter
being the inter-species scattering length, $a_{\mathrm{BF}} $. The
analysis is performed for three different regimes of the fermionic
component: polarized, BCS, and unitarity. Using the
dimensionally-reduced system, we consider spatial correlations
between densities of particles of both species. Considering
perturbations around the GS, we study the spatiotemporal correlation
between the species too.

The paper is organized as follows: In Sec.~\ref{sec-2}, we derive 1D and 2D
variational equations from the full 3D description. In Sec.~\ref{sec-3}, we
apply our equations to the numerical analysis of the GS in the BFM, and
consideration of the spatial correlation between the species. For the 1D
setting, we calculate spatiotemporal correlation in the form of the
so-called Pearson coefficient, considering perturbations around the GS.
Conclusions are presented in Sec. \ref{sec-4}.

\section{The variational approximation}
\label{sec-2}

We consider a diluted superfluid mixture formed by $N_{\mathrm{B}}$ bosonic
atoms of mass $m_{\mathrm{B}}$, and $N_{\mathrm{F}}
$ fermionic atoms of mass $m_{\mathrm{F}}$ and spin $s_{\mathrm{F}}$.
The atoms interact through the pseudopotential, $\delta (\mathbf{r})$ \cite%
{Bloch08}. We assume that bosons form a Bose-Einstein condensate (BEC),
described by the Gross-Pitaevskii equation \cite{Bloch08}, while the local
density approximation \cite{Bloch08} applies to the description of the
weakly interacting fermionic component. Accordingly, the dynamical equations
for the BFM can be derived from action $\mathcal{S}$,

\begin{equation}
\mathcal{S}=\int {dtd{\mathbf{r}}\left( \mathcal{L}_{\mathrm{B}}+\mathcal{L}%
_{\mathrm{F}}+\mathcal{L}_{\mathrm{BF}}\right) },  \label{E-1}
\end{equation}%
where $\mathcal{L}_{B}$ and $\mathcal{L}_{\mathrm{F}}$ are the Lagrangian
densities of the Bose and Fermi components, while $\mathcal{L}_{\mathrm{BF}}$
accounts for the interaction between the them:
\begin{eqnarray}
{\mathcal{L}_{B}} &=&\frac{{i\hbar }}{2}\left( {\Psi _{\mathrm{B}}^{\ast }%
\frac{{\partial {\Psi _{\mathrm{B}}}}}{{\partial t}}-{\Psi _{\mathrm{B}}}%
\frac{{\partial \Psi _{\mathrm{B}}^{\ast }}}{{\partial t}}}\right) -\frac{{{%
\hbar ^{2}}}}{{2{m_{\mathrm{B}}}}}{\left\vert {\nabla {\Psi _{\mathrm{B}}}}%
\right\vert ^{2}}
-{U_{B}}{\left\vert {{\Psi _{\mathrm{B}}}}\right\vert ^{2}}-\frac{1}{2}{g_{%
\mathrm{B}}}{\left\vert {{\Psi _{\mathrm{B}}}}\right\vert ^{4}},  \label{E-2}
\end{eqnarray}

\begin{eqnarray}
{\mathcal{L}_{\mathrm{F}}} &=&\frac{{i\hbar }}{2\lambda _{1}}\left( {\Psi _{%
\mathrm{F}}^{\ast }\frac{{\partial {\Psi _{\mathrm{F}}}}}{{\partial t}}-{%
\Psi _{\mathrm{F}}}\frac{{\partial \Psi _{\mathrm{F}}^{\ast }}}{{\partial t}}%
}\right) -\frac{{{\hbar ^{2}}}}{{2\lambda _{2}{m_{\mathrm{F}}}}}{\left\vert {%
\nabla {\Psi _{\mathrm{F}}}}\right\vert ^{2}}-
{U_{\mathrm{F}}}{\left\vert {{\Psi _{\mathrm{F}}}}\right\vert ^{2}}-\frac{1}{2}{{g_{\mathrm{F}}}%
}{\left\vert {{\Psi _{\mathrm{F}}}}\right\vert ^{4}} \nonumber \\
&&-\frac{3%
}{5}\xi \frac{{{\hbar ^{2}}}}{{2{m_{\mathrm{F}}}}}{C_{\mathrm{F}}}{%
\left\vert {{\Psi _{\mathrm{F}}}}\right\vert ^{10/3}},  \label{E-2F}
\end{eqnarray}

\begin{equation}
{\mathcal{L}_{B\mathrm{F}}}=-\frac{1}{2}{g_{\mathrm{BF}}}{\left\vert {{\Psi
_{B}}}\right\vert ^{2}}{\left\vert {{\Psi _{\mathrm{F}}}}\right\vert ^{2}}.
\label{E-3}
\end{equation}%
Here ${C_{\mathrm{F}}}\equiv {\left[ {6{\pi ^{2}}/\left( {2{s_{\mathrm{F}}}+1%
}\right) }\right] ^{2/3}}$ is a constant that depends on spin $s_{\mathrm{F}%
} $; ${g_{B}}\equiv 4\pi {\hbar ^{2}}{a_{B}}/${$m_{\mathrm{B}}$}, ${g_{%
\mathrm{F}}}\equiv 4\pi {\hbar ^{2}}({a_{\mathrm{F}}}/{m_{\mathrm{F}}})[2S_{%
\mathrm{F}}/(2S_{\mathrm{F}}+1)]$, and ${g_{\mathrm{BF}}}\equiv 4\pi {\hbar
^{2}}${$a_{\mathrm{BF}}$}$/${$m_{\mathrm{BF}}$} are the three interaction
parameters of the mixture, with $a_{\mathrm{B}}$, $a_{\mathrm{F}}$ and $a_{%
\mathrm{BF}}$ being the respective scattering lengths; ${m_{\mathrm{BF}}}%
\equiv {m_{B}}{m_{\mathrm{F}}}/({m_{B}}+{m_{\mathrm{F}}})$ is the reduced
mass; and $U_{\mathrm{B}/\mathrm{F}}$ are external potentials acting on
bosons/fermions. Complex wave functions $\Psi _{\mathrm{B}/\mathrm{F}}\left(
\mathbf{r},t\right) $ are normalized to the numbers of particles, $N_{%
\mathrm{B}/\mathrm{F}}$. Parameters $\lambda _{1}$, $\lambda _{2}$, and $\xi
$ in the fermionic Lagrangian density (\ref{E-2F}), along with $s_{\mathrm{F}%
}$, determine three different regimes \cite{Manini05}-\cite{Ancilotto12}
listed in the Table 1.

\Table{\label{Table-1}$\protect\lambda _{1}$, $\protect\lambda _{2}$, $\protect\xi $, and
$s_{\mathrm{F}}$ for three different regimes of the Fermi gas.}
\br
Regime & $\lambda _{1}$ & $\lambda _{2}$ & $\xi $ & $s_{\mathrm{F}}$ \\
\mr
Polarized & 1 & 1 & 1 & 0 \\
BCS & 2 & 4 & 1 & 1/2 \\
Unitarity & 2 & 4 & 0.44 & 1/2 \\
\br
\end{tabular}
\end{indented}
\end{table}

We apply the formalism developed below to the ${}^{7}$Li-${}^{6}$Li mixture,
with the same scattering parameter for both species, $a_{\mathrm{B/F}}=5$nm.
The use of isotopes of the same alkali element is suggested by the
similarity of their electric polarizability, thus implying similar external
potentials induced by an optical trap. Unless specified otherwise,
in what follows below we consider configurations with fully polarized fermions.
Values of parameters for all fermionic regimes are collected in Table \ref{Table-1}.
The BCS and unitarity regimes involve more than one spin state of fermions, hence
the magnetic trap will split the respective spin energy levels. For this reason,
we assume the presence of the optical trap, which supports equal energy levels for
all the spin states, making it possible to discriminate different regimes of the
interaction in the Bose-Fermi mixture. In the BCS and unitarity regimes, we assume balanced
populations of the two spin components.

Finally, varying action $\mathcal{S}$ with respect to $\Psi _{\mathrm{B}%
}^{\ast }$ and to $\Psi _{\mathrm{F}}^{\ast }$, we derive the following
system of nonlinear Schr\"{o}dinger equations for bosons and fermions:

\begin{equation}
i\hbar {\partial _{t}}{\Psi _{B}}=\left[ {-\frac{{{\hbar ^{2}}}}{{2{m_{B}}}}{%
\nabla ^{2}}+{g_{B}}{{\left\vert {{\Psi _{B}}}\right\vert }^{2}}+{g_{\mathrm{%
BF}}}{{\left\vert {{\Psi _{\mathrm{F}}}}\right\vert }^{2}}+{U_{B}}}\right] {%
\Psi _{B},}  \label{E-5}
\end{equation}
\begin{eqnarray}
\nonumber \frac{{i\hbar }}{{{\lambda _{1}}}}{\partial _{t}}{\Psi _{\mathrm{F}}} &=&%
\left[ -\frac{{{\hbar ^{2}}}}{{2{\lambda _{2}}{m_{\mathrm{F}}}}}{\nabla ^{2}%
}+{{g_{\mathrm{F}}}{{\left\vert {{\Psi _{\mathrm{F}}}}\right\vert }%
^{2}}+{g_{\mathrm{BF}}}{{\left\vert {{\Psi _{B}}}\right\vert }^{2}}}+{U_{\mathrm{F}}}\right. \\
&&\left.+\frac{{{\hbar ^{2}}}}{{2{m_{j}}}}\xi {C_{\mathrm{F}}}{{%
\left\vert {{\Psi _{\mathrm{F}}}}\right\vert }^{4/3}}\right] {%
\Psi _{\mathrm{F}}}.  \label{E-6}
\end{eqnarray}

As is known from previous works, the mean-field description of the fermionic component,
based on the effective NLS equation (\ref{E-6}) is valid in the hydrodynamic approximation,
which implies that the Fermi component forms a correlated superfluid. In static situations, this
approximation is relevant for sufficiently smooth configurations, whose characteristic spatial scale is
much greater than the de Broglie wavelength at the Fermi surface \cite{Yang1,Yang2}. In various
dynamical settings, both pure fermionic and mixed Bose-Fermi ones, the hydrodynamic approach remains
relevant for slow quasi-adiabatic evolution \cite{Adhikari10}-\cite{Ancilotto12};
\cite{demix,AdhikariPhysicaD}.

While numerical integration of this system in the 3D form system
is very heavy computationally, the effective dimension may be reduced to 1D or 2D
when the system is tightly confined by a trapping potential. To this end, the
variational method is employed, making use of the factorization of the 3D wave function,
which includes a Gaussian ansatz in the tightly confined transverse directions. The factorization
has been widely used for Bose and Fermi system separately, as it shown in Refs. \cite{Salasnich02} and \cite{Diaz12}, respectively. In next two sections we apply the variational technique to a Bose-Fermi mixture,
obtaining results that agree well with full 3D simulations. On the other hand, the applicability of the
Gaussian ansatz for approximating of the transverse part of the 3D wave function is determined by a
relation of the transverse-confinement strength, and the effective strength of the bosonic or fermionic
nonlinearity. In case the latter factor is stronger, the transverse wave function may be used in the
form of the Thomas-Fermi (TF) approximation \cite{AdhikariPhysicaD}.

\subsection{The one-dimensional variational system}

\label{sec-2-1}

The common form of the confinement is provided by the tight
harmonic-oscillator potential acting in the trapping dimensions. Thus, the
confinement to 1D (the cigar-shaped configuration along the $z$-axis)
corresponds to the following potentials, which include weak axial
components, $V_{1\mathrm{D},{\mathrm{B}/\mathrm{F}}}$:
\begin{equation}
V_{\mathrm{B}/\mathrm{F}}\left( {{\mathbf{r}},t}\right) =\frac{1}{2}m_{%
\mathrm{B}/\mathrm{F}}\omega _{t,{\mathrm{B}/\mathrm{F}}}^{2}\left( {{x^{2}}+%
{y^{2}}}\right) +V_{1\mathrm{D},{\mathrm{B}/\mathrm{F}}}\left( {z,t}\right) .
\label{E-7}
\end{equation}%
Assuming that the transverse trapping potential is strong enough, the
dimensional reduction is carried out by means of the usual factorized ansatz
for the wave functions, $\Psi _{\mathrm{B}/\mathrm{F}}\left( {{\mathbf{r}},t}%
\right) =v_{\mathrm{B}/\mathrm{F}}\left( {x,y;\sigma _{\mathrm{B}/\mathrm{F}%
}\left( {z,t}\right) }\right) f_{\mathrm{B}/\mathrm{F}}\left( {z,t}\right) $%
, where the transverse ground-state Gaussians, $v_{\mathrm{B}/\mathrm{F}}$%
,with widths $\sigma _{\mathrm{B}/\mathrm{F}}$, and the axial functions, $f_{%
\mathrm{B}/\mathrm{F}}$, are normalized to $1$ and $N_{\mathrm{B}/\mathrm{F}%
} $ respectively:

\begin{equation}
{v_{\mathrm{B}/\mathrm{F}}}\left( {x,y;{\sigma _{\mathrm{B}/\mathrm{F}}}%
\left( {z,t}\right) }\right) =\frac{1}{{{\pi ^{1/2}}{\sigma _{\mathrm{B}/%
\mathrm{F}}}\left( {z,t}\right) }}{\exp }\left( -\frac{x^{2}+y^{2}}{{2{%
\sigma _{\mathrm{B}/\mathrm{F}}^{2}}{{\left( {z,t}\right) }}}}\right) .
\label{E-8}
\end{equation}%
For both species, we define the axial density as ${%
n_{1D,{\mathrm{B}/\mathrm{F}}}}\equiv {\left\vert f_{\mathrm{B}/\mathrm{F}%
}\right\vert ^{2}}$. The relevance of the use of the Gaussian form in the
confined directions for bosons and fermions was demonstrated in in Refs.
\cite{Salasnich02} and \cite{Diaz12}, respectively, by showing that the
respective factorization produces GSs close to their exact 3D counterparts.
As demonstrated below, the Pauli principle makes the transverse width of the
Fermi gas larger, in comparison with the Bose gas in the presence of a
similar external potential.

Replacing $\Psi _{\mathrm{B}/\mathrm{F}}$ by the factorized density in the
Lagrangian density and integrating in the transverse plane of $(x,y)$, the
expression for action (\ref{E-1}) is cast into takes the following form:

\begin{equation}
\mathcal{S}=\int {dtdz\left( {{\mathcal{L}_{1\mathrm{D,B}}}+\mathcal{L}_{1%
\mathrm{D},\mathrm{F}}+\mathcal{L}_{1\mathrm{D},\mathrm{BF}}}\right) },
\label{E-9}
\end{equation}%
where the effective 1D densities are

\begin{equation}
{{\mathcal{L}_{1\mathrm{D,B}}}}=i\frac{\hbar }{2}\left( {f_{\mathrm{B}%
}^{\ast }{\partial _{t}}{f_{\mathrm{B}}}-{f_{\mathrm{B}}}{\partial _{t}}f_{%
\mathrm{B}}^{\ast }}\right) -{V{_{1\mathrm{D,B}}}n{_{1\mathrm{D,B}}}}-{e{_{1%
\mathrm{D,B}}}},
\end{equation}

\begin{equation}
{\mathcal{L}_{1\mathrm{D},\mathrm{F}}}=i\frac{\hbar }{2\lambda _{1}}\left( {%
f_{\mathrm{F}}^{\ast }{\partial _{t}}{f_{\mathrm{F}}}-{f_{\mathrm{F}}}{%
\partial _{t}}f_{\mathrm{F}}^{\ast }}\right) -{V_{1\mathrm{D},\mathrm{F}}}{%
n_{1d,\mathrm{F}}}-{e_{1\mathrm{D},\mathrm{F}}},
\end{equation}
\begin{equation}
{\mathcal{L}_{1\mathrm{D,BF}}}=-\frac{1}{\pi }\frac{{{g_{\mathrm{BF}}}}}{{%
\sigma _{\mathrm{B}}^{2}+\sigma _{\mathrm{F}}^{2}}}{n{_{1\mathrm{D,B}}}n_{1%
\mathrm{D},\mathrm{F}}},  \label{E-12}
\end{equation}%
and $e{_{1\mathrm{D},\mathrm{B}}}$ and $e{_{1\mathrm{D},\mathrm{F}}}$ are
the 1D energy densities of the boson and fermion species, respectively:
\begin{equation}
{e{_{1\mathrm{D,B}}}}=\frac{{{\hbar ^{2}}}}{{2{m_{\mathrm{B}}}}}{\left\vert {%
{\partial _{z}}{f_{\mathrm{B}}}}\right\vert ^{2}}+
\left[ {\frac{{{g_{\mathrm{B}}}}}{{4\pi \sigma _{\mathrm{B}}^{2}}}{n_{1%
\mathrm{D,B}}}+\frac{{{\hbar ^{2}}}}{{2{m_{B}}\sigma _{\mathrm{B}}^{2}}}+%
\frac{1}{2}{m_{B}}\omega _{t,\mathrm{B}}^{2}\sigma _{\mathrm{B}}^{2}}\right]
{n{_{1\mathrm{D,B}}}},  \label{E-13}
\end{equation}
\begin{eqnarray}
\nonumber {e_{1\mathrm{D},\mathrm{F}}}&=&\frac{{{\hbar ^{2}}}}{{2\lambda _{2}{m_{\mathrm{%
F}}}}}{\left\vert {{\partial _{z}}{f_{\mathrm{F}}}}\right\vert ^{2}}+
\left[ {\frac{{{g_{\mathrm{F}}}}}{{4\pi \sigma _{\mathrm{F}}^{2}}}{n_{1%
\mathrm{D},\mathrm{F}}}+\frac{{{\hbar ^{2}}}}{{2\lambda _{2}{m_{\mathrm{F}}}%
\sigma _{\mathrm{F}}^{2}}}+\frac{1}{2}{m_{\mathrm{F}}}\omega _{t,\mathrm{F}%
}^{2}\sigma _{\mathrm{F}}^{2}}\right] n_{1\mathrm{D},\mathrm{F}} \\
&&+\frac{{{\hbar ^{2}}}}{{2{m_{\mathrm{F}}}}}\frac{3\xi }{{5\sigma _{\mathrm{F}}^{4/3}}}%
{C_{1\mathrm{D},\mathrm{F}}}n_{1\mathrm{D},\mathrm{F}}^{5/3},
\label{E-14}
\end{eqnarray}%
with $C{_{1\mathrm{D},\mathrm{F}}}\equiv (3/5){(6\pi (2{s_{\mathrm{F}}}%
+1))^{2/3}}$ in Eq. (\ref{E-14}). Next, varying action $\mathcal{S}$ given
by Eq.~(\ref{E-9}) with respect to $f _{B}^{\ast }$ and $f_{\mathrm{F}%
}^{\ast }$ yields the respective Euler-Lagrange equations, i.e., the motion
equations for the BFM in the 1D approximation:

\begin{eqnarray}
\nonumber i\hbar {\partial _{t}}{f_{\mathrm{B}}}&=&\left[ {-\frac{{{\hbar ^{2}}}}{{2{m_{%
\mathrm{B}}}}}\partial _{Z}^{2}+{V{_{1\mathrm{D,B}}}}+\frac{1}{%
\pi }\frac{{{g_{\mathrm{BF}}}}}{{\sigma _{\mathrm{B}}^{2}+\sigma _{\mathrm{F}%
}^{2}}}{{\left\vert {{f_{\mathrm{F}}}}\right\vert }^{2}}}
+{\frac{{{g_{\mathrm{B}}}}}{{2\pi \sigma _{\mathrm{B}}^{2}}}{{%
\left\vert {{f_{\mathrm{B}}}}\right\vert }^{2}}} \right.\\
&&+\left.\frac{{{\hbar ^{2}}}}{{2{%
m_{B}}\sigma _{\mathrm{B}}^{2}}}+\frac{1}{2}{m_{B}}\omega _{t,\mathrm{B}%
}^{2}\sigma _{\mathrm{B}}^{2}\right] {f_{\mathrm{B}},}  \label{E-15}
\end{eqnarray}
\begin{eqnarray}
\nonumber i\frac{\hbar }{{{\lambda _{1}}}}{\partial _{t}}{f_{\mathrm{F}}}&=&\left[ {-%
\frac{{{\hbar ^{2}}}}{{2{\lambda _{2}}{m_{\mathrm{F}}}}}\partial _{Z}^{2}+{%
V_{1d,\mathrm{F}}}+\frac{1}{\pi }\frac{{{g_{\mathrm{BF}}}}}{{\sigma _{%
\mathrm{B}}^{2}+\sigma _{\mathrm{F}}^{2}}}{{\left\vert {{f_{B}}}\right\vert }%
^{2}}}+\frac{{{g_{\mathrm{F}}}}}{{2\pi \sigma _{\mathrm{F}}^{2}%
}}{{\left\vert {{f_{\mathrm{F}}}}\right\vert }^{2}} \right.\\
&&+\left.\frac{{{\hbar ^{2}}\xi }}{{2{m_{\mathrm{F}}}}}\frac{{{C_{\mathrm{F},1%
\mathrm{D}}}}}{{\sigma _{\mathrm{F}}^{4/3}}}{{\left\vert {{f_{\mathrm{F}}}}%
\right\vert }^{4/3}}+ {\frac{{{\hbar ^{2}}}}{{2{m_{\mathrm{F}}}\lambda _{2}\sigma _{%
\mathrm{F}}^{2}}}+\frac{1}{2}{m_{\mathrm{F}}}\omega _{t,\mathrm{F}%
}^{2}\sigma _{\mathrm{F}}^{2}}\right] {f_{\mathrm{F}}}.  \label{E-16}
\end{eqnarray}%
Relationships between $\sigma _{\mathrm{B}/\mathrm{F}}$ and $f_{\mathrm{B}/%
\mathrm{F}}$ are obtained by varying the 1D action (\ref{E-9}) with respect
to $\sigma _{\mathrm{B}/\mathrm{F}}$:

\begin{equation}
{\chi _{I,\mathrm{B}}}\sigma _{\mathrm{B}}^{4}-\frac{{{\hbar ^{2}}}}{{{m_{%
\mathrm{B}}}}}-\frac{{{g_{B}}}}{{2\pi }}{n_{1\mathrm{D,B}}}=0,  \label{E-17}
\end{equation}
\begin{equation}
{\chi _{I,\mathrm{F}}}\sigma _{B}^{4}-\frac{2}{5}\frac{{{\hbar ^{2}}}}{{{m_{%
\mathrm{F}}}}}\xi {C_{\mathrm{F},1\mathrm{D}}}n_{1\mathrm{D},\mathrm{F}%
}^{2/3}\sigma _{\mathrm{F}}^{2/3}-\frac{\hbar ^{2}}{\lambda _{2}m_{\mathrm{F}%
}}-\frac{g_{\mathrm{F}}}{{2\pi }}{n_{1\mathrm{D},\mathrm{F}}}=0,
\label{E-18}
\end{equation}%
where $\chi _{I,\mathrm{B}/\mathrm{F}}\equiv m_{\mathrm{B}/\mathrm{F}}\omega
_{t,\mathrm{B}/\mathrm{F}}^{2}-2g_{\mathrm{BF}}n_{1\mathrm{D},\mathrm{F}/%
\mathrm{B}}/[\pi (\sigma _{\mathrm{B}}^{2}+\sigma _{\mathrm{F}}^{2})^{2}]$.

\begin{figure}[tbp]
\centering
\resizebox{0.8\textwidth}{!}{\includegraphics{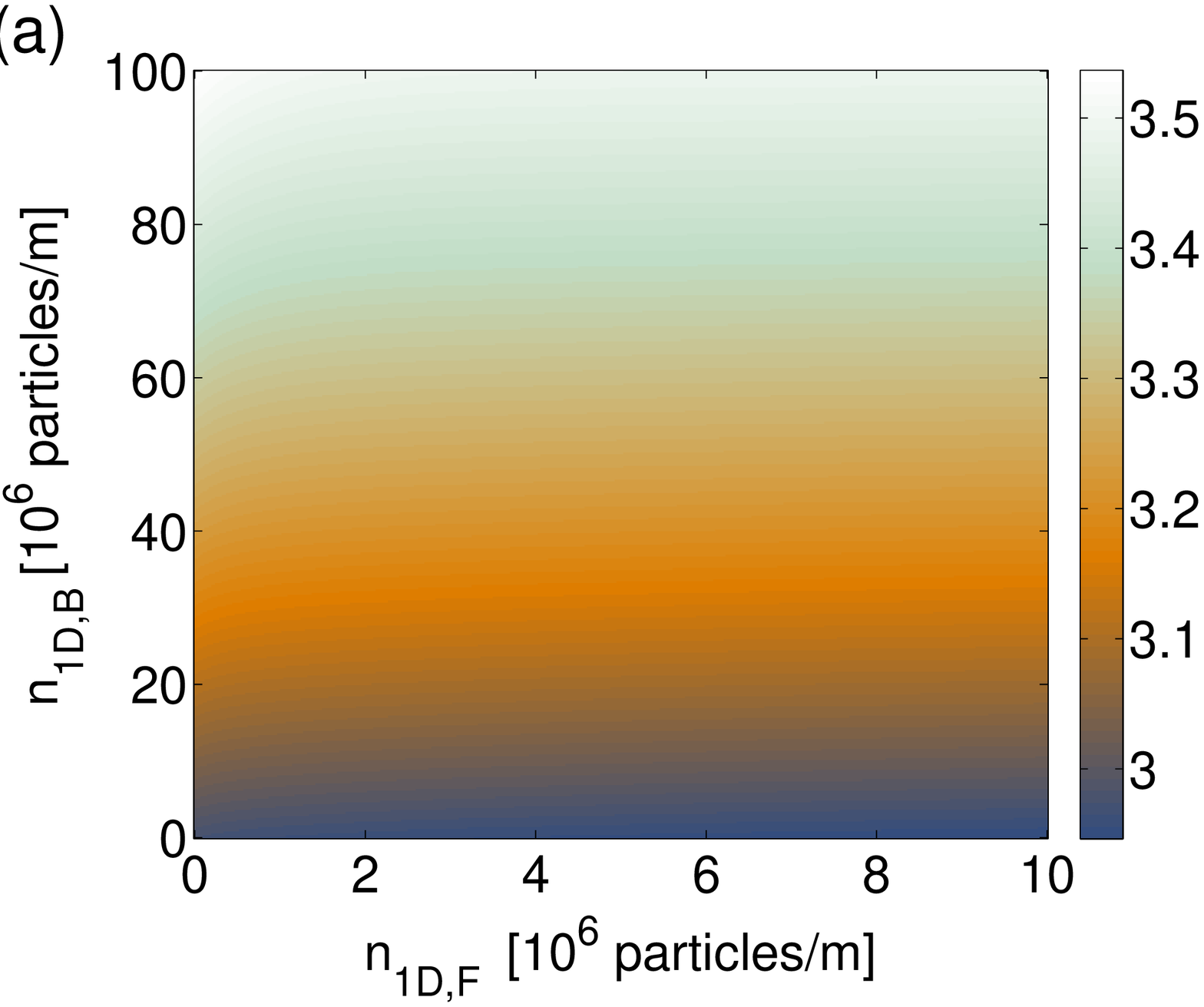}
\includegraphics{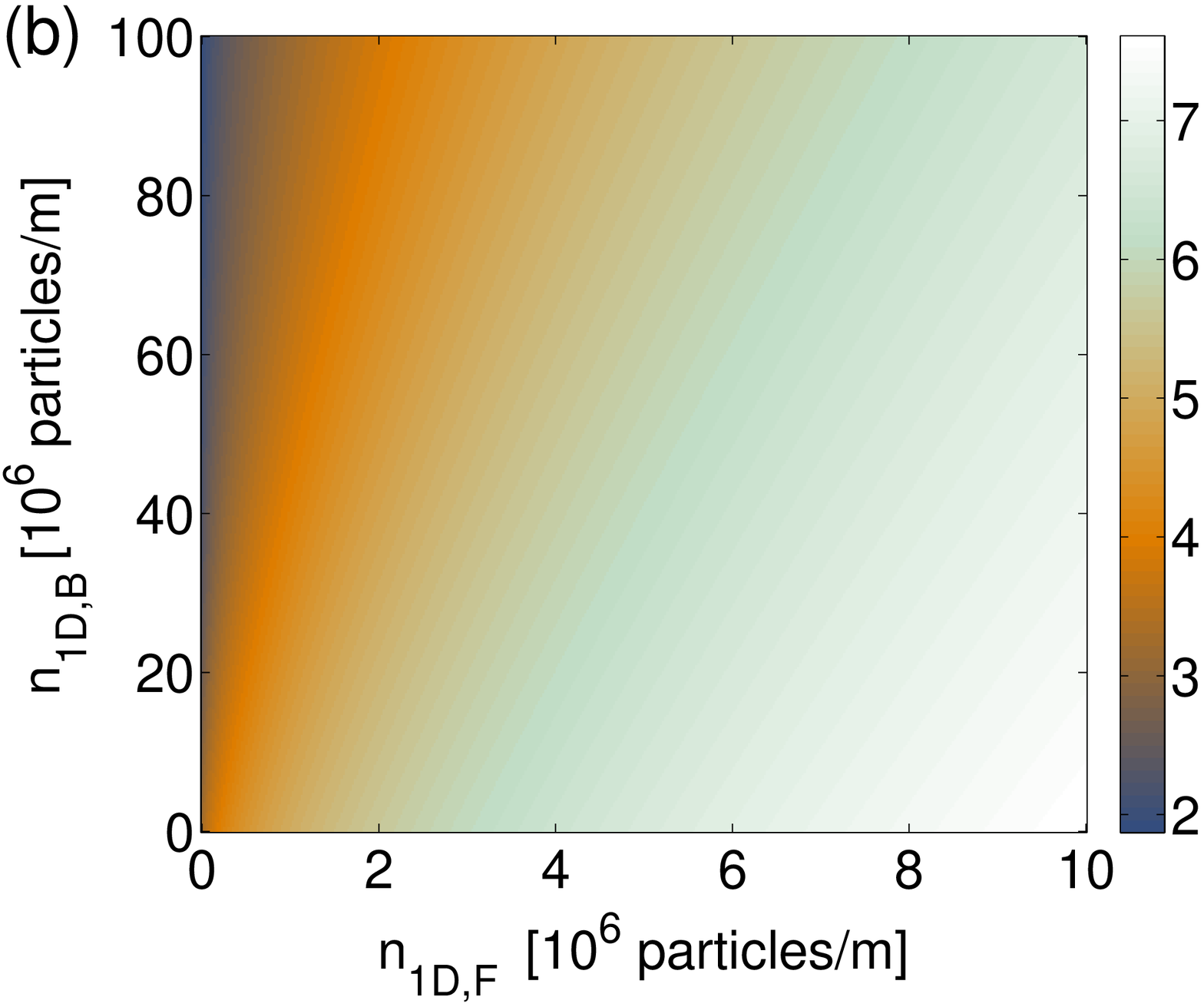}}
\resizebox{0.8\textwidth}{!}{\includegraphics{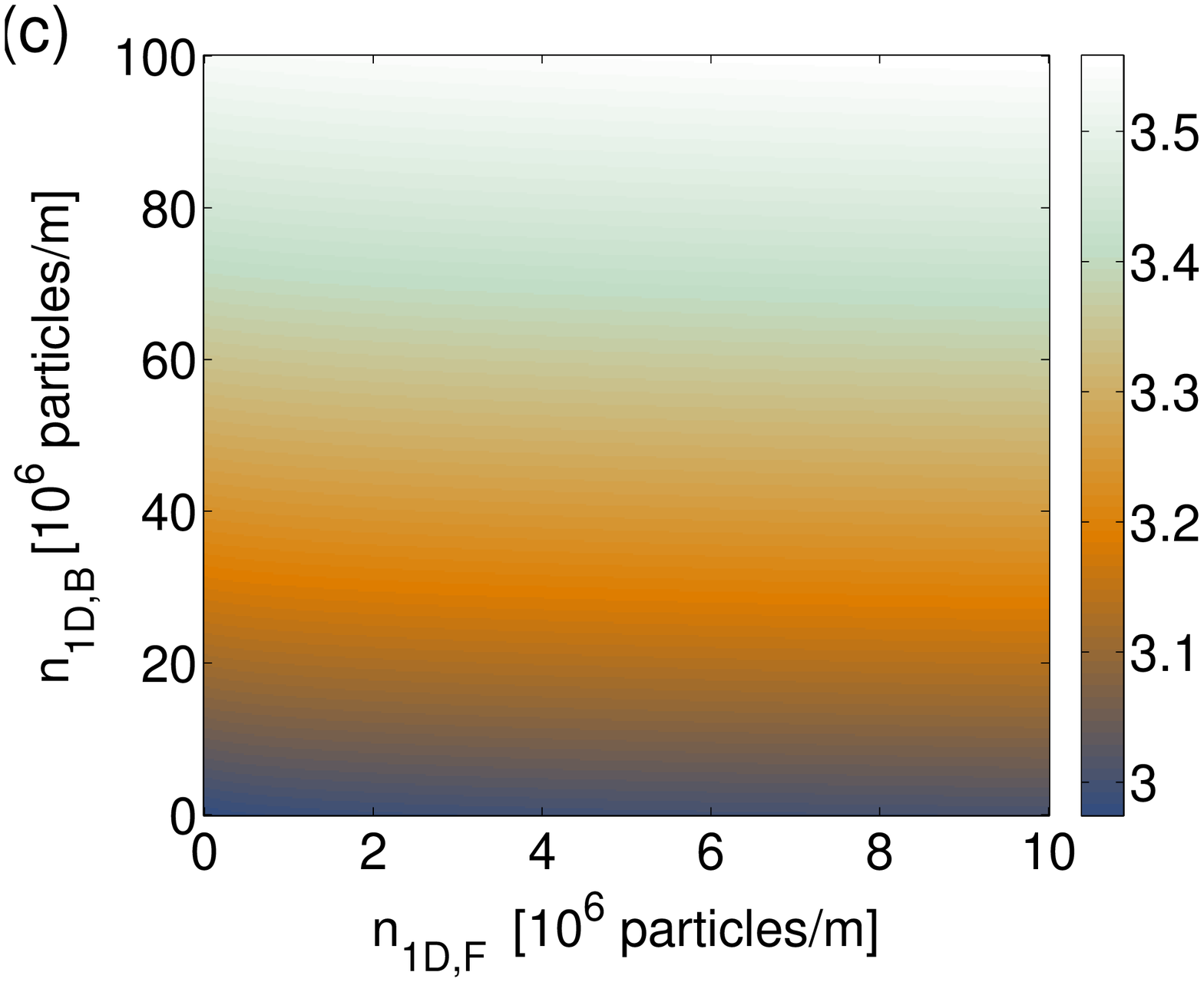}
\includegraphics{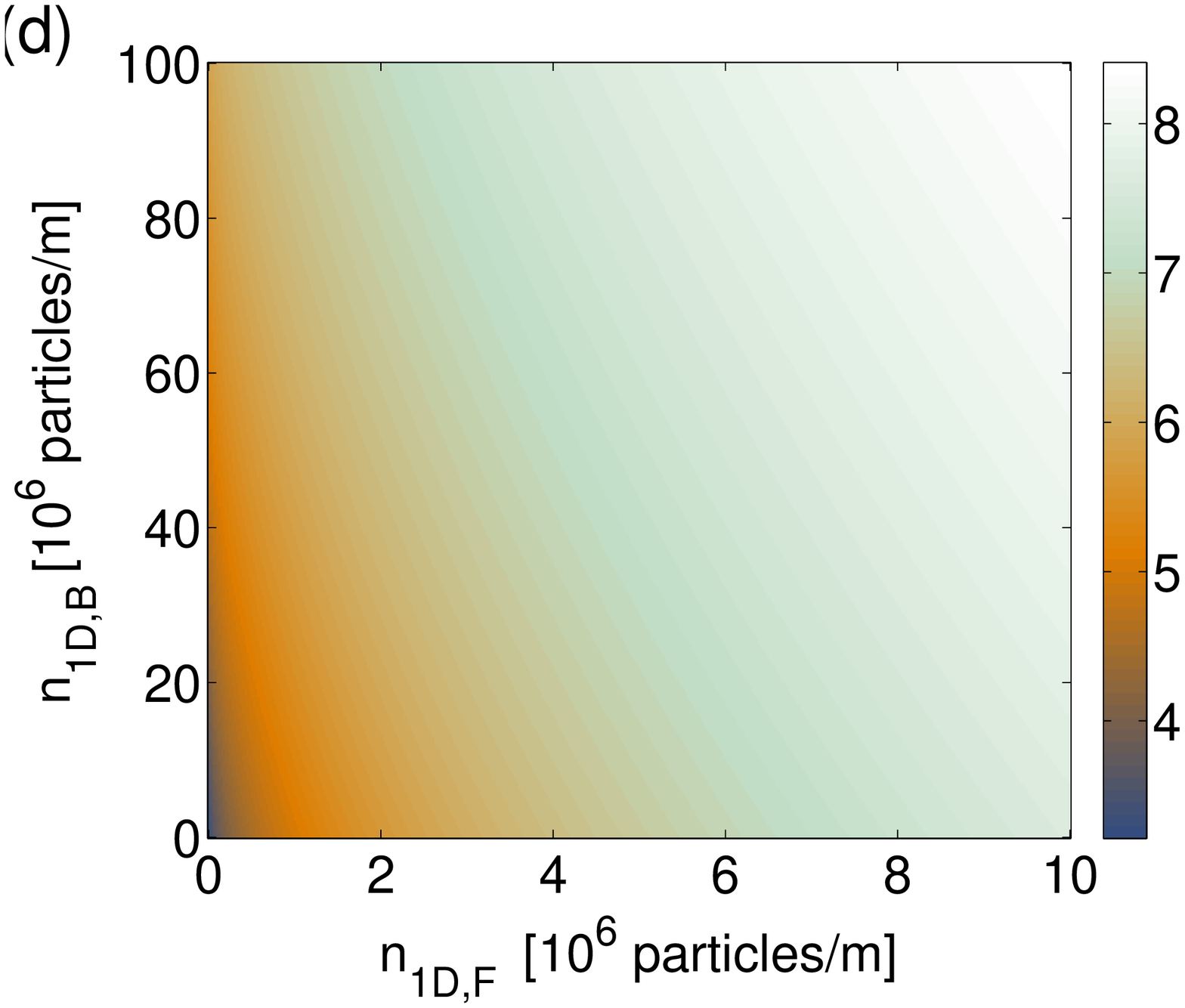}
}
\caption{(Color online) Color-coded charts for the transverse widths of the
bosonic and fermionic species, $\protect\sigma _{\mathrm{B}}$ and $\protect%
\sigma _{\mathrm{F}}$ (measured in units of ${\rm \mu} m$),
as functions of the bosonic and fermionic 1D
densities, $n_{1\mathrm{D,B/F}}$. (a) $\protect\sigma _{\mathrm{B}}$ and (b)
$\protect\sigma _{F}$ for $a_{\mathrm{BF}}=-25$ nm. (c) $\protect\sigma _{%
\mathrm{B}}$ and (d) $\protect\sigma _{\mathrm{F}}$ for $a_{\mathrm{BF}}=25$
nm. The other parameters are $a_{\mathrm{B/F}}=5$ nm and $\protect\omega _{t,%
\mathrm{B/F}}=1000$ Hz. Only fully polarized fermions are considered here.}
\label{fig:1}
\end{figure}

Thus, Eqs.~(\ref{E-15})-(~\ref{E-18}) constitute a system of four 1D coupled
equations produced by the reduction of the underlying 3D system (\ref{E-5}), (%
\ref{E-6}). A numerical solution of Eqs. (\ref{E-17}) and
(\ref{E-18}) by means
of the Newton's method produces the dependences of the transverse widths $%
\sigma _{\mathrm{B}}$ and $\sigma _{\mathrm{F}}$ on the 1D densities, $n_{1%
\mathrm{D,B}}$ and $n_{1\mathrm{D},\mathrm{F}}$, which are shown,
respectively, in the left and right columns of Fig.~\ref{fig:1}, for the BFM
with the attractive and repulsive interactions (the top and bottom rows,
respectively). In all the cases, the Fermi species has, generally, a greater
transverse width than its Bose counterpart.

We consider for the bosonic density a range greater than for its fermionic
counterpart because our calculations correspond to BFMs in which the number
of bosons exceeds the number of fermions. For the attractive mixture ($a_{%
\mathrm{BF}}=-25$ nm), it is observed that, for the bosons (see Fig. \ref%
{fig:1}(a)) the width, $\sigma _{\mathrm{B}}$, slightly increases with $n_{1%
\mathrm{D,B}}$ and $n_{1\mathrm{D},\mathrm{F}}$. For the fermions (see Fig. %
\ref{fig:1}(b)), the width, $\sigma _{\mathrm{F}}$, strongly increases with $%
n_{1\mathrm{D},\mathrm{F}}$, due to the Pauli principle, and decreases with $%
n_{1\mathrm{D,B}}$, because of the attractive boson-fermion interaction. In
the repulsive mixture ($a_{\mathrm{BF}}=25$ nm), the situation for bosons is
almost the same as in the case of the attraction, with the difference that $%
\sigma _{\mathrm{B}}$ slightly decreases with the fermionic density, $n_{1%
\mathrm{D},\mathrm{F}}$. For the fermions (see Fig. \ref{fig:1}(d)) the
increase in the boson density ($n_{1\mathrm{D,B}}$) generates a significant
increase in $\sigma _{\mathrm{F}}$ when the fermionic density ($n_{1\mathrm{D%
},\mathrm{F}}$) is low. These four diagrams demonstrate the significant
dependence of the widths of both species on the densities,\ thus showing the
importance of treating $\sigma _{\mathrm{B}}$ and $\sigma _{\mathrm{F}}$ as
the variational variables.

\begin{figure}[tbp]
\centering
\resizebox{0.8\textwidth}{!}{\includegraphics{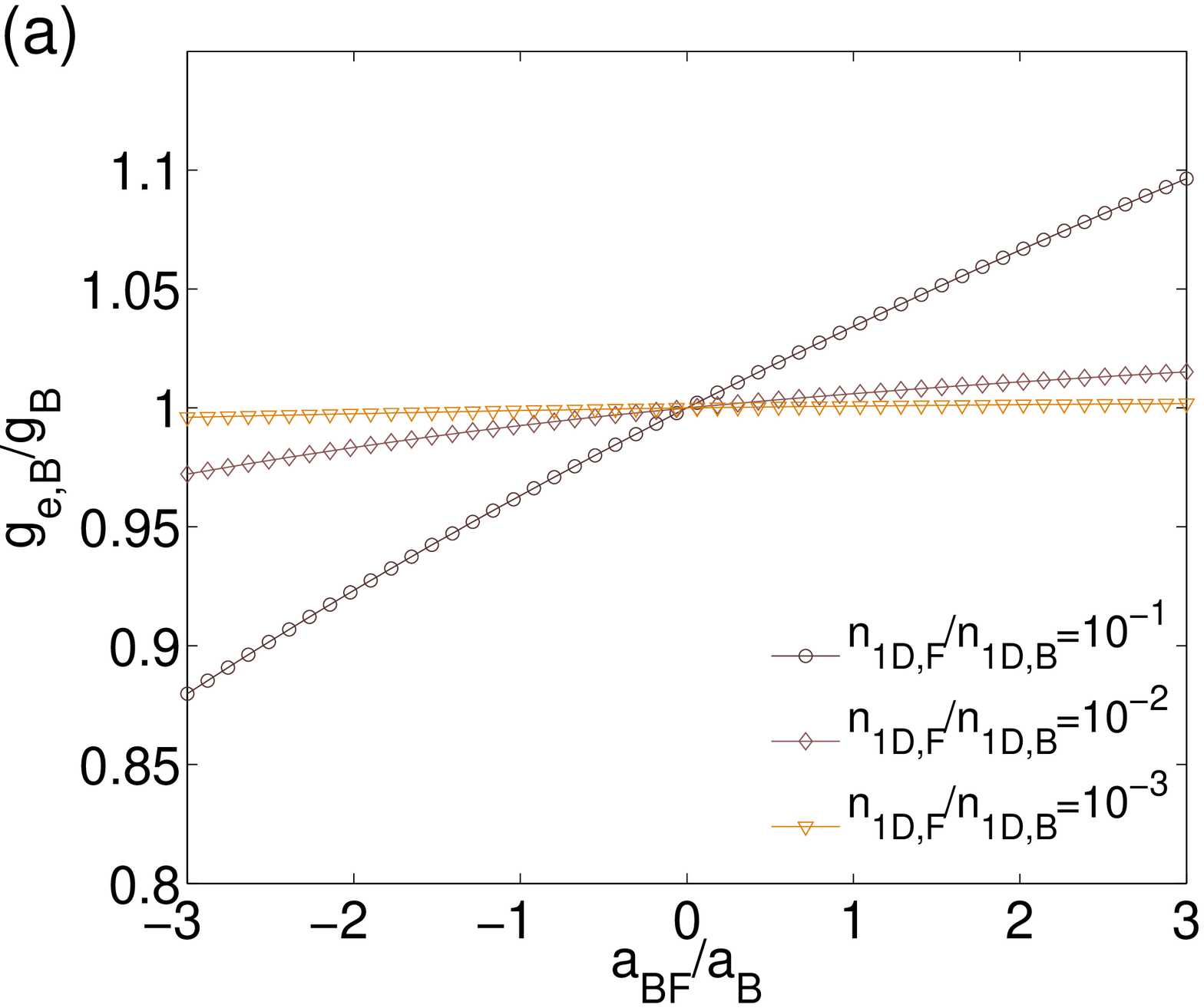}
\includegraphics{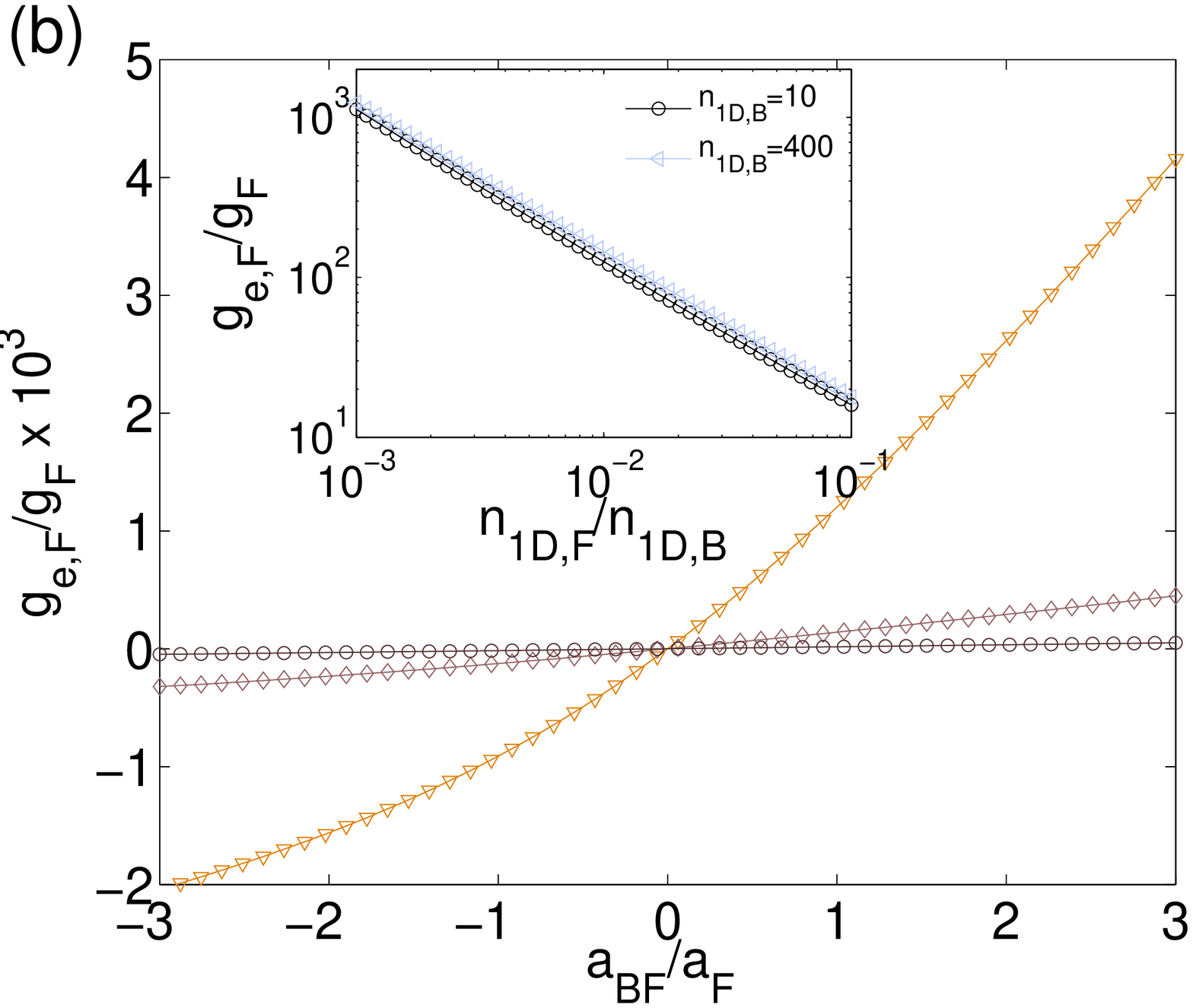}
}
\caption{(Color online) The variation of the effective scattering
parameters: (a) $g_{e,\mathrm{B}}/g_{\mathrm{B}}$ and (b) $g_{e,\mathrm{F}%
}/g_{\mathrm{F}}$ as a function of the interspecies scattering length $a_{%
\mathrm{BF}}/a_{\mathrm{B}}$ for three values of $n_{1\mathrm{D,F}}/n_{1%
\mathrm{D,B}}$. The parameters are $\protect\omega _{t,\mathrm{B/F}}=1000$
Hz, $a_{\mathrm{B/F}}=5$ nm and $n_{1\mathrm{D},\mathrm{B}}=100$. The inset
of the figure corresponds to $g_{\mathrm{BF}}/g_{\mathrm{B/F}}=1$ for two
values of $n_{1\mathrm{D,B}}$.}
\label{fig:2a}
\end{figure}

The above results can be explained by the dependence of the interaction
between the species on their densities. To do this, we define a single
scattering parameter,

\begin{equation}
{g_{e,\mathrm{B}}}={g_{\mathrm{B}}}\left[ {1+2\frac{{{g_{\mathrm{BF}}}}}{{{%
g_{\mathrm{B}}}}}\frac{{{n_{1\mathrm{D},\mathrm{F}}}}}{{{n_{1\mathrm{D},B}}}}%
{{\left( {1+{{\left( {\frac{{{\sigma _{\mathrm{F}}}}}{{{\sigma _{\mathrm{B}}}%
}}}\right) }^{2}}}\right) }^{-1}}}\right] ,  \label{E-19}
\end{equation}%
and its counterpart for the fermions, obtained by replacing subscript $%
\mathrm{B}$ with $\mathrm{F}$. Making use of the so defined interaction parameter \ref{E-19},
in Eqs. \ref{E-15} and \ref{E-16},
these equations assume the form given by

\begin{eqnarray}
 i\hbar {\partial _t}{f_{\rm{B}}} = \left[ { - \frac{{{\hbar ^2}}}{{2{m_{\rm{B}}}}}\partial _Z^2 + {V_{1{\rm{D}},{\rm{B}}}} + \frac{{{g_{{\rm{e}}{\rm{,B}}}}}}{{2\pi \sigma _{\rm{B}}^2}}{{\left| {{f_{\rm{B}}}} \right|}^2} + \frac{{{\hbar ^2}}}{{2{m_B}\sigma _{\rm{B}}^2}} + \frac{1}{2}{m_B}\omega _{t,{\rm{B}}}^2\sigma _{\rm{B}}^2} \right]{f_{\rm{B}}},  \label{E-19a}
\end{eqnarray}
and
\begin{eqnarray}
\nonumber i\frac{\hbar }{{{\lambda _1}}}{\partial _t}{f_{\rm{F}}}&=&\left[ { - \frac{{{\hbar ^2}}}{{2{\lambda _2}{m_{\rm{F}}}}}\partial _Z^2 + {V_{1D,{\rm{F}}}} + \frac{{{g_{{\rm{e}}{\rm{,F}}}}}}{{2\pi \sigma _{\rm{F}}^2}}{{\left| {{f_{\rm{F}}}} \right|}^2} + \frac{{{\hbar ^2}}}{{2{m_{\rm{F}}}{\lambda _2}\sigma _{\rm{F}}^2}} + \frac{1}{2}{m_{\rm{F}}}\omega _{t,{\rm{F}}}^2\sigma _{\rm{F}}^2} \right.\\
&&+ \left. {\frac{{{\hbar ^2}\xi }}{{2{m_{\rm{F}}}}}\frac{{{C_{{\rm{F}},1{\rm{D}}}}}}{{\sigma _{\rm{F}}^{4/3}}}{{\left| {{f_{\rm{F}}}} \right|}^{4/3}}} \right]{f_{\rm{F}}}.  \label{E-19b}
\end{eqnarray}%
where the Eqs. \ref{E-19a} and \ref{E-19b} represented the motion equations for bosons and fermions separately, with effective scattering parameter $g_{\rm{e,B}}$ and $g_{\rm{e,F}}$ respectively. Figure \ref{fig:2a} shows the variation of the effective scattering
parameters, $g_{e,\mathrm{B}}$ and $g_{e,\mathrm{F}}$, with the
inter-species scattering length, $a_{\mathrm{BF}}$. Figure \ref{fig:2a}(a)
shows that the interaction between the bosons is suppressed when the
interaction is attractive, and increases when it is repulsive. When the
fermionic density is much lower that the bosonic density (circles), the
influence of the variation of the interaction between the species on the
effective interaction strength is minimal. Figure \ref{fig:2a}(b) shows the
variation of the effective scattering parameter of the fermions ($g_{e,%
\mathrm{F}}$) under the same conditions as in Fig. \ref{fig:2a}(a),
demonstrating the strong influence of the inter-species interaction on the
fermions, due to the higher bosonic density. It is worthy to note that, for
the range of $a_{\mathrm{BF}}$ presented in Fig. \ref{fig:2a}(b), $g_{e,%
\mathrm{F}}$ is about three orders of magnitude higher than $g_{\mathrm{F}}$%
. The inset in Fig. \ref{fig:2a}(b) additionally shows the dependence of $%
g_{e,\mathrm{F}}$ on the fermionic density $n_{1\mathrm{D},\mathrm{F}}$
(keeping $n_{1\mathrm{D},\mathrm{B}}$ fixed). A much larger
number of bosons than that of fermions (the situation that that we address
in this work), in the case of an attractive mixture, implies that the fermionic profile will be close
to that of the bosonic gas. Accordingly, the Gaussian ansatz is appropriate for in both species,
which is confirmed below by the comparison between
3D numerical and variational solutions.

\subsection{The two-dimensional variational system}

\label{sec-2-2}

To derive 2D equations for the disc-shaped gas, the shape of the confinement
potential is taken as (cf. Eq. (\ref{E-7}) for the cigar-shaped
configuration):

\begin{equation}
{V_{\mathrm{B}/\mathrm{F}}}\left( {\mathbf{r}}\right) ={V_{2\mathrm{D},%
\mathrm{B}/\mathrm{F}}}\left( {x,y}\right) +\frac{1}{2}{m_{\mathrm{B}/%
\mathrm{F}}}\omega _{z,\mathrm{B}/\mathrm{F}}^{2}{z^{2}},  \label{E-20}
\end{equation}%
where the second term corresponds to the strong harmonic trap acting along
the $z$ direction. The corresponding factorized ansatz is adopted as ${\Psi
_{\mathrm{B}/\mathrm{F}}}={\phi _{\mathrm{B}/\mathrm{F}}}\left( {x,y}\right)
{u_{\mathrm{B}/\mathrm{F}}}\left( {z;\xi _{\mathrm{B}/\mathrm{F}}\left( {%
x,y,t}\right) }\right) $, where $u_{\mathrm{B}/\mathrm{F}}$ and $\phi _{%
\mathrm{B}/\mathrm{F}}$ are normalized to $1$ and $N_{\mathrm{B}/\mathrm{F}}$%
, respectively, and $u_{\mathrm{B}/\mathrm{F}}$ is represented by the
Gaussian wave function,

\begin{equation}
{u_{\mathrm{B}/\mathrm{F}}}\left( {z;{\xi _{\mathrm{B}/\mathrm{F}}}}\right) =%
\frac{1}{{{\pi ^{1/4}}\xi _{\mathrm{B}/\mathrm{F}}^{1/2}}}{\exp }\left( -%
\frac{{{z^{2}}}}{{2\xi _{\mathrm{B}/\mathrm{F}}^{2}}}\right) ,  \label{E-21}
\end{equation}%
${\xi _{\mathrm{B}/\mathrm{F}}}\left( {x,y,t}\right) $ being the widths of
the gas in the confined direction. Substituting the factorized ansatz (\ref%
{E-21}) into action (\ref{E-1}) and integrating over $z$, we arrive at the
following expression for the effective 2D action:

\begin{equation}
\mathcal{S}=\int {dtdxdy\left( \mathcal{L}_{2D\mathrm{,B}}+\mathcal{L}_{2%
\mathrm{D},\mathrm{F}}+\mathcal{L}_{2\mathrm{D,BF}}\right) },  \label{E-22}
\end{equation}

\begin{equation}
\mathcal{L}_{2\mathrm{D,B}}=i\frac{\hbar }{2}\left( {\phi _{\mathrm{B}%
}^{\ast }{\partial _{t}}{\phi _{\mathrm{B}B}}-{\phi _{\mathrm{B}}}{\partial
_{t}}\phi _{\mathrm{B}}^{\ast }}\right) -{V_{2D\mathrm{,B}}}{n_{2\mathrm{D,B}%
}}-{e_{2\mathrm{D,B}}},
\end{equation}

\begin{equation}
{\mathcal{L}_{2\mathrm{D},\mathrm{F}}}=i\frac{\hbar }{{2{\lambda _{1}}}}%
\left( {\phi _{\mathrm{F}}^{\ast }{\partial _{t}}{\phi _{\mathrm{F}}}-{\phi
_{\mathrm{F}}}{\partial _{t}}\phi _{\mathrm{F}}^{\ast }}\right) -{V_{2%
\mathrm{D},\mathrm{F}}}{n_{2\mathrm{D},\mathrm{F}}}-{e_{2\mathrm{D},\mathrm{F%
}}},
\end{equation}%
\begin{equation}
{\mathcal{L}_{2\mathrm{D,BF}}}=-\frac{1}{{{\pi ^{1/2}}}}\frac{{{g_{B\mathrm{F%
}}}}}{\sqrt{\xi _{\mathrm{B}}^{2}+\xi _{\mathrm{F}}^{2}}}{n_{2\mathrm{D},%
\mathrm{B}}}{n_{2\mathrm{D},\mathrm{F}}},  \label{E-25}
\end{equation}%
where $n_{2\mathrm{D,B/F}}\equiv \left\vert {\phi _{\mathrm{B}/\mathrm{F}}}%
\left( {x,y}\right) \right\vert ^{2}$ are the 2D particle densities of the
boson and fermion species, and ${e_{2\mathrm{D,B}}}$ and ${e_{2\mathrm{D},%
\mathrm{F}}}$ are their energy densities:

\begin{equation}
{e_{2\mathrm{D,B}}}=\frac{{{\hbar ^{2}}}}{{2{m_{\mathrm{B}}}}}{\left\vert {{%
\nabla _{\bot }}{\phi _{\mathrm{B}}}}\right\vert ^{2}}+\left[ {\frac{{{g_{%
\mathrm{B}}}}}{{{  \sqrt{8\pi }   }{\xi _{\mathrm{B}}}}}{n}_{2%
\mathrm{D,B}}+\frac{{{\hbar ^{2}}}}{{4{m_{\mathrm{B}}}\xi _{\mathrm{B}}^{2}}}%
}\right.
+\left. {\frac{1}{4}{m_{\mathrm{B}}}\omega _{z,\mathrm{B}}^{2}\xi _{\mathrm{B%
}}^{2}}\right] {n_{2\mathrm{D},\mathrm{B}}},  \label{E-26}
\end{equation}%
\begin{eqnarray}
\nonumber {e_{2\mathrm{D},\mathrm{F}}}&=&\frac{{{\hbar ^{2}}}}{{2{\lambda _{2}}{m_{%
\mathrm{F}}}}}{\left\vert {{\nabla _{\bot }}{\phi _{\mathrm{F}}}}\right\vert
^{2}}+\left[ {\frac{{{g_{\mathrm{F}}}}}{{{ \sqrt{8\pi } }{\xi _{\mathrm{F}}}}}{n_{2%
\mathrm{D},\mathrm{F}}}+\frac{{{\hbar ^{2}}}}{{4\lambda _{2}{m_{\mathrm{F}}}%
\xi _{\mathrm{F}}^{2}}}+\frac{1}{4}{m_{\mathrm{F}}}\omega _{z,\mathrm{F}%
}^{2}\xi _{\mathrm{F}}^{2}}\right. \\
&&+\left.{\frac{{{\hbar ^{2}}}}{{2{m_{\mathrm{F}}}}}\xi \frac{3}{{5\xi _{%
\mathrm{F}}^{2/3}}}{C_{2\mathrm{D},\mathrm{F}}}n_{2\mathrm{D},\mathrm{F}%
}^{2/3}} \right] {n_{2\mathrm{D},\mathrm{F}}},
\label{E-27}
\end{eqnarray}%
with $C_{2\mathrm{D},\mathrm{F}}\equiv {(3/5)^{1/2}}{(6/(2s_{\mathrm{F}%
}+1))^{2/3}}\pi $ in Eq. (\ref{E-27}).

The motion equations of the 2D system are obtained by the variation of the
action $S$ given by Eq. (\ref{E-22}) with respect to variables $\phi _{%
\mathrm{B}}$ and $\phi _{\mathrm{F}}$:

\begin{eqnarray}
\nonumber i\hbar {\partial _{t}}{\phi _{\mathrm{B}}}&=&\left[ {-\frac{{{\hbar ^{2}}}}{{2{%
m_{B}}}}\nabla _{\bot }^{2}+{V_{2\mathrm{D,B}}}+\frac{1}{{{\pi ^{1/2}}}}%
\frac{{{g_{\mathrm{BF}}}}}{\sqrt{\xi _{\mathrm{B}}^{2}+\xi _{\mathrm{F}}^{2}}%
}{n_{2\mathrm{D},\mathrm{F}}}}+\frac{{{g_{\mathrm{B}}}}}{{\sqrt{2\pi }{\xi _{\mathrm{B}}}}}{{%
\left\vert {{\phi _{\mathrm{B}}}}\right\vert }^{2}}\right. \\
&&+\left.\frac{{{\hbar ^{2}}}}{{4{m_{\mathrm{B}}}\xi _{\mathrm{B}}^{2}}}+\frac{1}{4}%
{m_{\mathrm{B}B}}\omega_{z,\mathrm{B}}^{2}\xi _{\mathrm{B}}^{2}\right]{\phi%
_{\mathrm{B}},}\label{E-28}
\end{eqnarray}
\begin{eqnarray}
\nonumber i\frac{\hbar }{{{\lambda _{1}}}}{\partial _{t}}{\phi _{\mathrm{F}}} &=&\left[
{-\frac{{{\hbar ^{2}}}}{{2{\lambda _{2}}{m_{\mathrm{F}}}}}\nabla _{\bot
}^{2}+{V_{2D,\mathrm{F}}}+\frac{1}{{{\pi ^{1/2}}}}\frac{{{g_{\mathrm{BF}}}}}{%
\sqrt{\xi _{\mathrm{B}}^{2}+\xi _{\mathrm{F}}^{2}}}{n_{2\mathrm{D,B}}}}%
+\frac{{{g_{\mathrm{F}}}}}{{\sqrt{2\pi }{\xi _{\mathrm{F}}%
}}}{\left\vert {{\phi _{\mathrm{F}}}}\right\vert ^{2}}\right. \\
&&+ \left. \frac{{{\hbar ^{2}}}}{{2{m_{\mathrm{F}}}}}\xi \frac{1}{{\xi _{\mathrm{F}%
}^{2/3}}}{C_{2\mathrm{D},\mathrm{F}}}{\left\vert {{\phi _{\mathrm{F}}}}%
\right\vert ^{4/3}}+{\frac{{{\hbar ^{2}}}}{{4\lambda _{2}{m_{\mathrm{F}}}\xi _{\mathrm{%
F}}^{2}}}+\frac{1}{4}{m_{\mathrm{F}}}\omega _{z,\mathrm{F}}^{2}\xi _{\mathrm{%
F}}^{2}}\right] {\phi _{\mathrm{F}}.}  \label{E-29}
\end{eqnarray}

\begin{figure}[tbp]
\centering
\resizebox{0.8\textwidth}{!}{\includegraphics{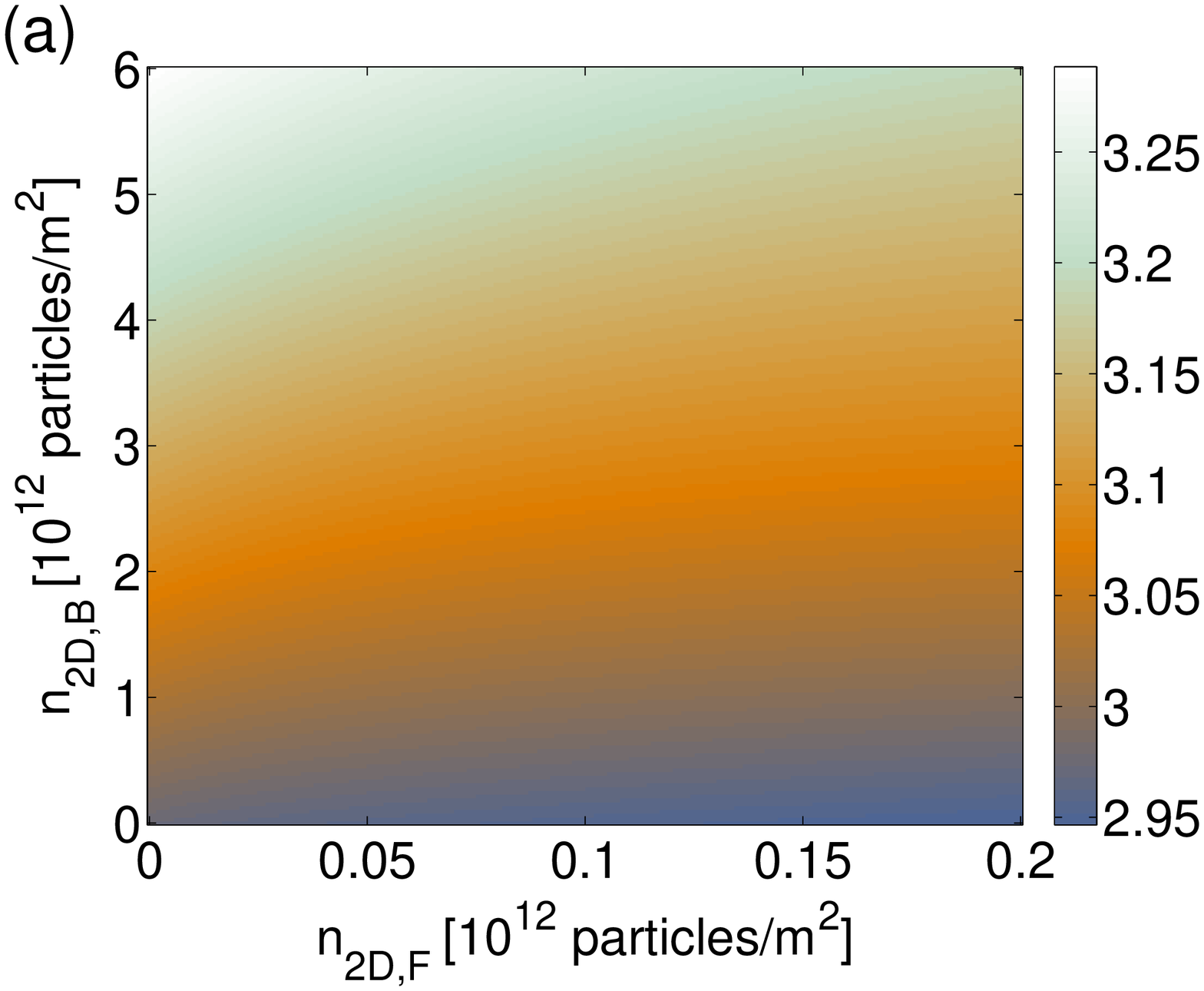}
\includegraphics{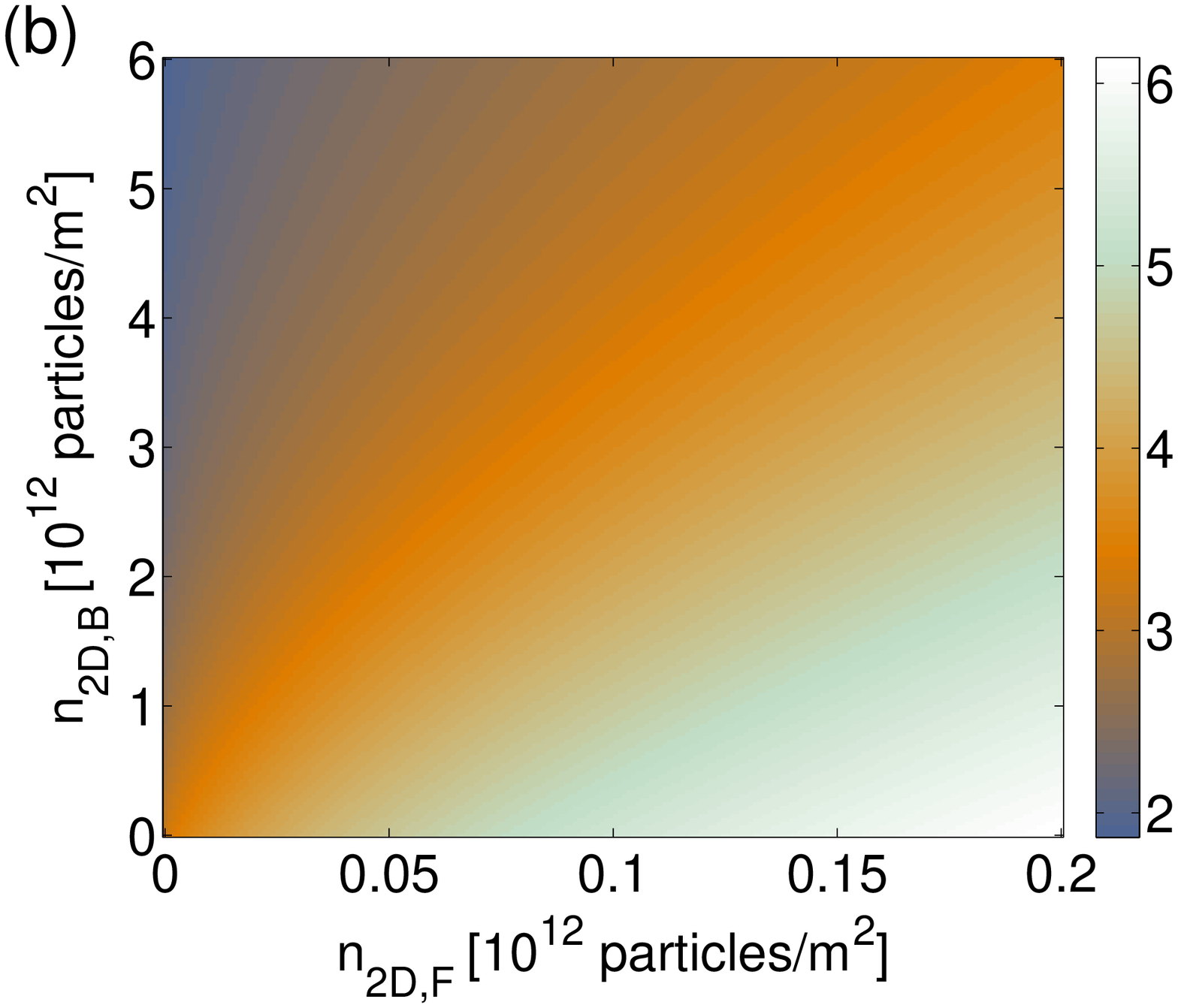}
}
\resizebox{0.8\textwidth}{!}{\includegraphics{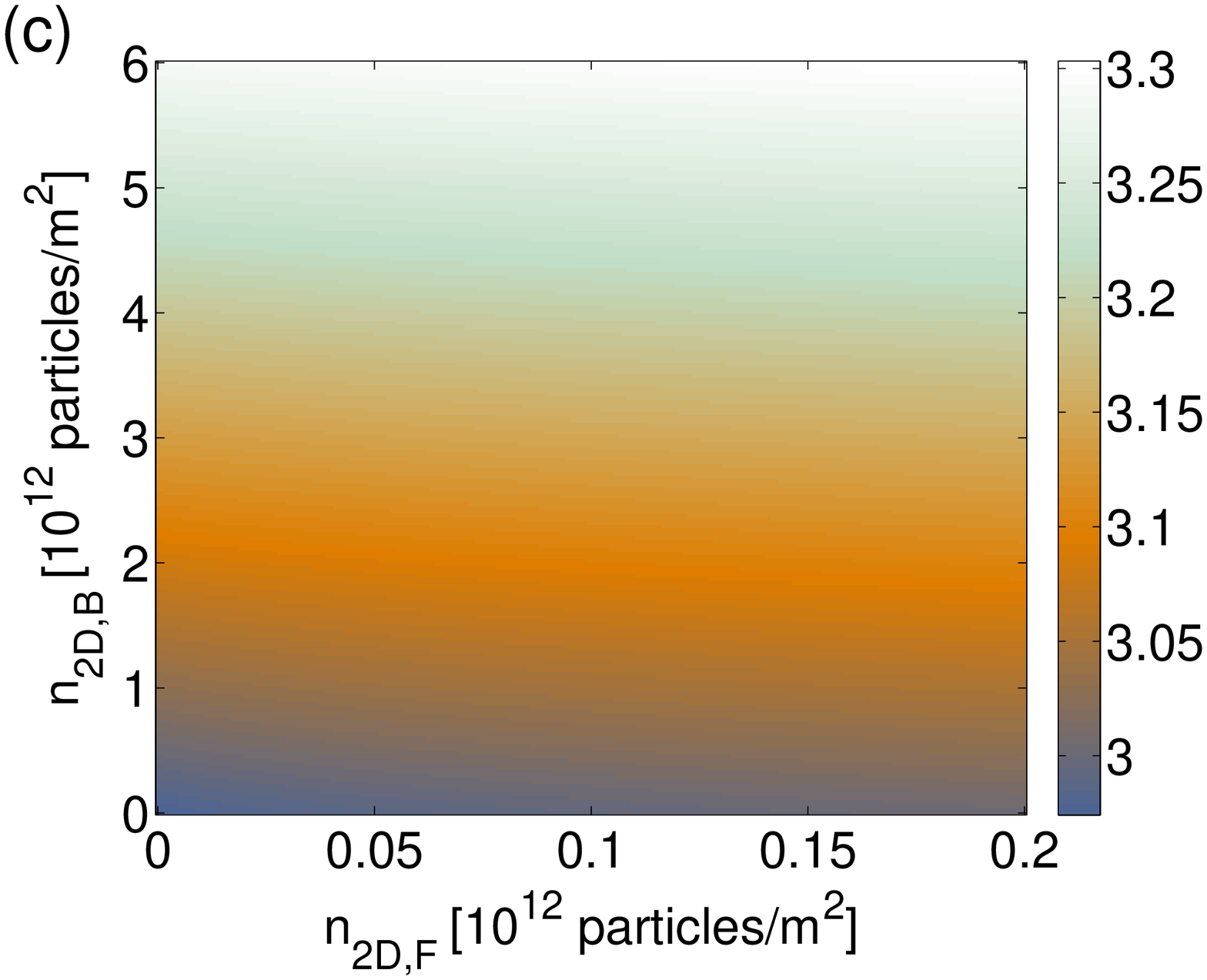}
\includegraphics{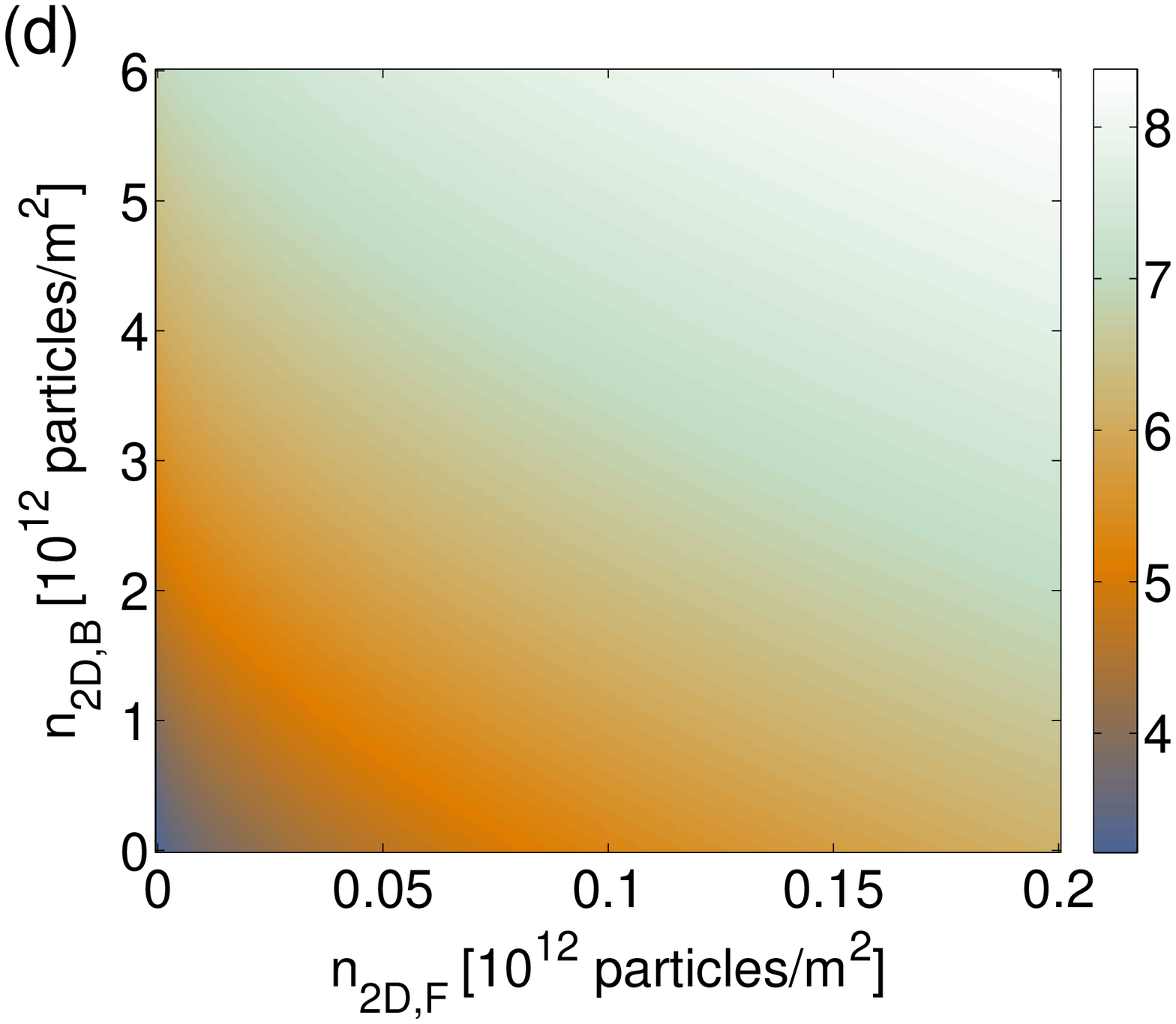}
}
\caption{(Color online) Color-coded charts for the transverse widths of the
bosonic and fermionic species, $\protect\xi _{\mathrm{B}}$ and $\protect\xi %
_{\mathrm{F}}$ (measured in units of ${\rm \mu} m$), as
functions of the bosonic and fermionic 2D densities, $n_{2%
\mathrm{D,B/F}}$. (a) $\protect\xi _{\mathrm{B}}$ and (b) $\protect\xi _{%
\mathrm{F}}$ for $a_{\mathrm{BF}}=-25$ nm. (c) $\protect\xi _{\mathrm{B}}$
and (d) $\protect\xi _{\mathrm{F}}$ for $a_{\mathrm{BF}}=25$ nm. The other
parameters are $a_{\mathrm{B/F}}=5$ nm and $\protect\omega _{z,\mathrm{B/F}%
}=1000$ Hz. Only fully polarized fermions are considered here.}
\label{fig:2}
\end{figure}

Relations between $\xi _{\mathrm{B}/\mathrm{F}}$ and $\phi _{\mathrm{B}/%
\mathrm{F}}$ are produced by the Euler-Lagrange equations associated to $\xi
_{\mathrm{B}/\mathrm{F}}$:
\begin{equation}
{\kappa _{I,B}}\xi _{\mathrm{B}}^{4}-\frac{{{g_{B}}}}{\sqrt{2\pi }}{n_{2%
\mathrm{D,B}}}{\xi _{\mathrm{B}}}-\frac{{{\hbar ^{2}}}}{{{m_{\mathrm{B}}}}}%
=0,  \label{E-30}
\end{equation}%
\begin{equation}
{\kappa _{I,\mathrm{F}}}\xi _{\mathrm{F}}^{4}-\frac{{2{\hbar ^{2}}}}{{5{m_{%
\mathrm{F}}}}}\xi {C_{2\mathrm{D},\mathrm{F}}}n_{2\mathrm{D},\mathrm{F}%
}^{2/3}\xi _{\mathrm{F}}^{4/3}-\frac{g_{\mathrm{F}}}{\sqrt{2\pi }}{n_{2%
\mathrm{D},\mathrm{F}}}{\xi _{\mathrm{F}}}
-\frac{{{\hbar ^{2}}}}{\lambda _{2}m_{\mathrm{F}}}=0  \label{E-31}
\end{equation}%
where ${\kappa _{I,\mathrm{F}}}\equiv {m_{\mathrm{F}}}\omega _{z,\mathrm{F}%
}^{2}+2{g_{\mathrm{BF}}}{n_{2\mathrm{D,B}}}/[{{\pi ^{1/2}}{{({\xi _{\mathrm{B%
}}^{2}+\xi _{\mathrm{F}}^{2}})}^{3/2}}}]$. Equations (\ref{E-30}) and (\ref%
{E-31}) for $\xi _{\mathrm{B}/\mathrm{F}}$ were solved numerically. The four
graphs in Fig.~\ref{fig:2} show the dependence of $\xi _{\mathrm{B}}$ and $%
\xi _{\mathrm{F}}$ (in the left and right columns, respectively) on $n_{2%
\mathrm{D,B/F}}$, for the attractive and repulsive mixtures (top and bottom
panels, respectively). The analysis for the attractive case can be divided
into two parts: when the boson density is high, the fermions gas tends to
compress, so that at low densities its width is found to be even less than
that of the boson gas, a situation that tends to reverse for high fermionic
densities; for low bosonic densities, the width of the boson gas rapidly
decreases with the increase of the fermionic density, until becoming nearly
constant, while, on the contrary, the width of the fermion gas increases
progressively when its density is higher. Conversely, when the mixture is
repulsive, the situation is similar to that outlined above for the 1D
setting: low densities of the mixtures cause the gas to compress, while high
densities cause the mixture to expand, this trend being much stronger for
the fermion component.

To determine the influence that one species exerts on the other in 2D, we
define the effective scattering parameters, cf. a similar definition (\ref%
{E-19}) adopted in the 1D case:

\begin{equation}
{g_{e,\mathrm{B}}}={g_{\mathrm{B}}}\left[ 1+\sqrt{2}\frac{g_{\mathrm{BF}}}{%
g_{\mathrm{B}}}\frac{n_{2\mathrm{D},\mathrm{F}}}{n_{2\mathrm{D},B}}\left(
1+\left( \frac{\xi _{\mathrm{F}}}{\xi _{\mathrm{B}}}\right) ^{2}\right)
^{-1/2}\right] ,  \label{E-32}
\end{equation}%
$g_{e,\mathrm{F}}$ being produced by replacing subscript $\mathrm{B}$ with $%
\mathrm{F}$. As shown in detail below, our calculations were performed
assuming the same order of magnitude for the number of particles for both 1D
and 2D cases (for this reason that the effective 1D density is about $10$
times the magnitude of its 2D counterpart). With regard to this, conclusions
concerning the effective interaction strength, which follow in the 1D
setting from Fig. \ref{fig:2a}, apply to 2D case as well.

\section{Numerical Results}

\label{sec-3}

Our analysis was developed for the GS and dynamics of perturbations around
it in the presence of the 3D harmonic-oscillator trap of the following form:
\begin{equation}
{V_{\mathrm{B/F}}}\left( {\mathbf{r}}\right) =\frac{1}{2}{m_{\mathrm{B/F}}}%
\left( {\omega _{x,\mathrm{B/F}}^{2}{x^{2}}+\omega _{y,\mathrm{B/F}}^{2}{%
y^{2}}+\omega _{z,\mathrm{B/F}}^{2}{z^{2}}}\right) ,  \label{E-33}
\end{equation}%
In particular, for the GS of the 1D and 2D settings we focused on
determining the spatial correlation $C_{s}$ between the spatial particle
densities in both species, defined as

\begin{equation}
{C_{s}}\left( {{\bar{n}_{B}},{\bar{n}_{\mathrm{F}}}}\right) =\frac{{%
\left\langle {{\bar{n}_{B}}{\bar{n}_{\mathrm{F}}}}\right\rangle }}{\sqrt{%
\left\langle {\bar{n}_{B}^{2}}\right\rangle \left\langle {\bar{n}_{\mathrm{F}%
}^{2}}\right\rangle }},  \label{E-34}
\end{equation}%
where ${{\bar{n}}_{\mathrm{B}/\mathrm{F}}}={n_{\mathrm{B}/\mathrm{F}}}%
-\left\langle {{n_{\mathrm{B}/\mathrm{F}}}}\right\rangle $, $\left\langle
{}\right\rangle $ standing for the spatial average. For dynamical
perturbations around the GS, a spatiotemporal correlation, which is defined
by replacing the spatial average with the spatiotemporal average, is known
as the Pearson coefficient $C_{s-t}$ \cite{Bragard04}.
The GSs in the 1D and 2D cases were found by means of the imaginary-time
integrations of Eqs.~(\ref{E-15}), (\ref{E-16}) and (\ref{E-28}),~(\ref%
{E-29}), respectively.

\subsection{Results in one dimension.}

\label{sec-3-1}

\subsubsection{Accuracy of the variational method as a function of
scattering parameter $a_{\mathrm{BF}}$ for the ground states.}

\label{sec-3-1-1}

\begin{figure}[tbp]
\centering
\resizebox{0.8\textwidth}{!}{
\includegraphics{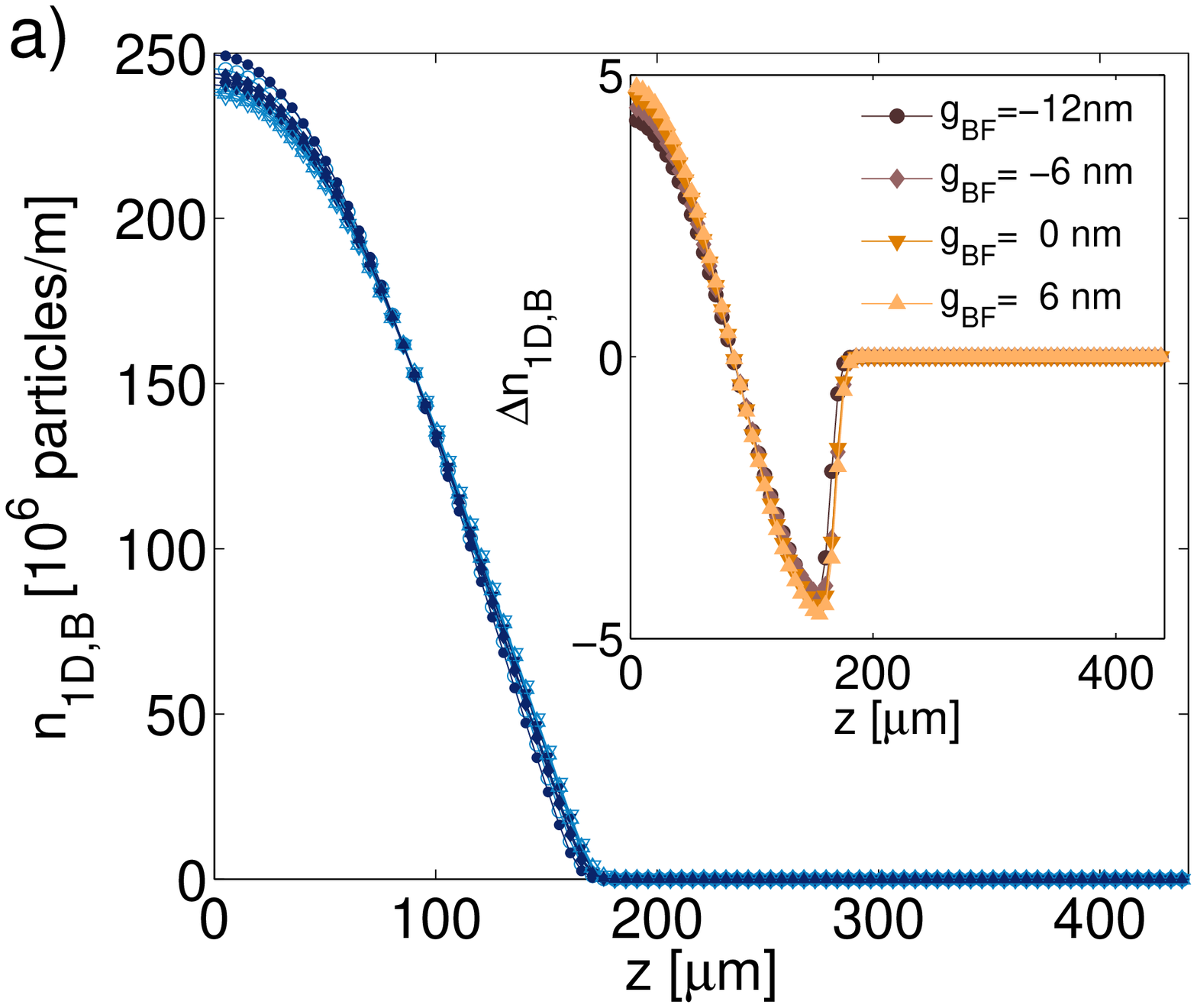}
\includegraphics{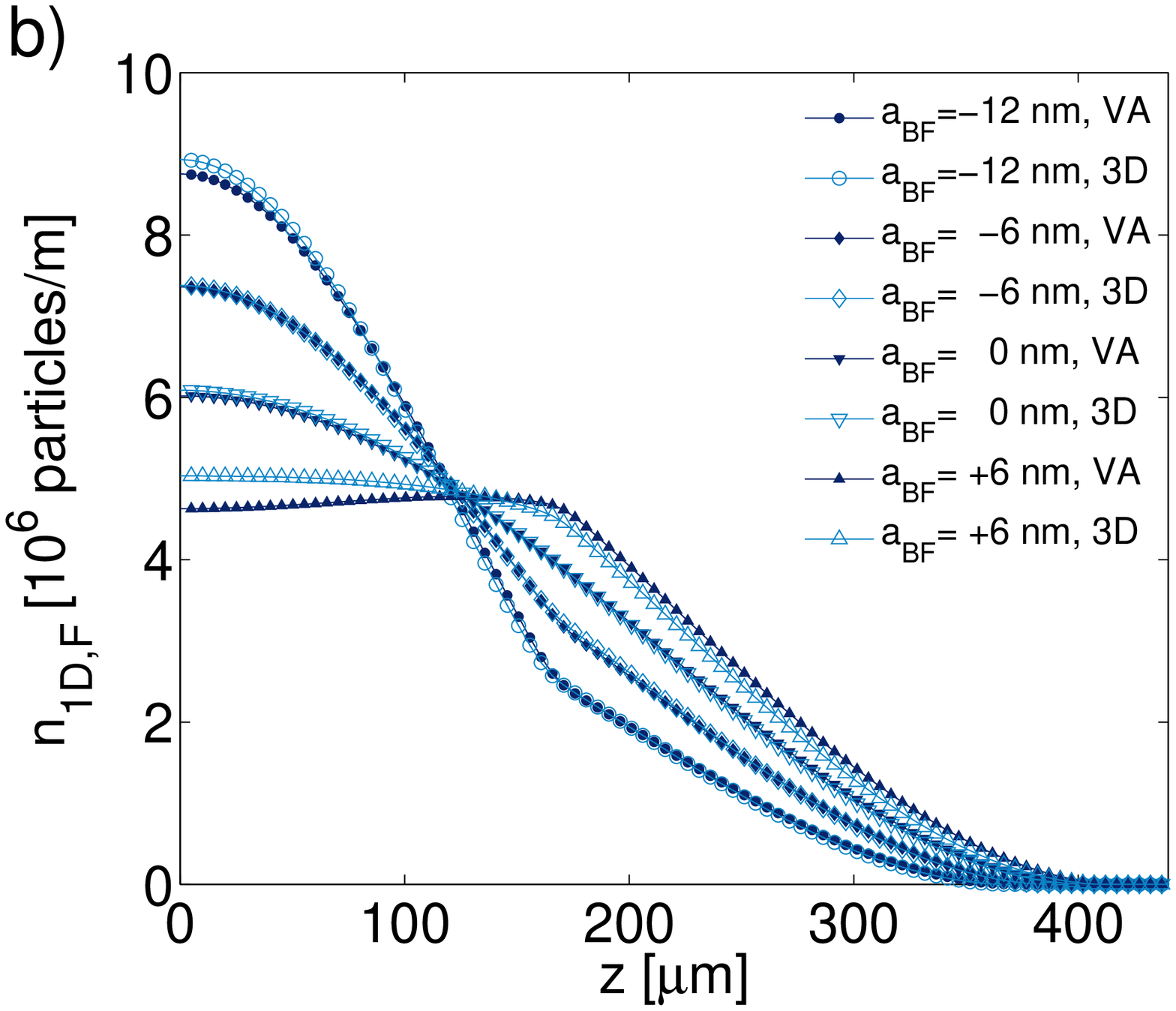}
}
\resizebox{0.8\textwidth}{!}{\includegraphics{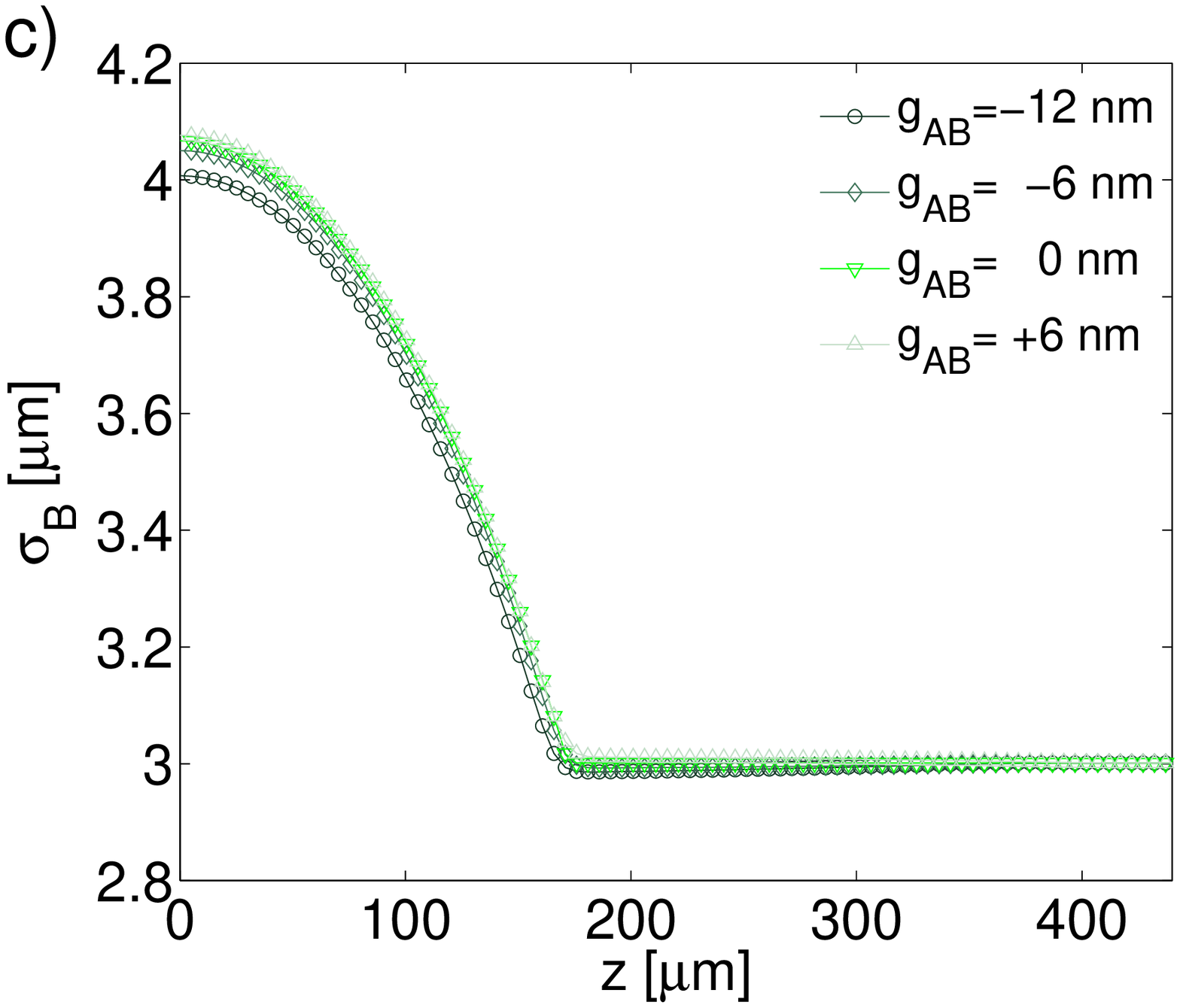}
\includegraphics{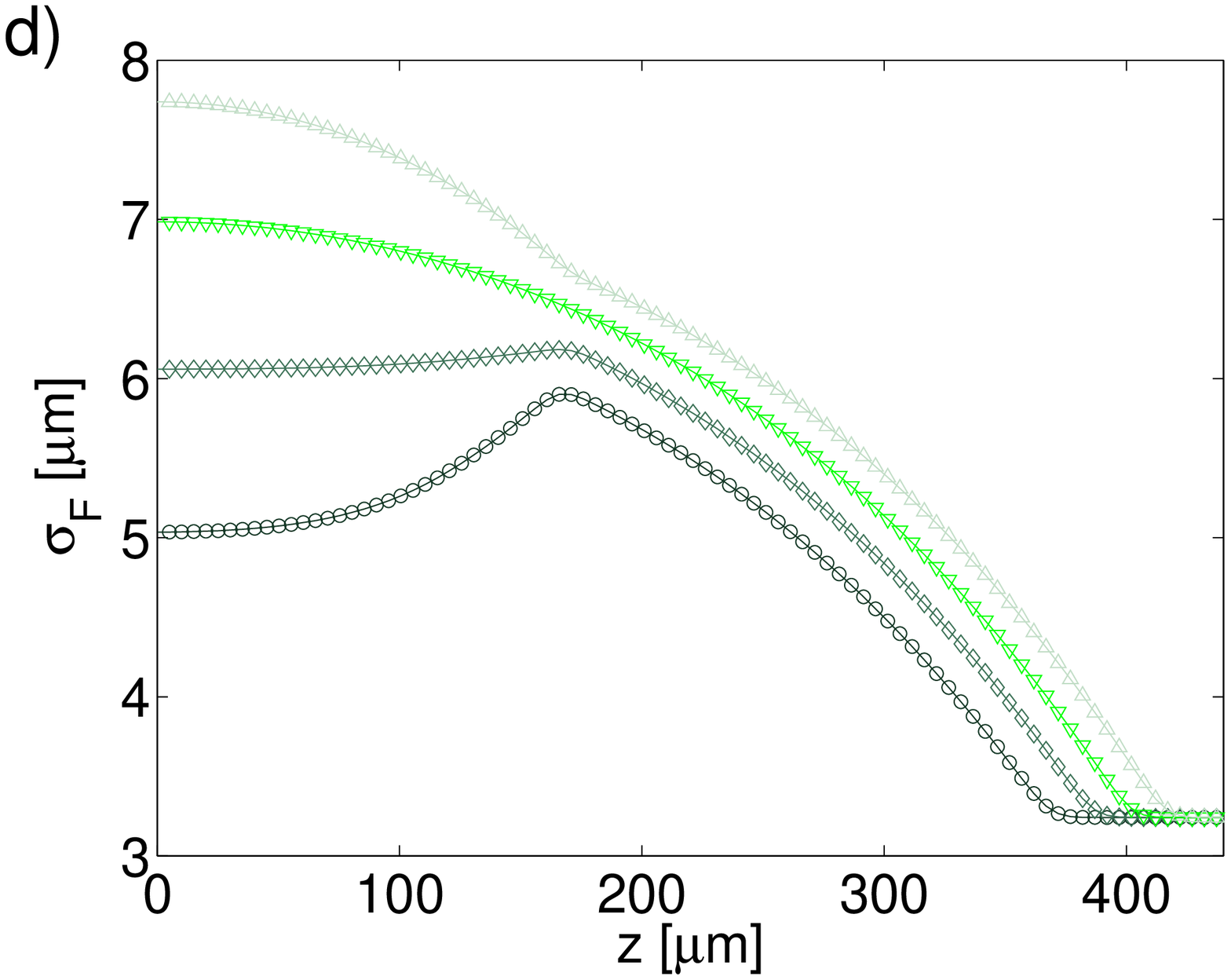}
}
\caption{(Color online) Profiles of the particle density (as produced
by the VA and 3D calculations, see inset in panel (b)), and the width in the
confined direction, for the bosonic and fermionic components, for five
values of $a_{\mathrm{BF}}$, as defined in the panels. (a) $n_{1\mathrm{D,B}%
} $, (b) $n_{1\mathrm{D,F}}$, (c) $\protect\sigma _{\mathrm{B}}$, and (d) $%
\protect\sigma _{\mathrm{F}}$. The parameters are $N_{\mathrm{B}}=5\times
10^{4}$, $N_{\mathrm{F}}=2.5\times 10^{3}$, $a_{\mathrm{B/F}}=5$ nm, $%
\protect\omega _{z,\mathrm{B/F}}=30$ Hz and $\protect\omega _{t,\mathrm{B/F}%
}=1000$ Hz. The inset in (a) shows the difference between both methods, VA and 3D,
by means of $\Delta n_{\mathrm{1D,B}}=\rho_{\mathrm{1D,B}}-n_{\mathrm{1D,B}}$.}
\label{fig:3}
\end{figure}

\begin{figure}[tbp]
\centering
\resizebox{0.5\textwidth}{!}{
\includegraphics{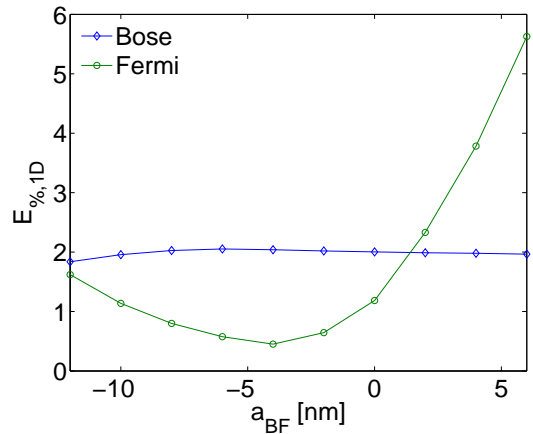}
}
\caption{(Color online) One-dimension overall percentage error for both species
(see the definition in the text). The parameters are the same as in Fig.~\ref{fig:3}}
\label{fig:5-n}
\end{figure}

As a starting point in the 1D case with free coordinate $z$, we analyze the
effect of the magnitude and sign of the interaction parameter on the spatial
profile of both species, and the accuracy of the variational
method (see Eqs.~\ref{E-15}~-~\ref{E-18}) compared to the 3D
solution, by varying the scattering length $a_{\mathrm{D,B}}$. The plots
in Fig.~\ref{fig:3} shows this situation, with panels~\ref{fig:3}(a) and ~%
\ref{fig:3}(b) corresponding to the profiles of $n_{1\mathrm{D,B}}$ and $n_{1%
\mathrm{D},\mathrm{F}}$, respectively, where the inset in
panel~\ref{fig:3}(b) specifies whether the profile was produced by the VA or
by means of the 3D solution,
while \ref{fig:3}(c) and ~\ref{fig:3}(d) correspond to the profiles of
$\sigma _{\mathrm{B}}$ and $\sigma _{\mathrm{F}}$.
The mixture with many more bosons than fermions is considered
here: $N_{\mathrm{B}}=5\times 10^{4}$, $N_{\mathrm{F}}=2.5\times 10^{3}$.
Because of this condition, the bosonic profile is mainly determined by its
self-interaction and the external potential. First, we deal
with the influence of the interspecies interaction ($g_{\mathrm{BF}}$) on
the density profile, and then address an error resulting from the variational
approach. Figure \ref{fig:3}(a) shows that variations of the bosonic density
profile are very small in comparison to the significant changes of the
inter-species scattering length. The situation is opposite for the fermionic
species. As the repulsive scattering length increases, the fermions
tend to be pushed to the periphery of the
bosonic-gas density profile. This phenomenon is known as \textit{demixing}
\cite{Adhikari08,Salasnich07a,Adhikari07,demix}. On the other hand, for the
attractive case, fermions are, naturally, concentrated in the same region
where the bosons are located. Figures Fig. \ref{fig:3}(c) and \ref{fig:3}(d)
show that the width of the bosonic profile (Fig. \ref{fig:3}(c)) slightly
increases as going from the inter-species attraction to repulsion. A similar
trend is observed for fermions, see Fig. \ref{fig:3}(d), but amplified in
the spatial zone of the interaction with the bosons, where the gas is
compressed in the case of the attraction and expands in the case of the
repulsion. It is noted that the fermionic component expands in the confined
direction much more than its bosonic counterpart, and that the fermionic
width fluctuates markedly with changes in density. Now, to
analyze the accuracy of the method we solved Eqs.~\ref{E-5} and~\ref{E-6}
and obtained the 1D density associated through of $\rho_{1D,%
\mathrm{B/F}}=\int\int \left\vert \Psi _{\mathrm{B/F}}\right\vert ^{2} dxdy$.
The inset in panel~\ref{fig:3}(a) shows that the difference between the
bosonic profiles obtained by both methods (VA or 3D) is $\sim 2\%$
of the maximum density for all cases (the fact that the error changes very
little with variations in $a_{\mathrm {BF}} $ is a consequence of the greater
number of bosons). Figure~\ref{fig:3}(b) shows that, for the case of the attractive mixture,
the variational profile is very close to the 3D result, in
particular for the case of $a_{\mathrm{BF}}=-6nm$. For the repulsive mixture,
it is observed that the error increases, which is a consequence of the lower
fermionic density at the center of the 3D harmonic potential,
dominated for the bosons, hence a monotonously decreasing function in the
transverse direction, such as the Gaussian, is not a sufficiently good approximation.
We define the global error for the variational method as $E_{\%,%
\mathrm{1D}}=\int \left\vert \rho_{\mathrm{1D}}-n_{\mathrm{1D}}\right\vert dz$
(for both species). Figure~\ref{fig:5-n} shows the global error for both species
as a function of the scattering parameter, $a_{\mathrm{BF}}$. This picture demonstrates
that the error for the bosonic species is around $2\%$, and it does not change much (as already noted in the
comment to Fig.~\ref{fig:3}(a)). For the fermions, the error attains a minimum value
$\sim 0.5\%$ at $a_{\mathrm{BF}}\approx 4$nm, thus corroborating that the
Gaussian is a very good approximation. This is a consequence of the fact that,
for this value of $a_{\mathrm{BF}}$, the interspecies interaction term practically
compensates the Pauli repulsion term, making the dynamics of the fully polarized Fermi gas
close to that giverned by the Sch\"{o}dinger equation (recall that the Gaussian
is the solution for the ground state). When the mixture becomes
more attractive, the fermionic dynamics is dominated by the bosons, producing a similar
error for both species, while for the repulsive mixture the Gaussian approximation is not
appropriate. Interestingly, for the non-interacting mixture, the error for the fermions is smaller
than for the bosons, because the fermionic density is very low, making the self-interaction
terms weak in comparison to the external potential, therefore it is appropriate
to use the Gaussian ansatz for describing the 1D dynamics. Using the TF profile
may produce, in principle, better results, but the respective analysis would
be cumbersome.

\subsubsection{Spatial correlations in the ground state.}

\begin{figure}[tbp]
\centering
\resizebox{0.5\textwidth}{!}{\includegraphics{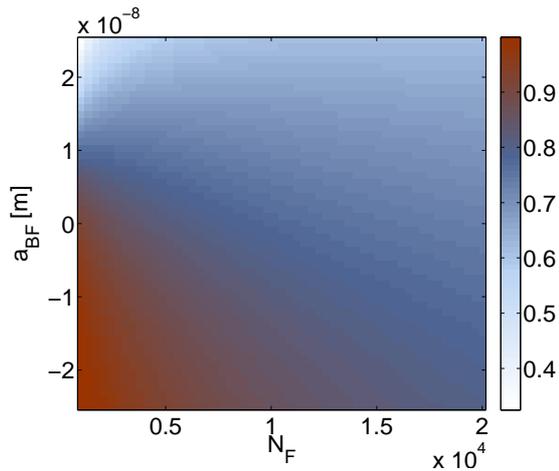}
}
\caption{(Color online) The color-coded plot of the spatial correlation $%
C_{s}$, defined as per Eq. (\protect\ref{E-34}), as a function of the
fermionic number, $N_{\mathrm{F}}$, and the scattering length of the
inter-species interaction, $a_{\mathrm{BF}}$. Other parameters are $N_{%
\mathrm{B}}=5\times 10^{4}$, $a_{\mathrm{B/F}}=5$ nm, $\protect\omega _{z,%
\mathrm{B/F}}=30$ Hz and $\protect\omega _{t,\mathrm{B/F}}=1000$ Hz.}
\label{fig:4}
\end{figure}

As the first application, we calculate the spatial correlation between
densities of the two species ($C_s$), as defined by Eq.~(\ref{E-34}) for
a wide range of parameters, ranging from very attractive to
to very repulsive mixtures. We keep in mind that the variational calculation
produces a minor error for attractive mixtures, therefore the values obtained
for the repulsive case will be considered as help to asses the validity
of our analysis. The eventual objective is to produce a parameter
which determines the mutual influence of both species.
In the Fig.~\ref{fig:4}, the spatial correlation $C_{s}$ is shown versus
$N_{\mathrm{F}}$ and $a_{\mathrm{BF}}$ , which shows that, for an attractive
interaction, $C_{s}$ reaches values greater than $0.8$ (in the mixed state),
whereas for the repulsive interaction $C_{s}$ decreases until reaching
values around $C_{s}=0.3$ (in the demixed state). It is also noted that the
number of fermions strongly affects the dependence of the correlation on the
inter-species scattering length. When $N_{\mathrm{F}}$ increases, the
difference between the maximum and the minimum value of $C_{s}$ decreases,
as $a_{\mathrm{BF}}$ varies from $-25$ nm to $+25$ nm, tending to reach
values close to $C_{s}=0.8$, which is the value attained in the absence of
the interaction, depending only on the presence of the same trap acting on
both components.

\subsubsection{Dynamics near the ground state}
\label{sec-3-1-3}

\begin{figure}[tbp]
\centering
\resizebox{1.\textwidth}{!}{
\includegraphics{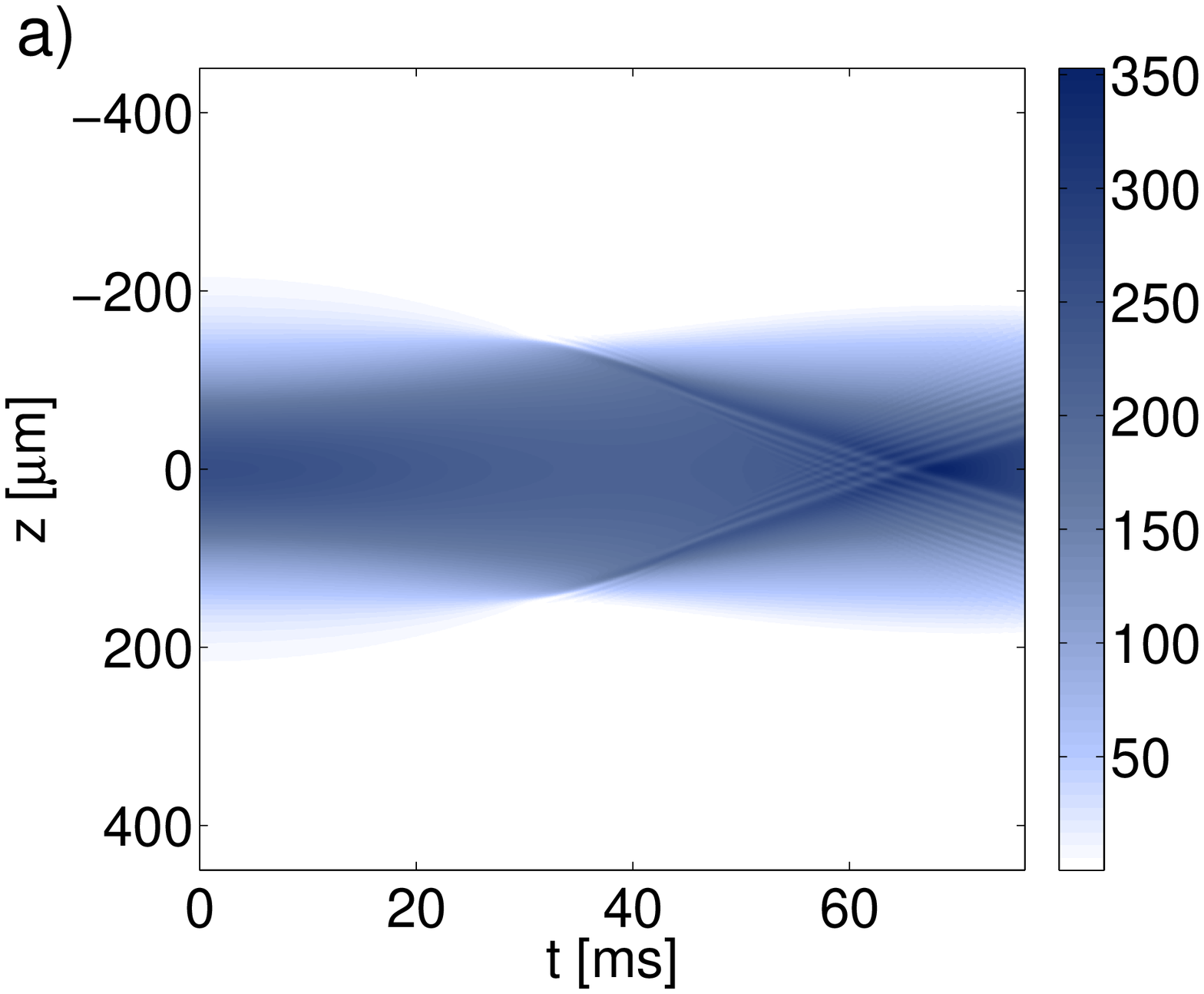}
\includegraphics{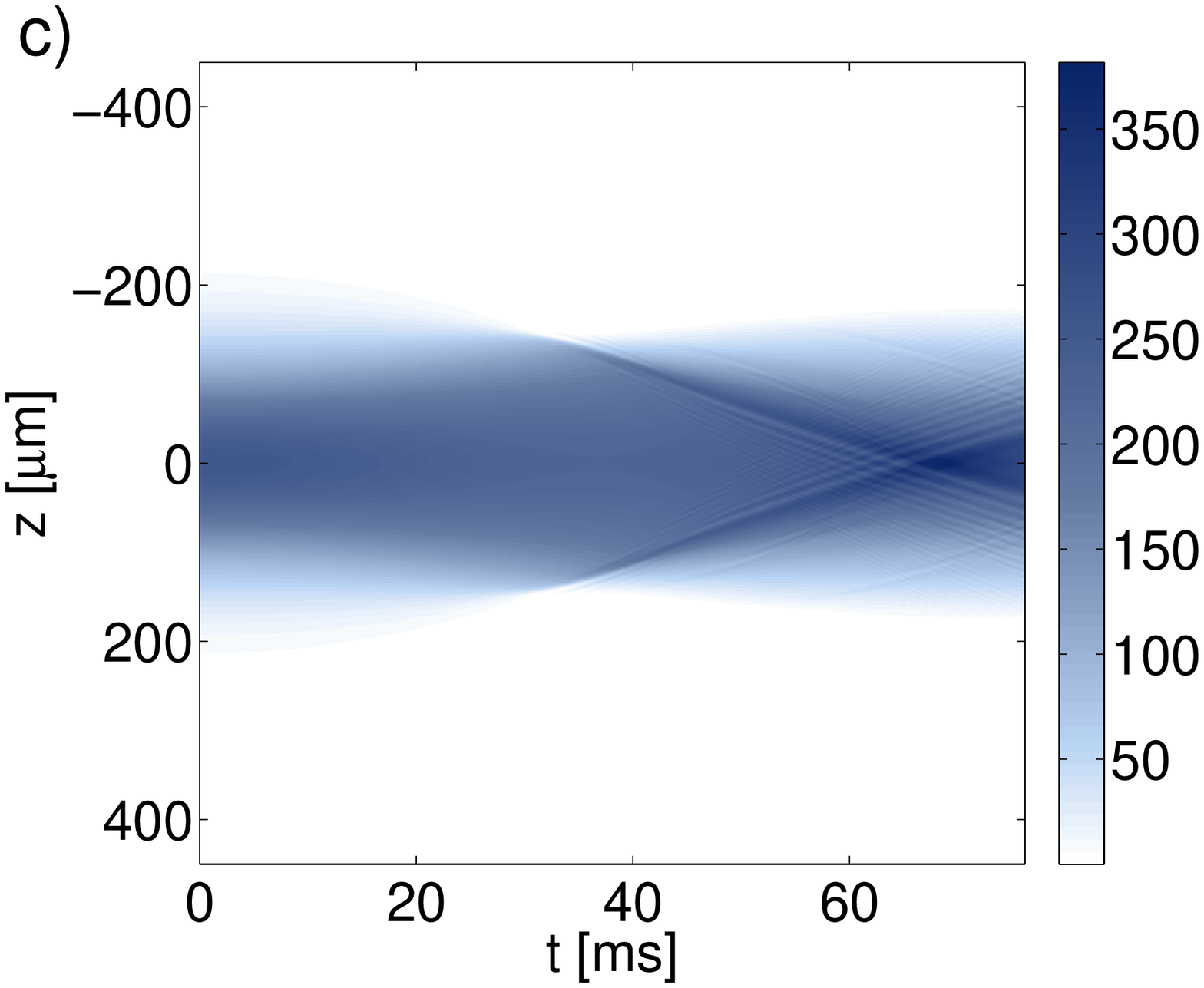}
\includegraphics{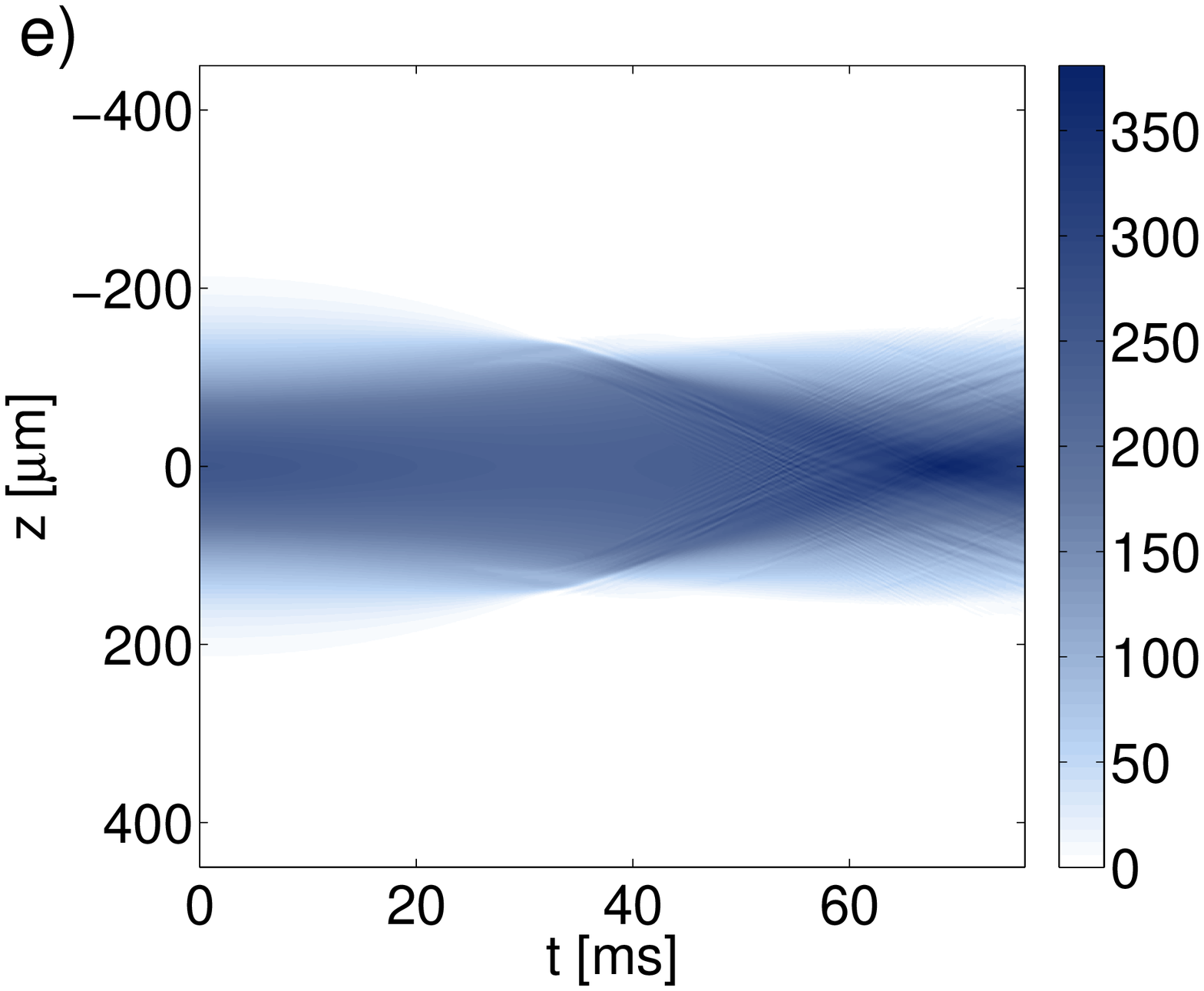}
}
\resizebox{1.\textwidth}{!}{
\includegraphics{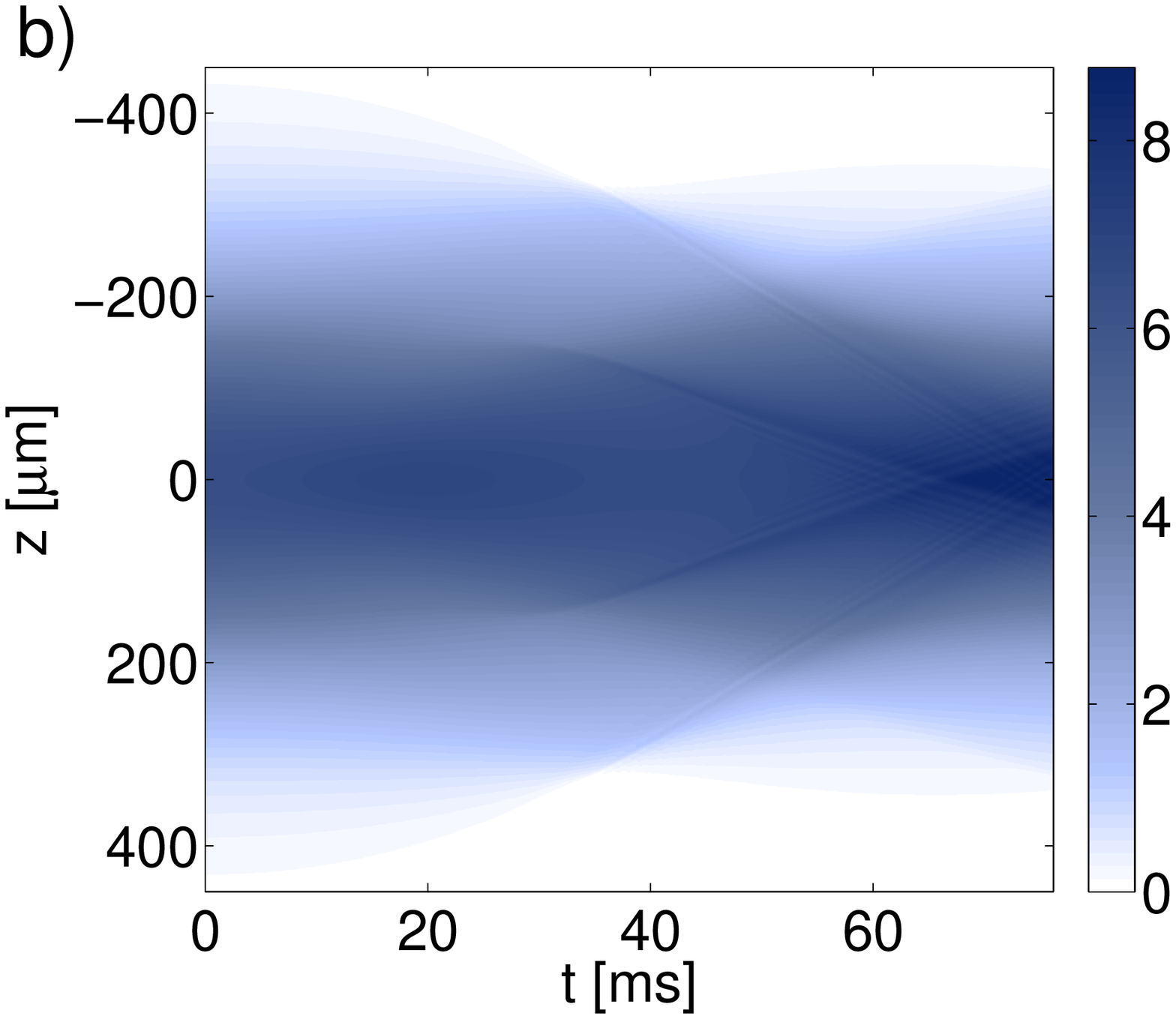}
\includegraphics{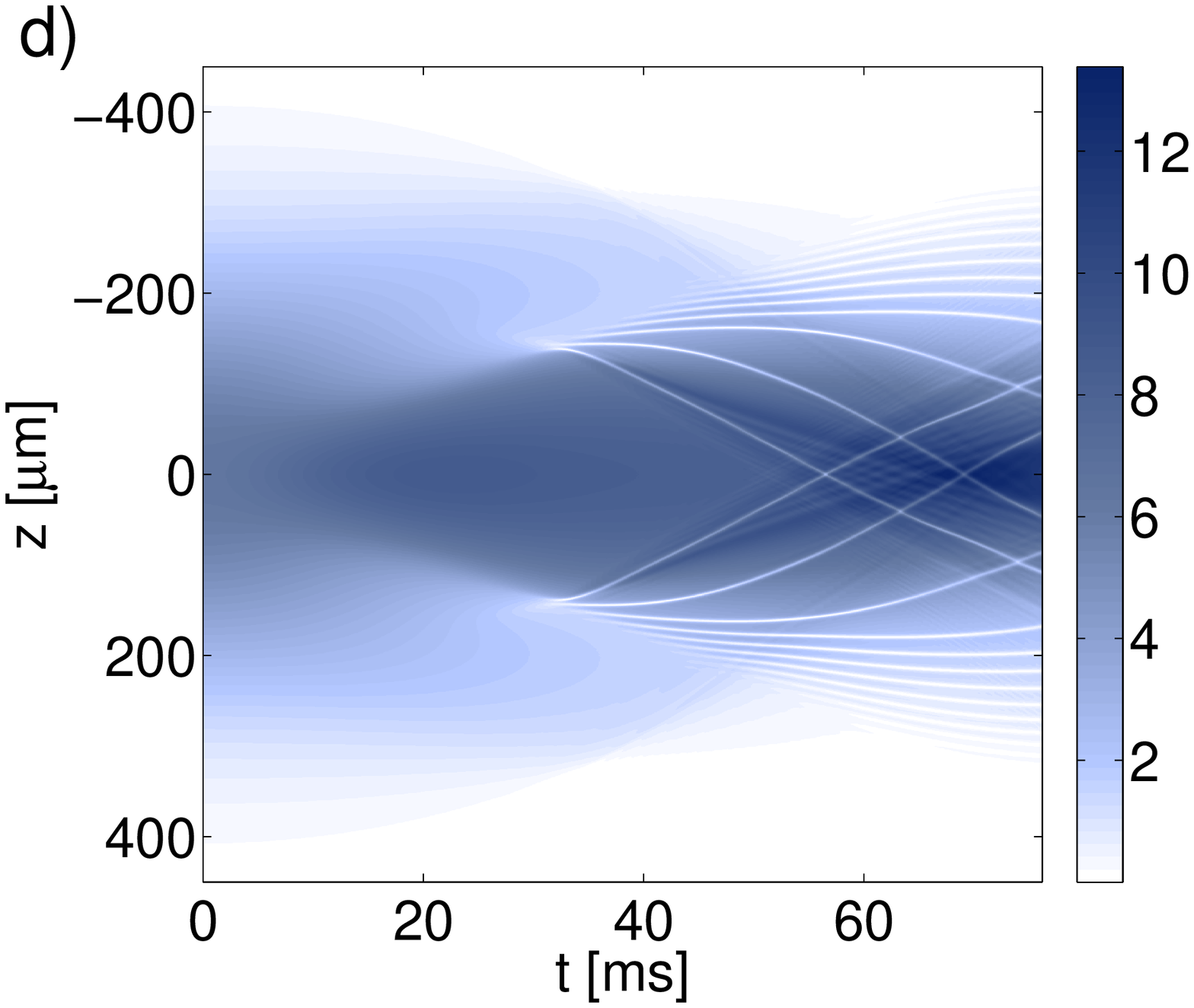}
\includegraphics{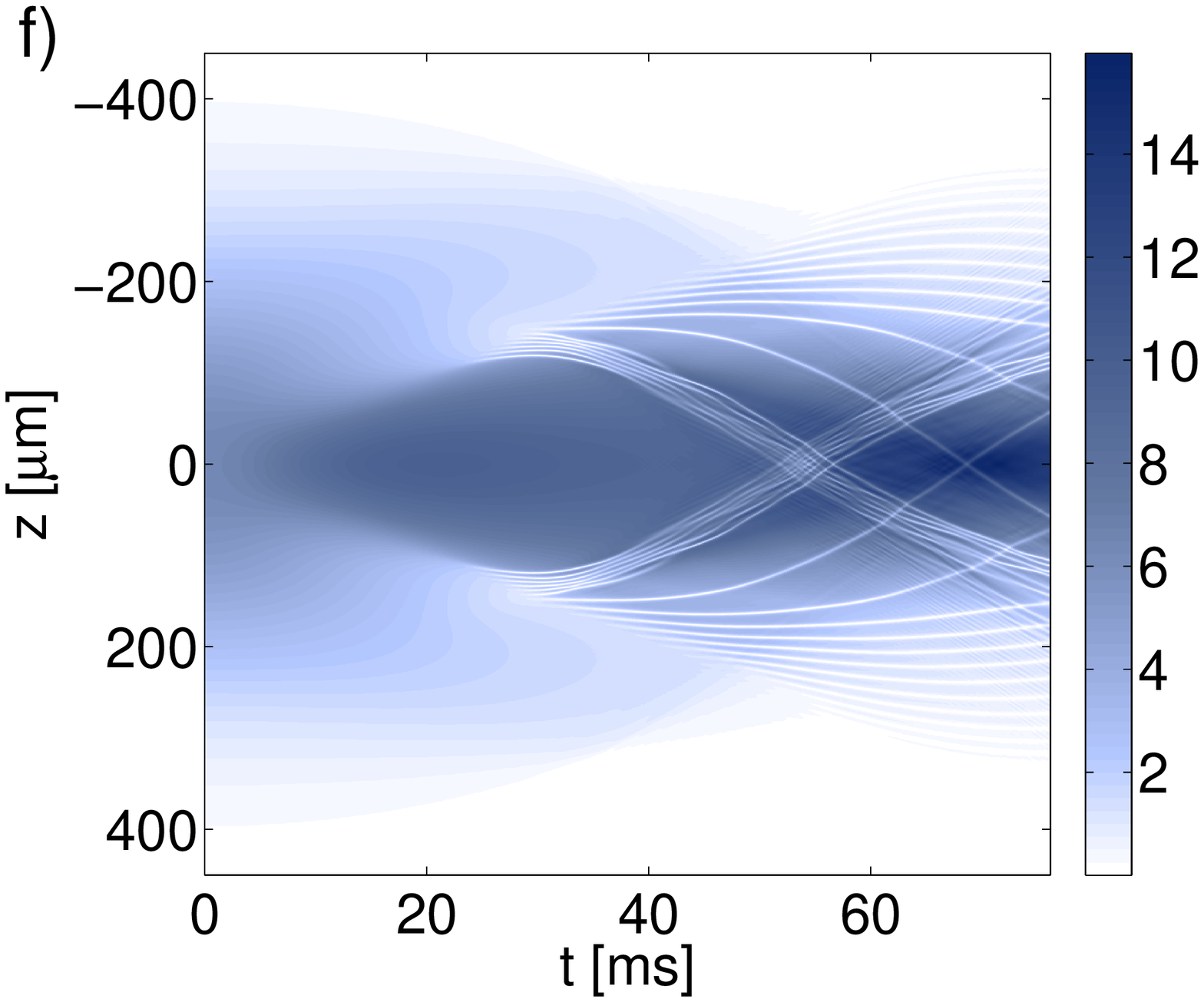}
}
\caption{(Color online) Space-time diagrams for the conservative dynamics of bosons (above) and fermions (bottom), for three different values of the interspecies scattering parameter: (a)$-$(b) $a_{\mathrm{BF}}=-6nm$, (c)$-$(d) $a_{\mathrm{BF}}=-12nm$, and (e)$-$(f) $a_{\mathrm{BF}}=-14nm$. The initial conditions are the same in all cases and are described in the text. Other parameters are the same as in Fig.~\ref{fig:3}}
\label{fig:7-n}
\end{figure}

We begin the study of conservative dynamics, considering a mixture with arbitrary initial conditions for the 1D fields: we assume a Gaussian shape along the $z$-axes with widths (standard deviation) of $100\mu$m and $200\mu$m for bosons and fermions, respectively. Figure~\ref{fig:7-n} shows three cases of the temporal evolution with these initial conditions for $a_{\mathrm{BF}}=-6nm$, $a_{\mathrm{BF}}=-12nm$, and $a_{\mathrm{BF}}=-14nm$. In the first case (panels~\ref{fig:7-n}(a) and ~\ref{fig:7-n}(b)), it is observed that the densities converge towards the center of the potential in an expected pattern of oscillations around the potential minimum; in addition, the fermions are affected by bosons, as can be seen for the mark left by the bosons in the fermionic density. For the second case (panels~\ref{fig:7-n}(c) and ~\ref{fig:7-n}(d)), it is observed that the increase in the magnitude of the attractive interaction generates dark solitons  in the fermionic density, some of which shows oscillatory dynamics very similar to that observed, experimentally and theoretically, in Refs. \cite{Cardoso13}-\cite{Donadello14}. Finally, the last case (panels~\ref{fig:7-n}(e) and ~\ref{fig:7-n}(f)) shows that further increase in the magnitude of the interspecies interaction generates a larger number of dark solitons. In other words, we show that the attractive interaction of fermions with bosons in a state different from the ground state eventually generates a gas of dark solitons.

\begin{figure}[tbp]
\centering
\resizebox{0.8\textwidth}{!}{\includegraphics{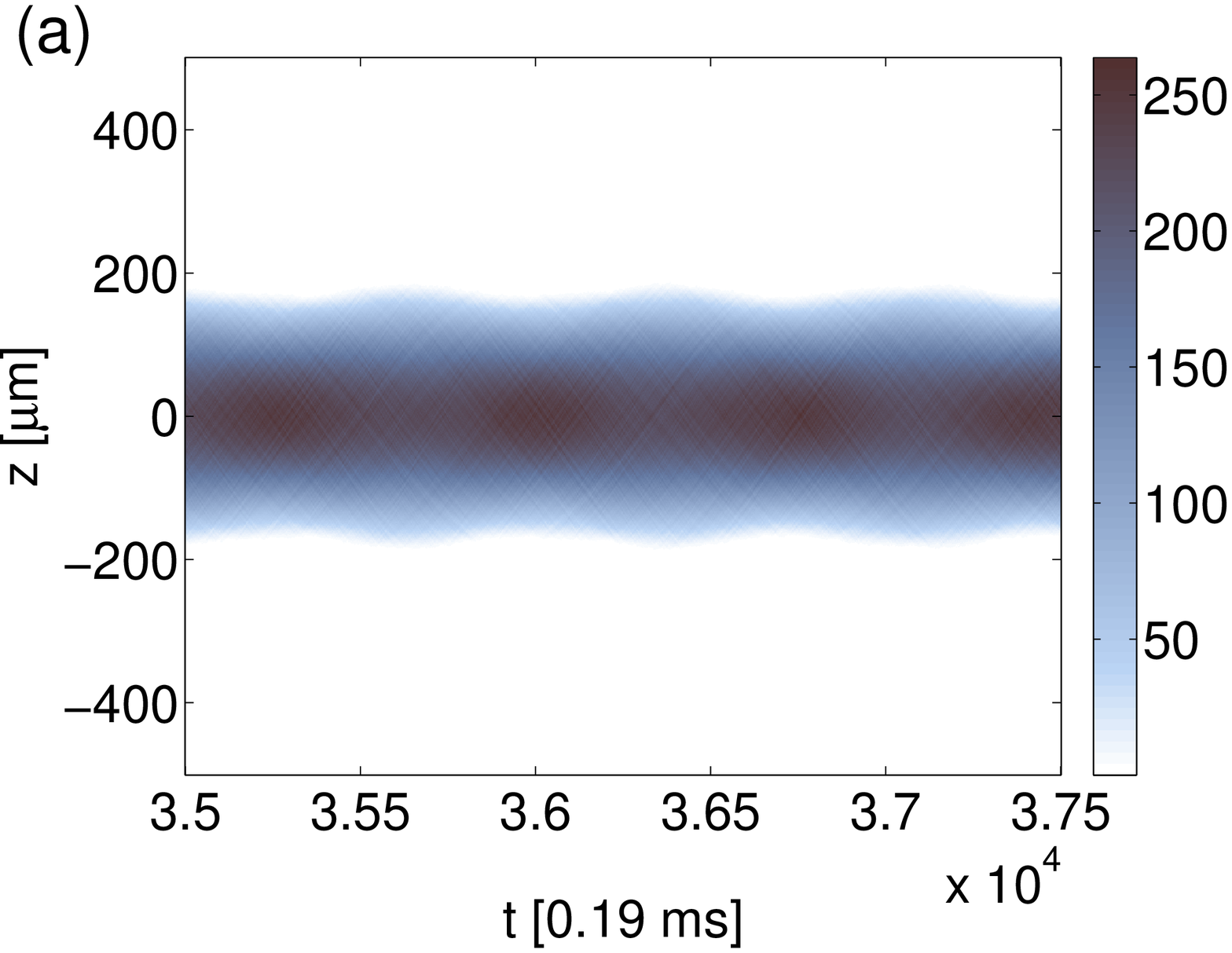}
\includegraphics{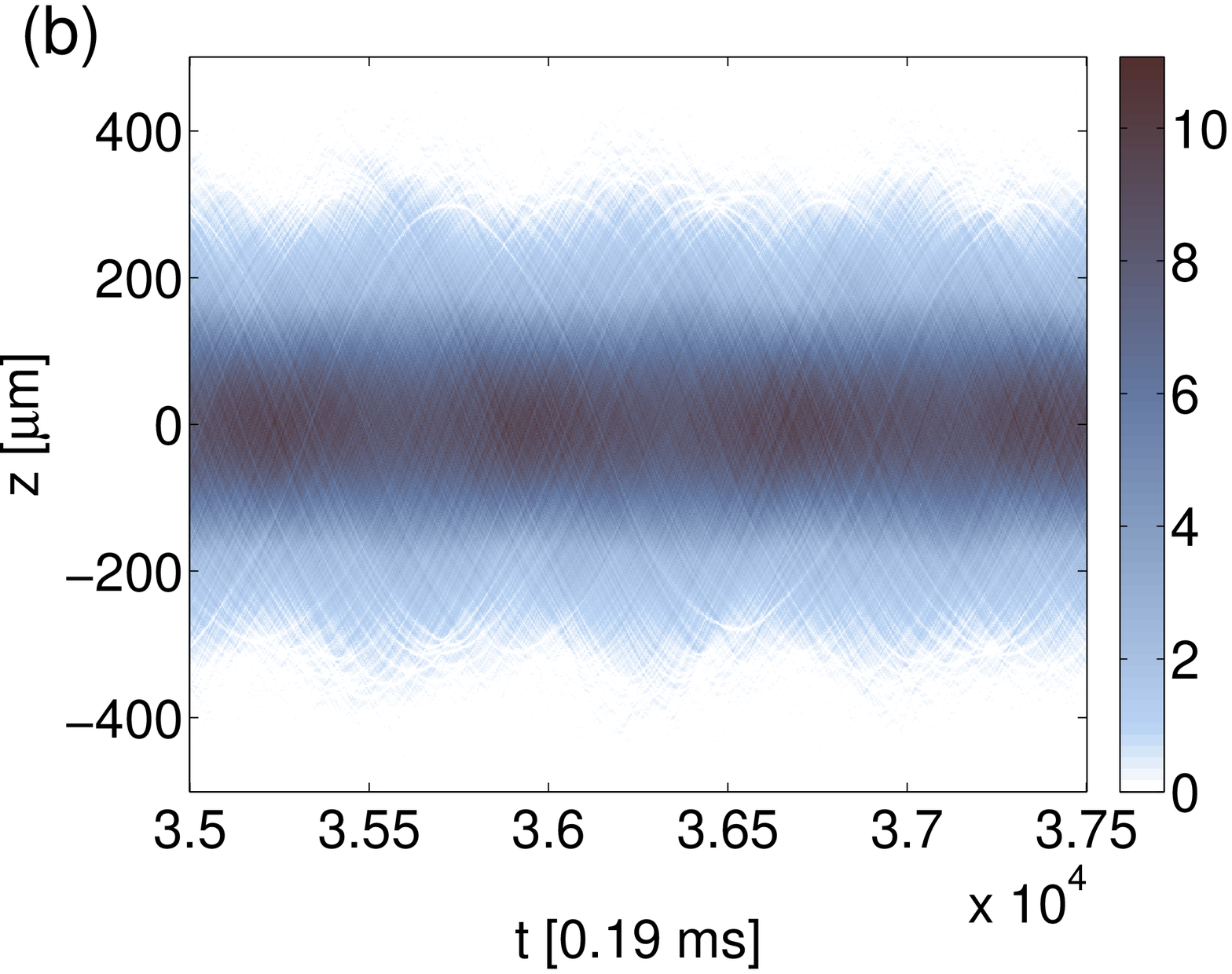}
}
\resizebox{0.8\textwidth}{!}{\includegraphics{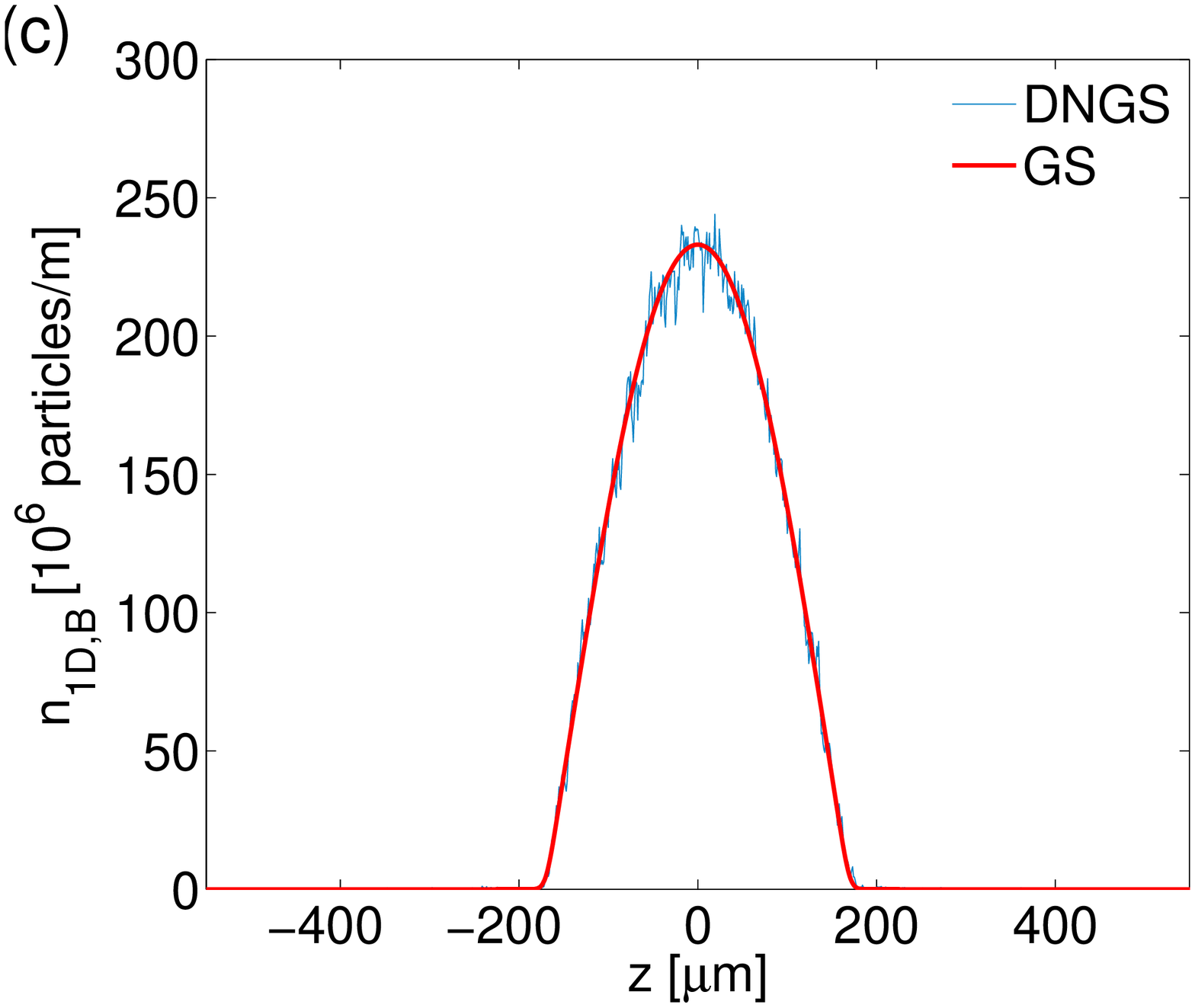}
\includegraphics{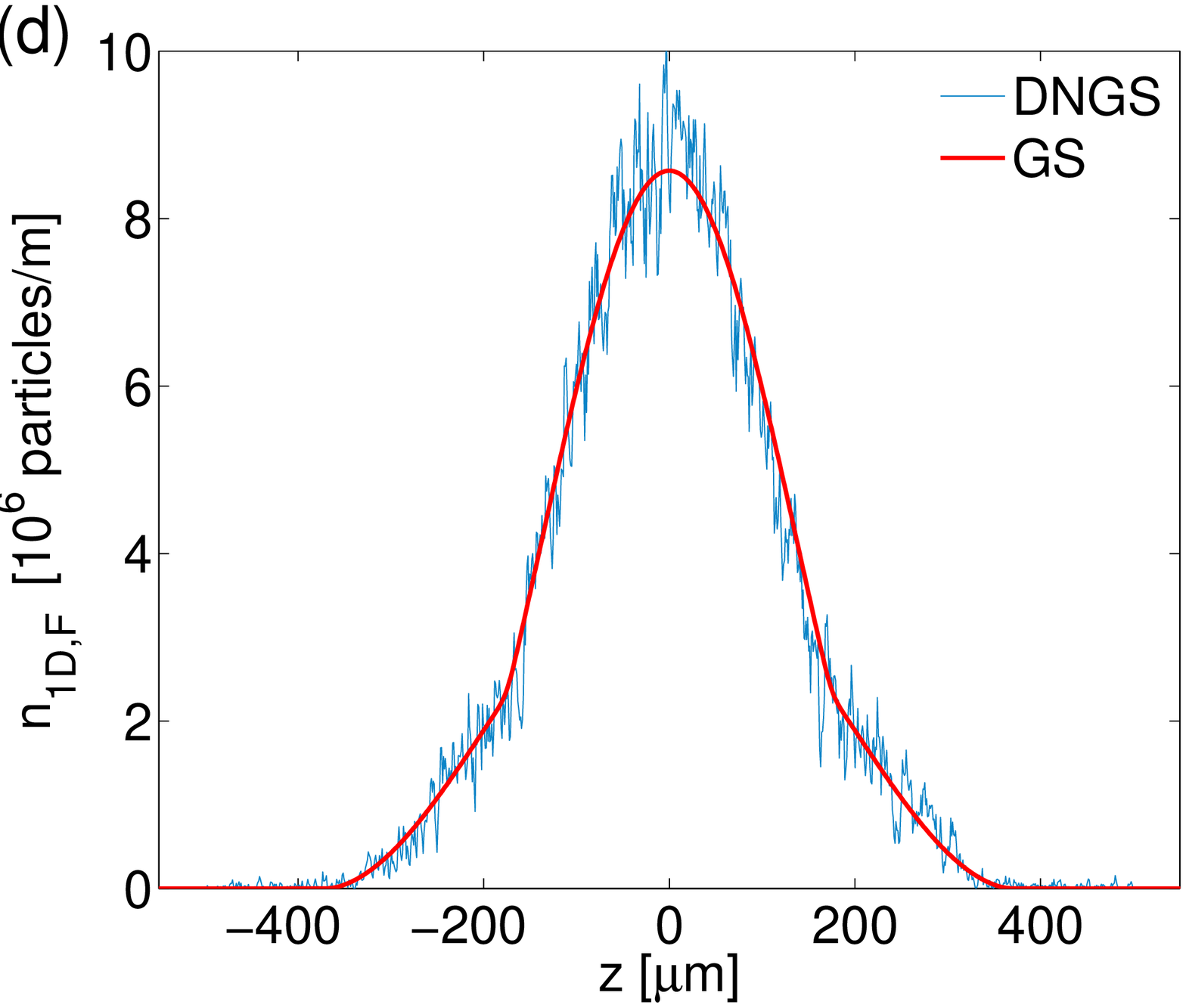}
}
\caption{(Color online) (a) and (b) Spatiotemporal evolution of the bosonic
and fermionic densities, respectively, in the case of suddenly switching the
inter-species interaction in the 1D ground state which corresponded to $a_{%
\mathrm{BF}}=0$. (c) and (d) Bosonic and fermionic density profiles
at the end of the simulations (the blue lines) compared to the
ground state (red
line), at the same values of physical parameters, which are $N_{\mathrm{B}%
}=5\times 10^{4}$, $N_{\mathrm{F}}=2.5\times 10^{3}$, $a_{\mathrm{B/F}}=5$
nm, $a_{\mathrm{BF}}=-15$ nm, $\protect\omega _{z,\mathrm{B/F}}=30$ Hz, and $%
\protect\omega _{t,\mathrm{B/F}}=1000$ Hz.} \label{fig:5}
\end{figure}

Now we aim to address the system's dynamics in the vicinity of the GS. We
start with the GS found in the absence of the inter-species interaction ($a_{%
\mathrm{BF}}=0$). Then, at $t=0$, we switch the interaction on, which may
imply the application of the magnetic field, that gives rise to $a_{\mathrm{%
BF}}\neq 0$ via the FR. Figure ~\ref{fig:5} shows an example of the ensuing
evolution in the attractive mixture with $a_{\mathrm{BF}}=-15$ nm, the
fermionic component being fully polarized. Panels ~\ref{fig:5}(a) and (b)
display space-time diagrams of oscillations of the bosonic and fermionic
densities in the dynamical state. Panels~\ref{fig:5}(c) and (d) compare the
density profiles at the end of the simulations with the GS of the
interacting mixture. It is seen that the dynamics amount to rapid
fluctuations around the GS. Figure~\ref{fig:7-n} makes it clear that the
spatiotemporal dynamics of fermions in panel~\ref{fig:5}(b) does not reduce
to noise around the ground state. It rather corresponds to a gas of dark
solitons oscillating in the harmonic potential. Moreover, the fact that the
spatiotemporal diagram has been produces for the time when the system is in
the stationary regime suggests that the dark solitons are fully stable. Note
that the noise observed in the bosonic component is clearly caused by the
fermionic dark solitons.

\begin{figure}[tbp]
\centering
\resizebox{0.5\textwidth}{!}{\includegraphics{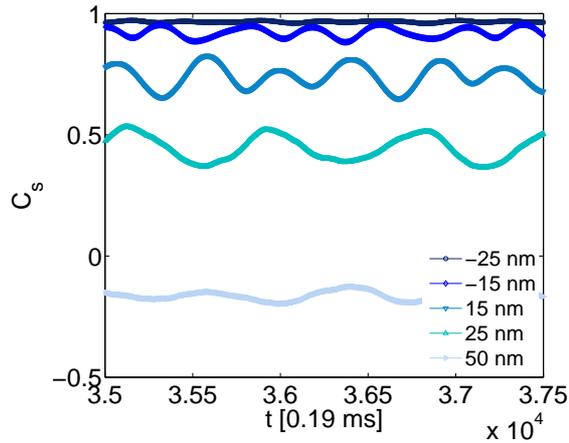}
}
\caption{(Color online) The evolution of
spatial correlation $C_{s}$ in the dynamical states near the ground
state for five values of $a_{\mathrm{BF}}$, as indicated in the
figure. The parameters are the same as in Fig.~\ref{fig:5}.} \label{fig:9-n}
\end{figure}

Now, we measure how correlated the densities remain in time. Similar to what
is observed in Fig.~\ref{fig:4}, Fig.~\ref{fig:9-n} shows that the
spatial-correlation parameter $C_{s}$, defined as per given Eq. (\ref{E-34})
rapidly decreases following the transition from negative to positive
values of $a_{\mathrm{BF}}$ (the mixing-demixing transition). When the
inter-species interaction is more attractive, it increases the correlations,
imposing spatiotemporal synchronization of density fluctuations. In the
case of strong repulsion, $a_{\mathrm{BF}}=50$ nm, correlation fluctuations
are suppressed because the system falls into a demixed state.

\begin{figure}[tbp]
\centering
\resizebox{0.5\textwidth}{!}{\includegraphics{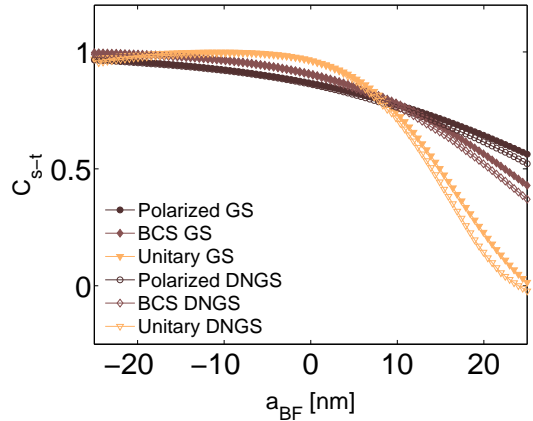}
}
\caption{(Color online) Spatiotemporal correlation $C_{s-t}$ versus $a_{%
\mathrm{BF}}$ for the ground state (GS) and in the dynamics near the
ground state (``DNGS") for three fermions regimes: polarized, BCS,
and unitarity. The parameters
are $N_{\mathrm{B}}=5\times 10^{4}$, $N_{F}=2.5\times 10^{3}$,
$a_{\mathrm{B/F}}=5$ nm, $\protect\omega _{z,B/F}=30$ Hz and $\protect\omega %
_{t,\mathrm{B/F}}=1000$ Hz. Table \ref{Table-1} provides parameters that
correspond to the different regimens in the Lagrangian density.}
\label{fig:6}
\end{figure}

Figure~\ref{fig:6} presents the analysis of the spatiotemporal
synchronization of the mixture, where three possible regimes are considered
for the fermions: fully polarized, BCS, and unitarity.
Parameters of the the Lagrangian density for each fermionic
regime are given in Table \ref{Table-1}.
The synchronization was calculated through the Person coefficient $C_{s-t}$
defined above. For each of the three cases, the spatial correlation, $Cs,$
was calculated too. When the interaction is attractive, there is no discrepancy
between the synchronization and correlation curves. On the other hand, in all
the three regimes it is observed that, as the interaction becomes more
repulsive, the synchronization of the mixture decreases in comparison to
the spatial correlation. This trend enhances as the system approaches the
demixing transition. Another generic feature is that stronger correlations
between the species near the GS imply a stronger synchronization on the
spatiotemporal dynamics too.

The correlation properties are different for the distinct fermionic regimes.
$C_{s-t}$ is always positive for the system with fully polarized fermions
and in the BCS regime, decreasing as the interaction gets more repulsive. On
the other hand, the unitarity regime features significant differences: at
first, $C_{s-t}$ grows to a maximum value at $a_{\mathrm{BF}}=-10$ nm; then
it decreases, reaching negative values of $C_{s-t}/C_{s}$ in the demixed
state. In all the cases, no significant difference are observed between $%
C_{st}$ and $C_{s}$ in the case of the attractive Bose-Fermi interaction,
i.e., a highly correlated GS supports the dynamical synchronization too.

\subsubsection{Accuracy of the variational method for the spatiotemporal dynamics}
\label{sec-3-1-4}

\begin{figure}[tbp]
\centering{
\resizebox{1.\textwidth}{!}{
\includegraphics{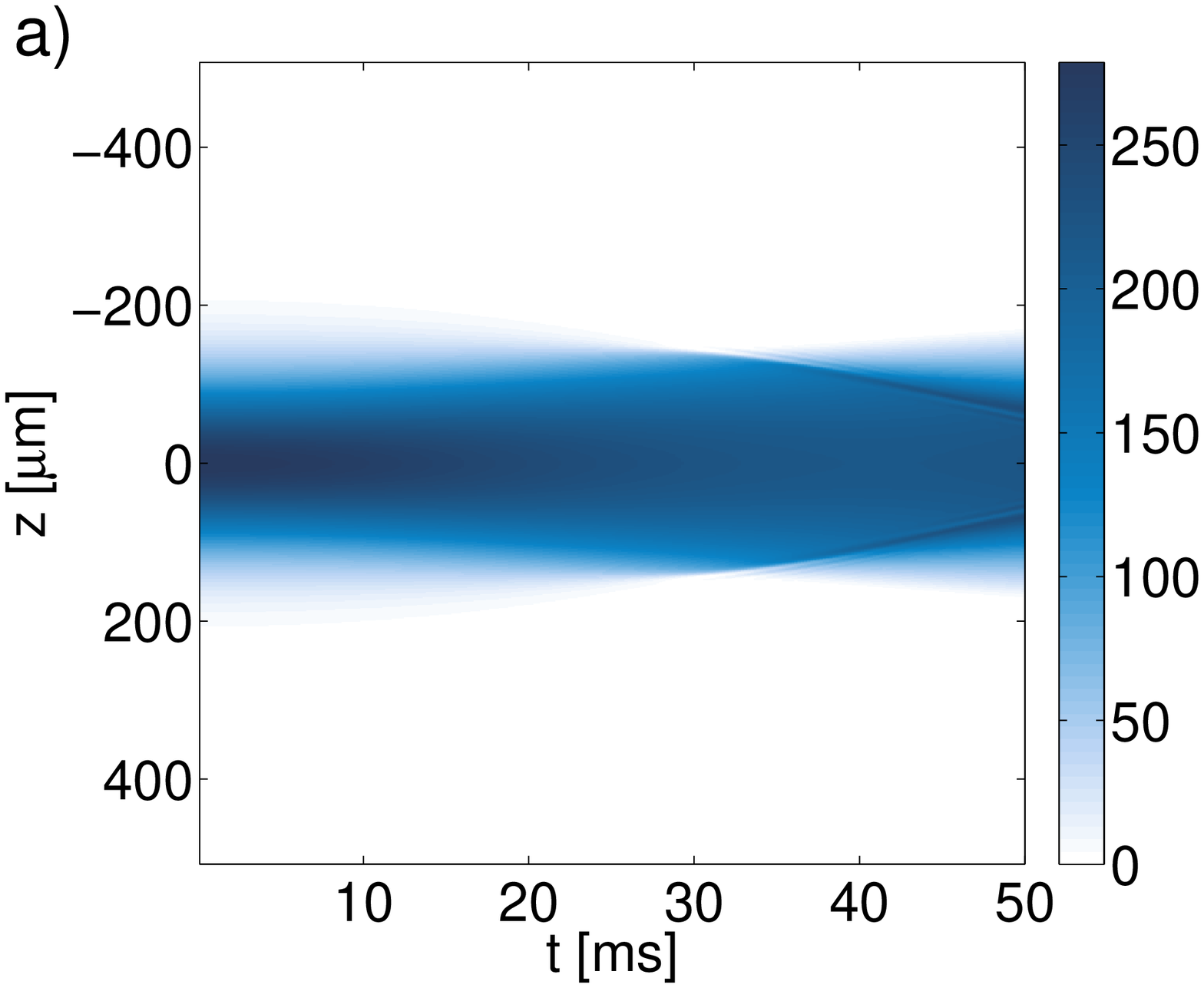}
\includegraphics{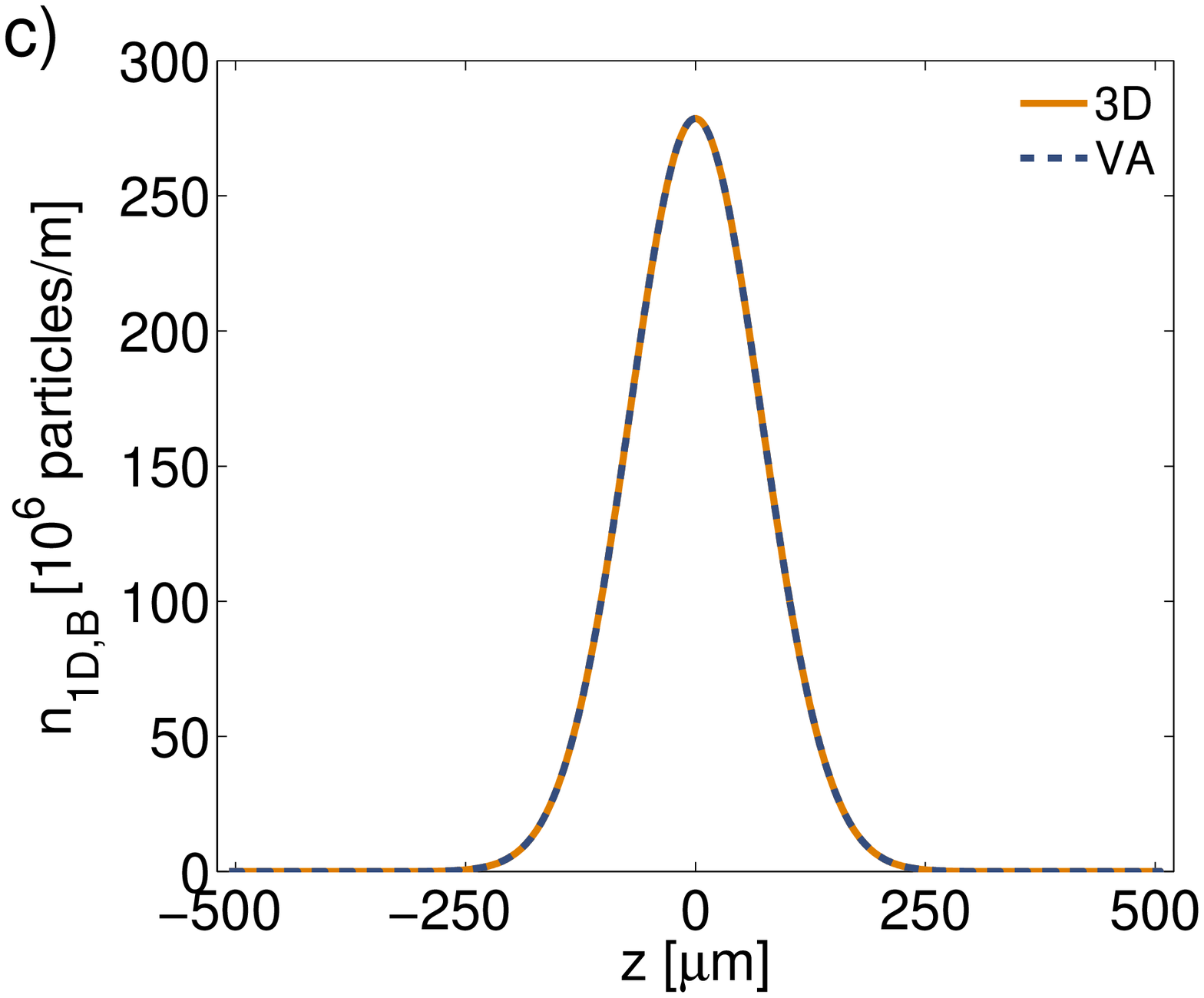}
\includegraphics{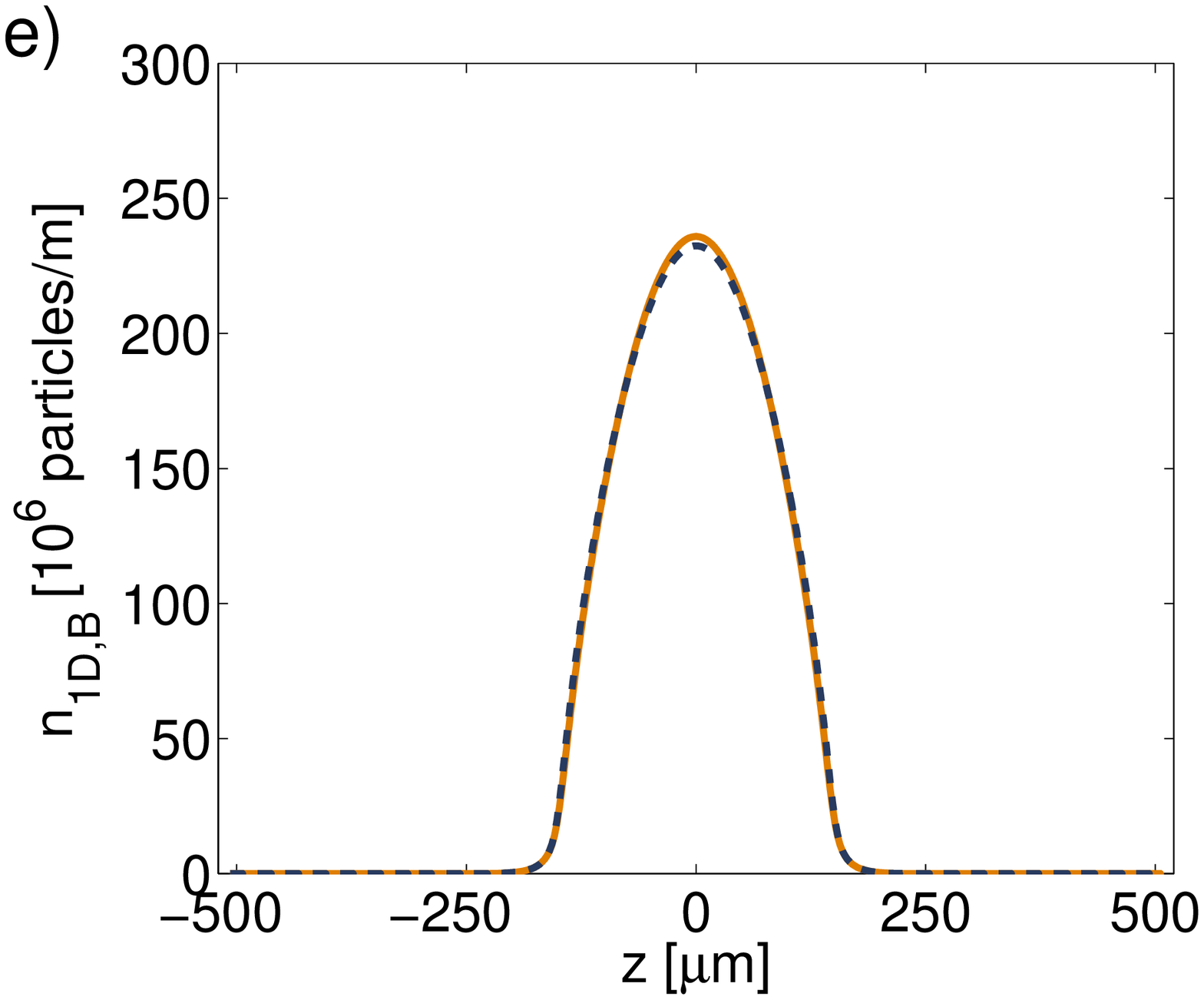}
\includegraphics{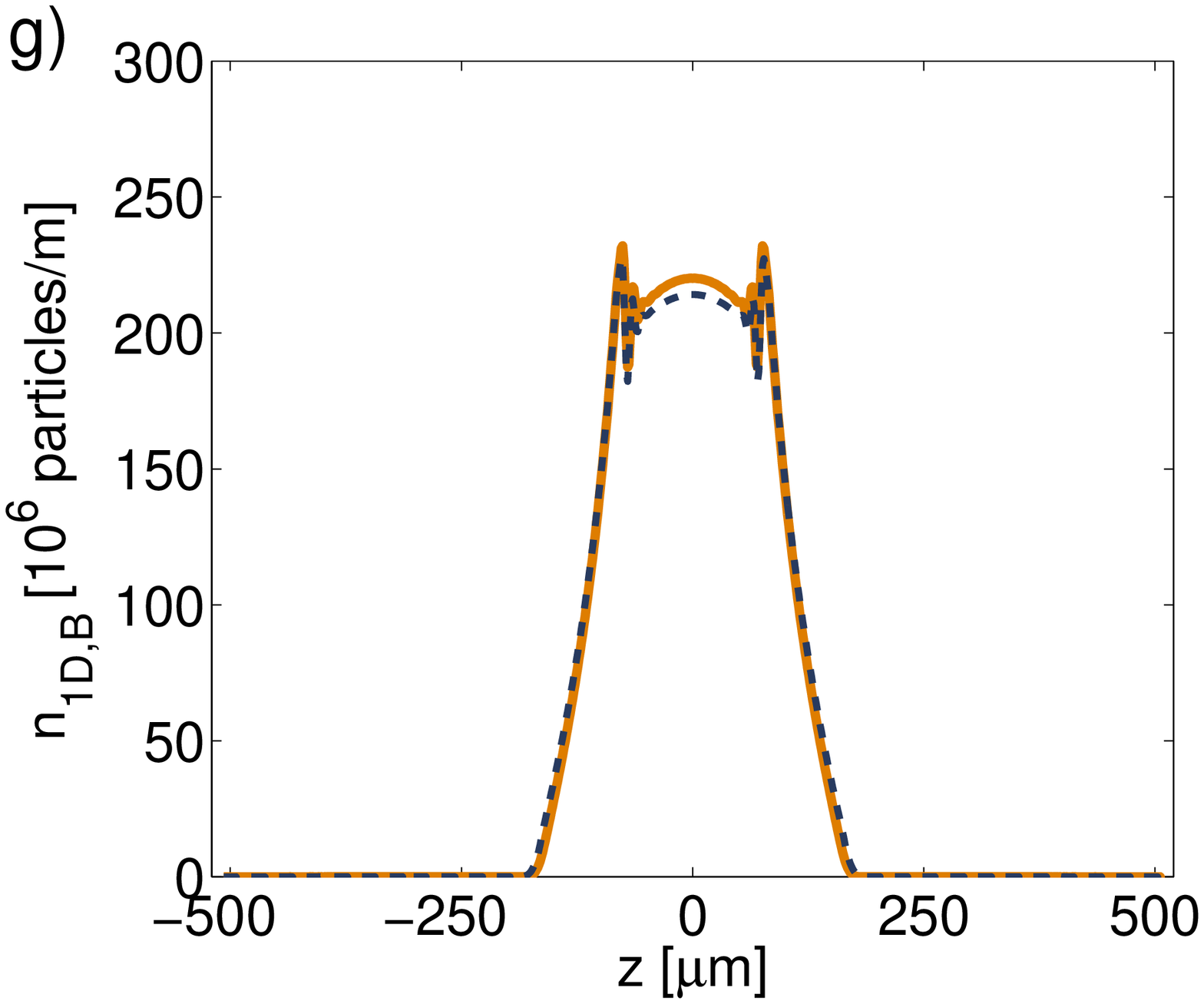}
}
\resizebox{1.\textwidth}{!}{
\includegraphics{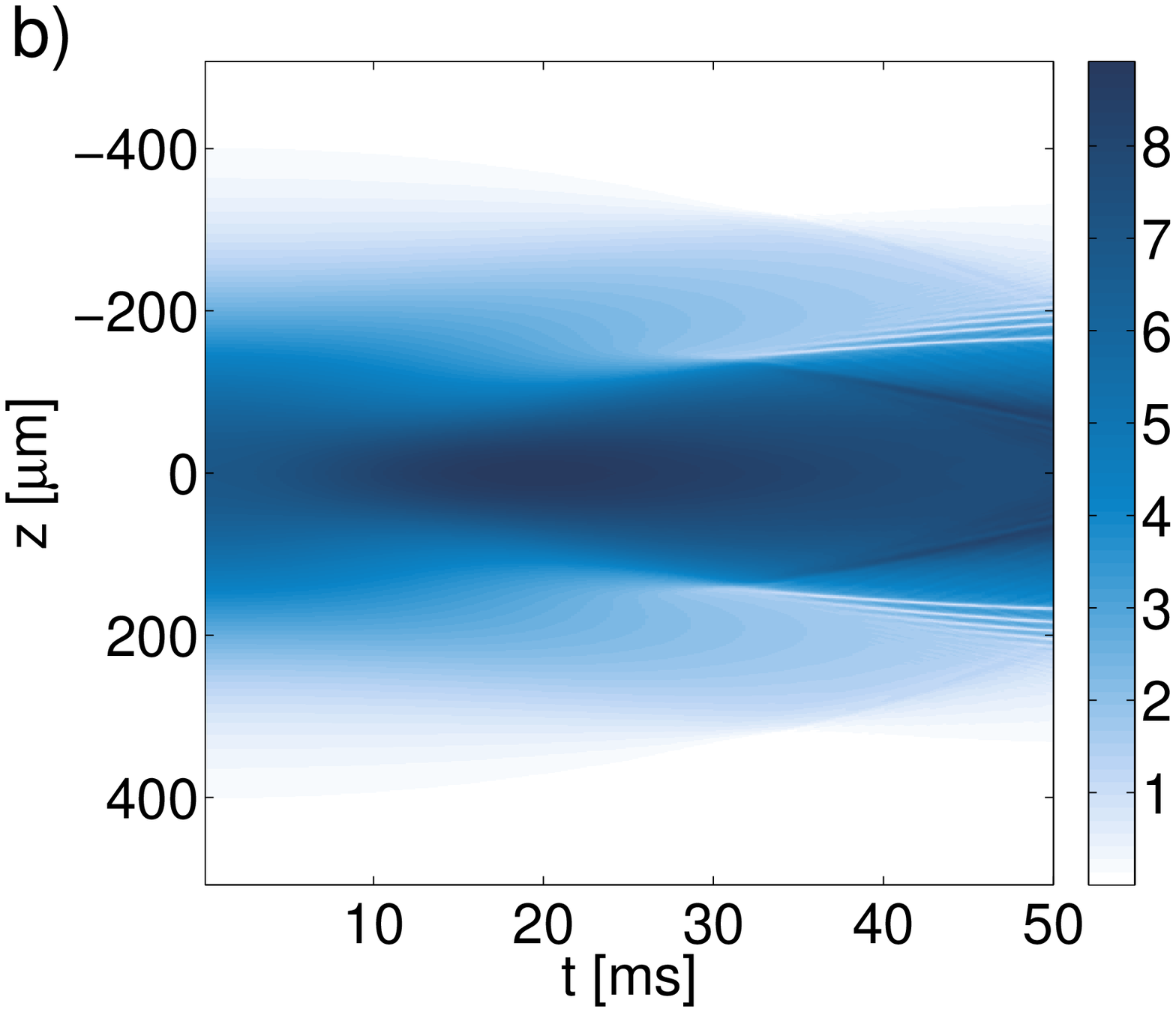}
\includegraphics{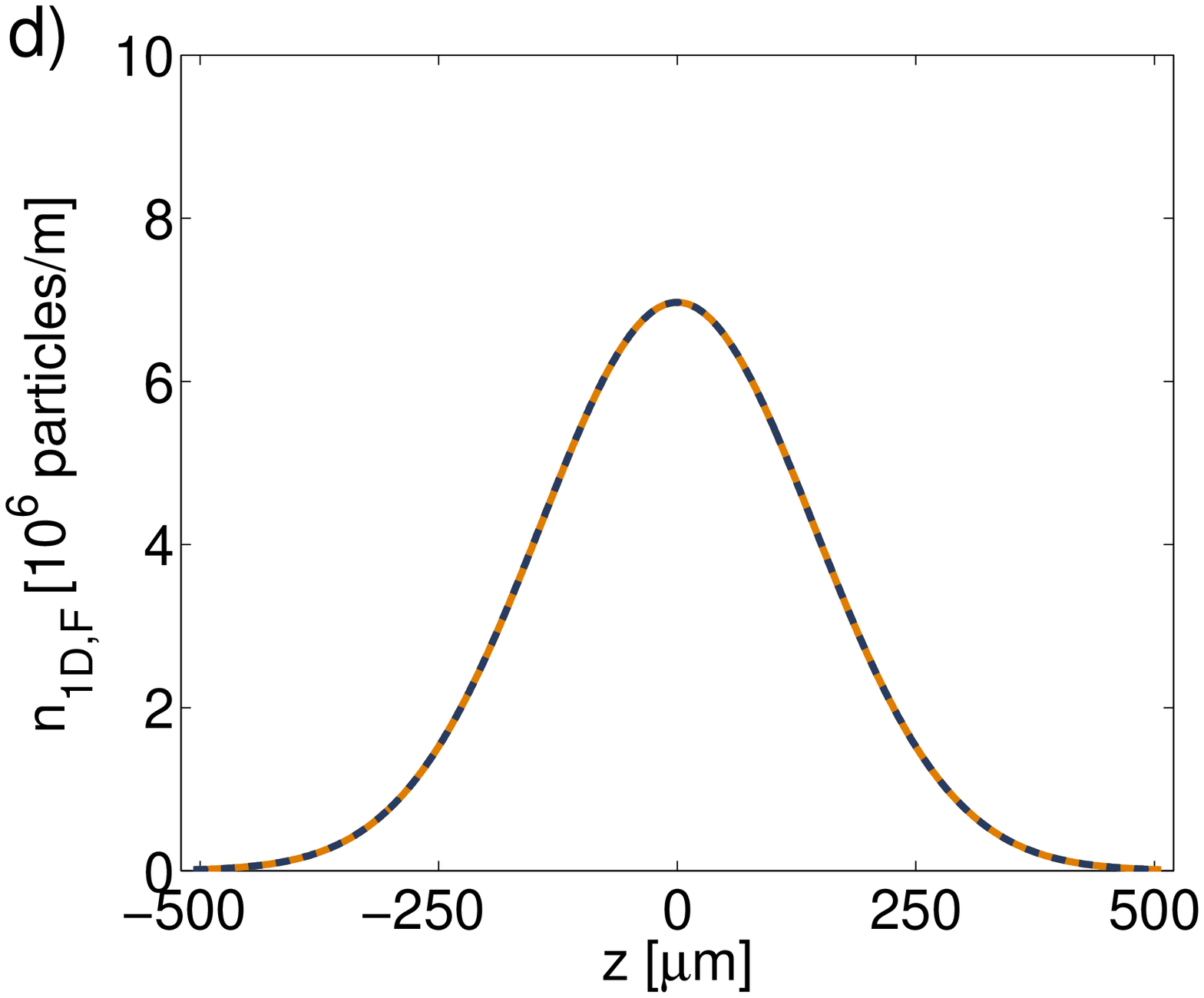}
\includegraphics{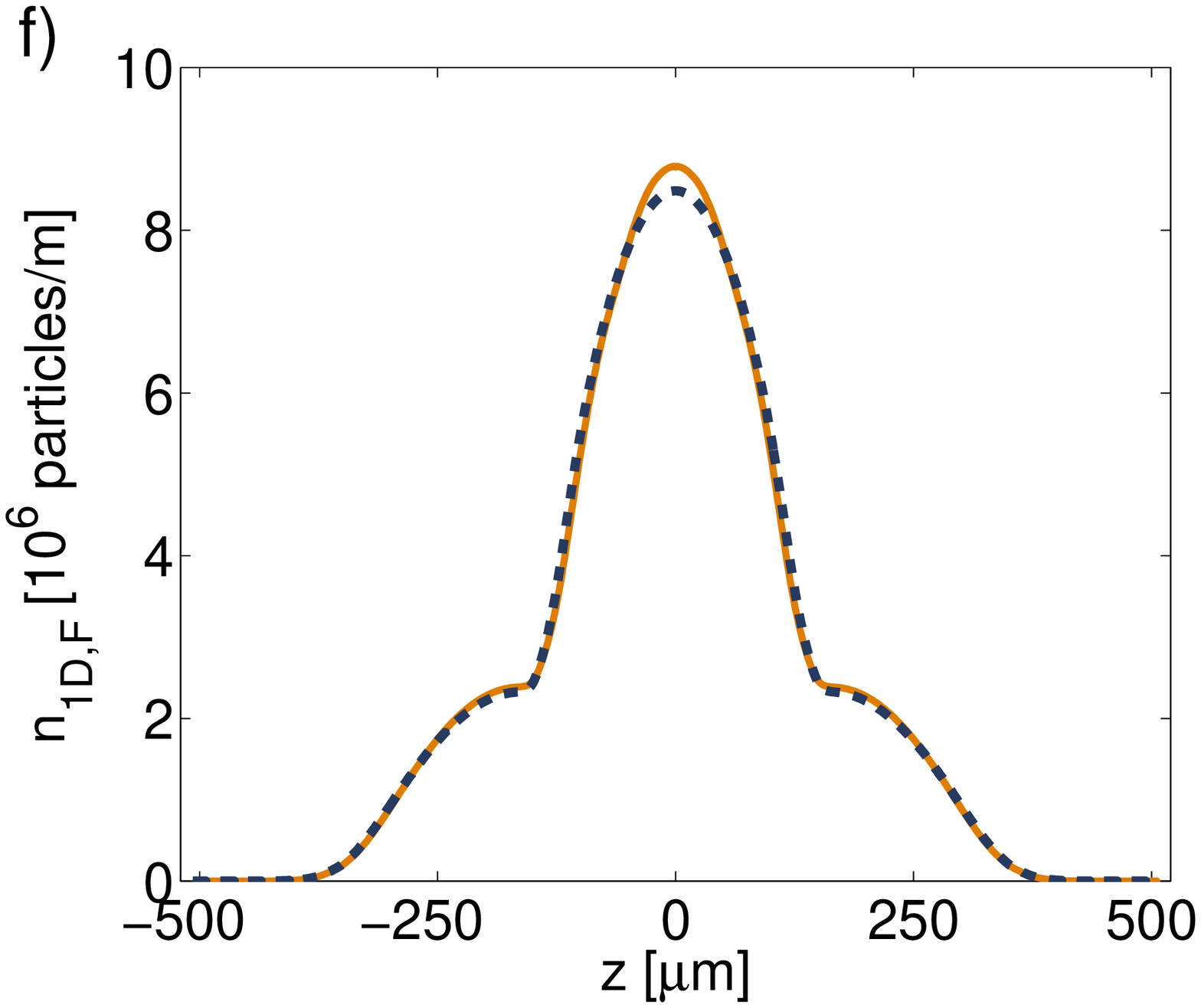}
\includegraphics{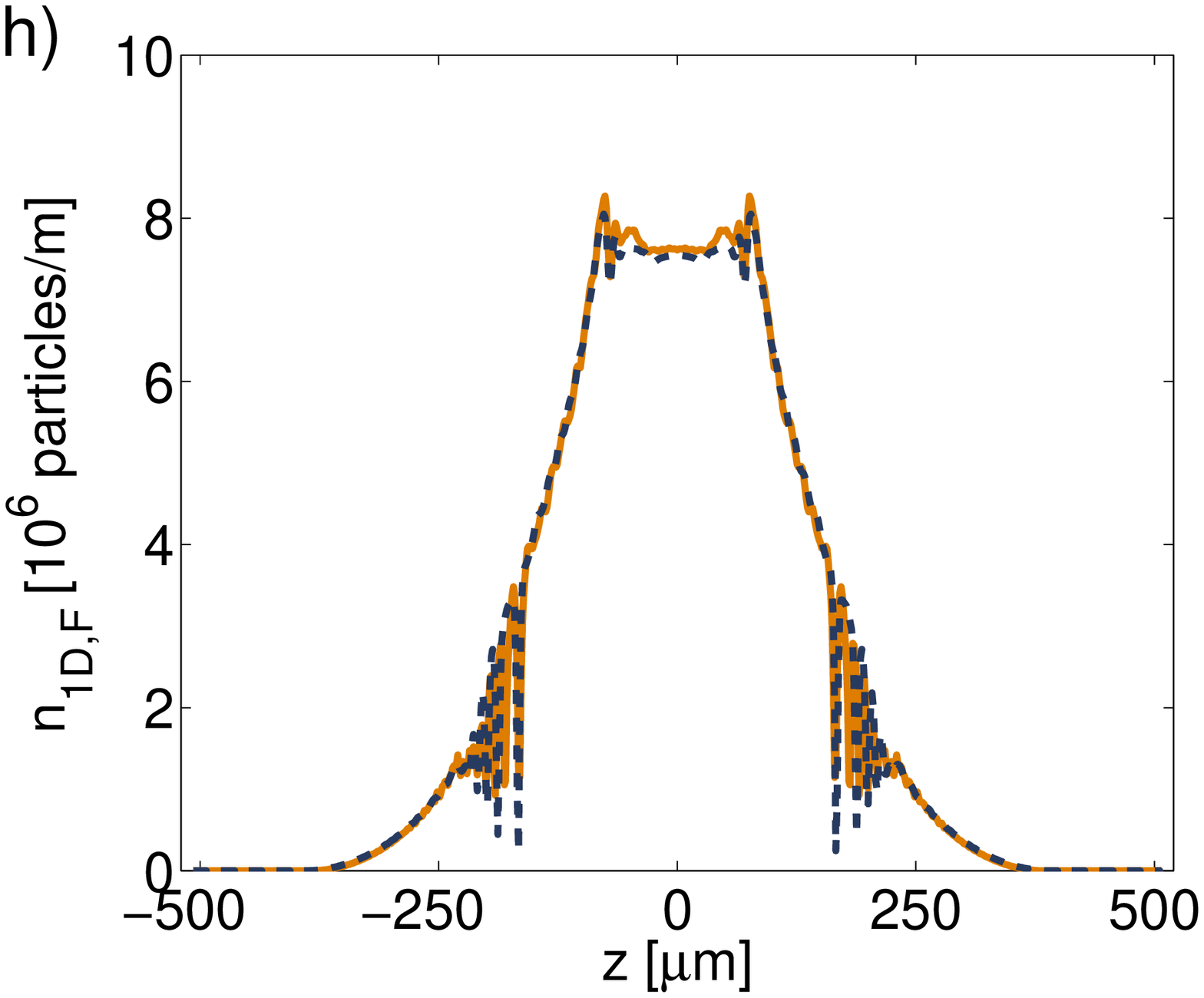}
}}
\caption{(Color online) Comparison of the dynamics, as obtained from the 1D variational approximation, and from the 3D simulations. Spatiotemporal diagrams for bosons (a) and fermions (b) are obtained from the 3D simulations. The other panels show spatial profiles for: (c)$-$(d) $t=0ms$, (e)$-$(f) $t=25ms$, and (e)$-$(f) $t=50ms$. Here $a_{\mathrm{BF}}=-10nm$, the initial conditions and other fixed parameters being the same as in Fig.~\ref{fig:7-n}.}
\label{fig:11-n}
\end{figure}

In the Sec.~\ref{sec-3-1-3} some examples of spatio-temporal dynamics were presented without discussing the accuracy of the results. In Fig.~\ref{fig:11-n} the spatio-temporal dynamics of the 1D density is displayed as produced by the solution of the 3D equations~\ref{E-5} and ~\ref{E-6}, using the same procedure as in Section 3, for initial conditions similar to those used  in Fig.~\ref{fig:7-n} but with $a_{\mathrm{BF}}=-10nm$.
The initial conditions for the 3D dynamics are given by the ansatz in Eq.~\ref{E-8}, with the Gaussian profile along the $z$-axes, similar to those used in the Fig.~\ref{fig:7-n}. Panels~\ref{fig:11-n}(a) and ~\ref{fig:11-n}(b) show spatio-temporal diagrams of the bosonic and fermionic density, respectively, making the emergence of dark solitons obvious, see also Fig.~\ref{fig:7-n}. This result corroborates that the dark solitons will emerge too in the 3D dynamics, which is approximated by the present 1D model.
The other panels show a comparison of the 1D spatial profiles, as obtained from 3D simulations and the 1D VA, for three instants of time: $t=0ms$ (panels~\ref{fig:11-n}(c) and ~\ref{fig:11-n}(d)), $t=25ms$ (panels~\ref{fig:11-n}(e) and ~\ref{fig:11-n}(f)), and $t=50ms$ (panels~\ref{fig:11-n}(g) and ~\ref{fig:11-n}(h)). The results demonstrate that the VA profiles are very similar to their counterparts produced by the 3D simulations, hence the present approximation provides good accuracy and allows one to study dynamical features of the Bose-Fermi mixture in a sufficiently simple form.

\subsection{Results in two dimensions.}

\label{sec-3-2}

\subsubsection{Accuracy of the variational method as a function of scattering parameter $a_{\mathrm{BF}}$ for the ground state.}
\label{sec-3-2-1}

\begin{figure}[tbp]
\centering
\resizebox{0.8\textwidth}{!}{
\includegraphics{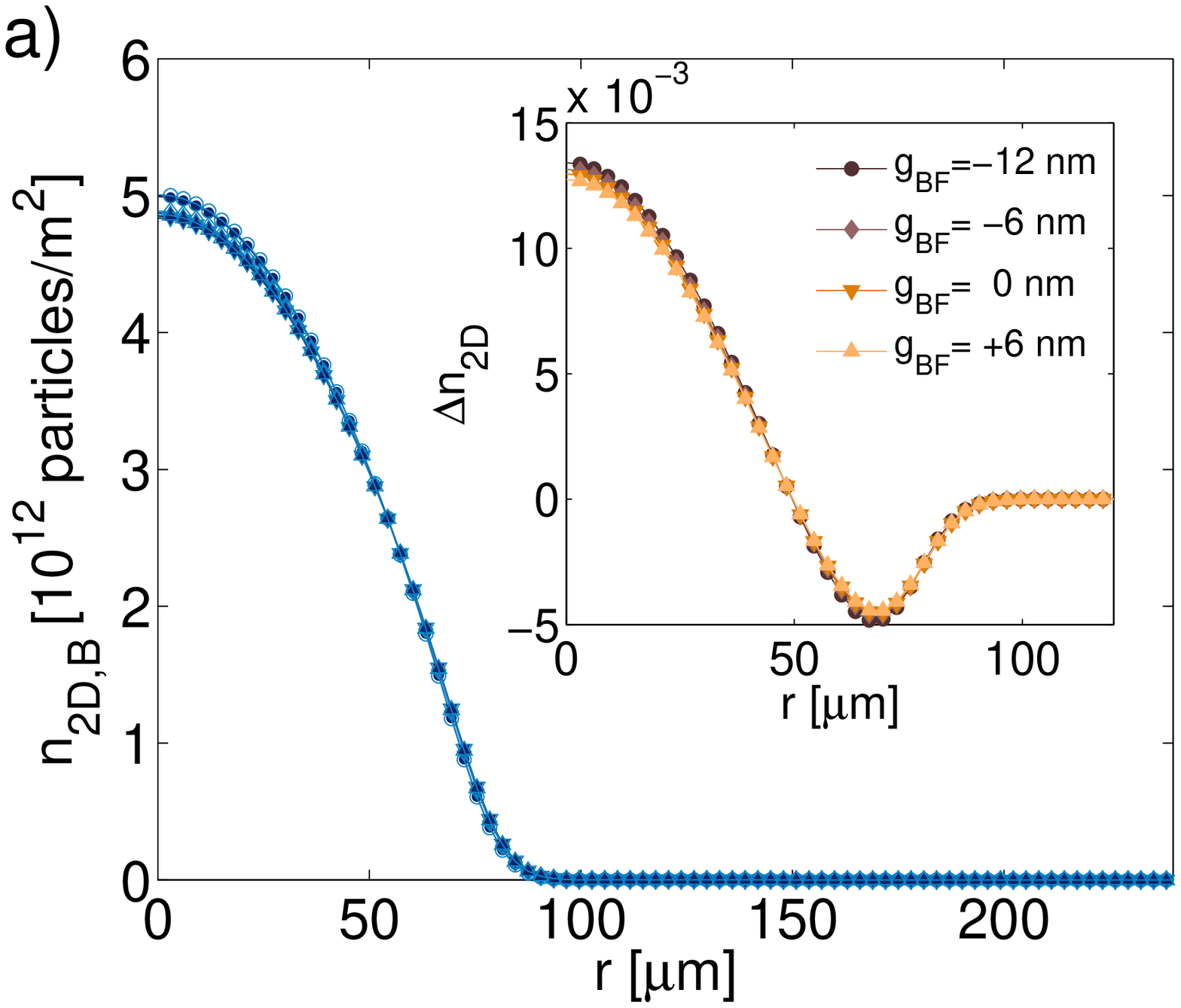}
\includegraphics{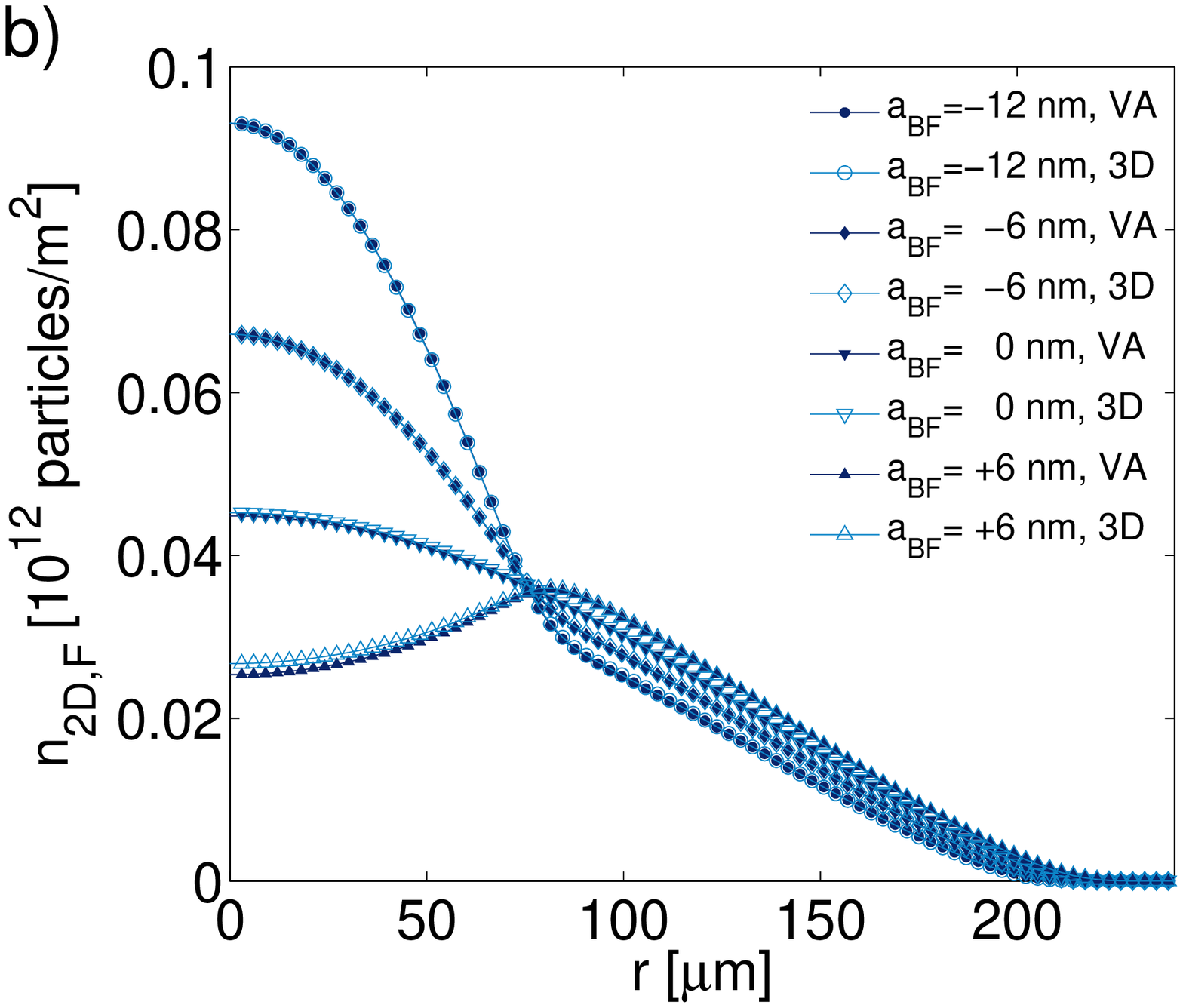}
}
\resizebox{0.8\textwidth}{!}{\includegraphics{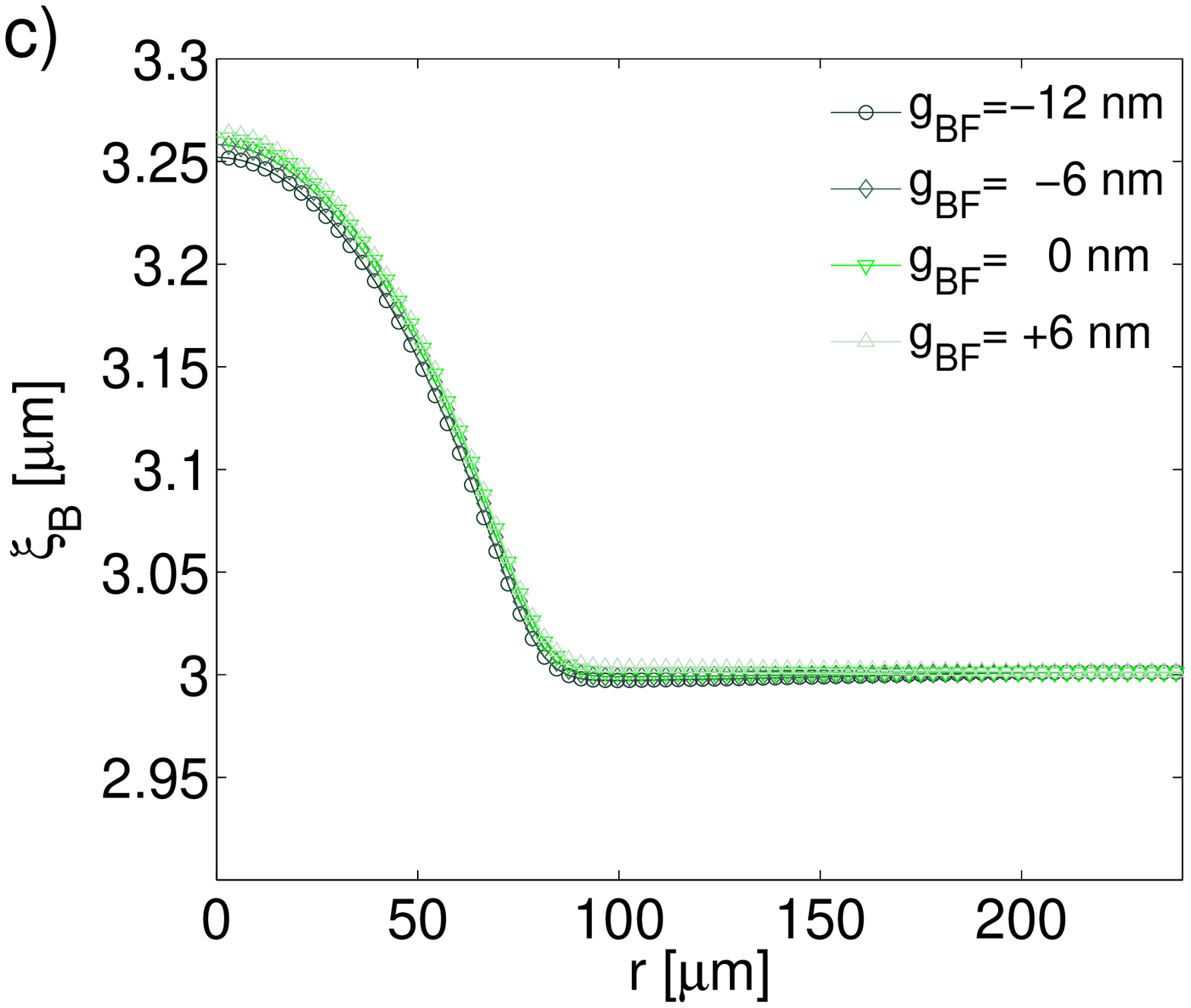}
\includegraphics{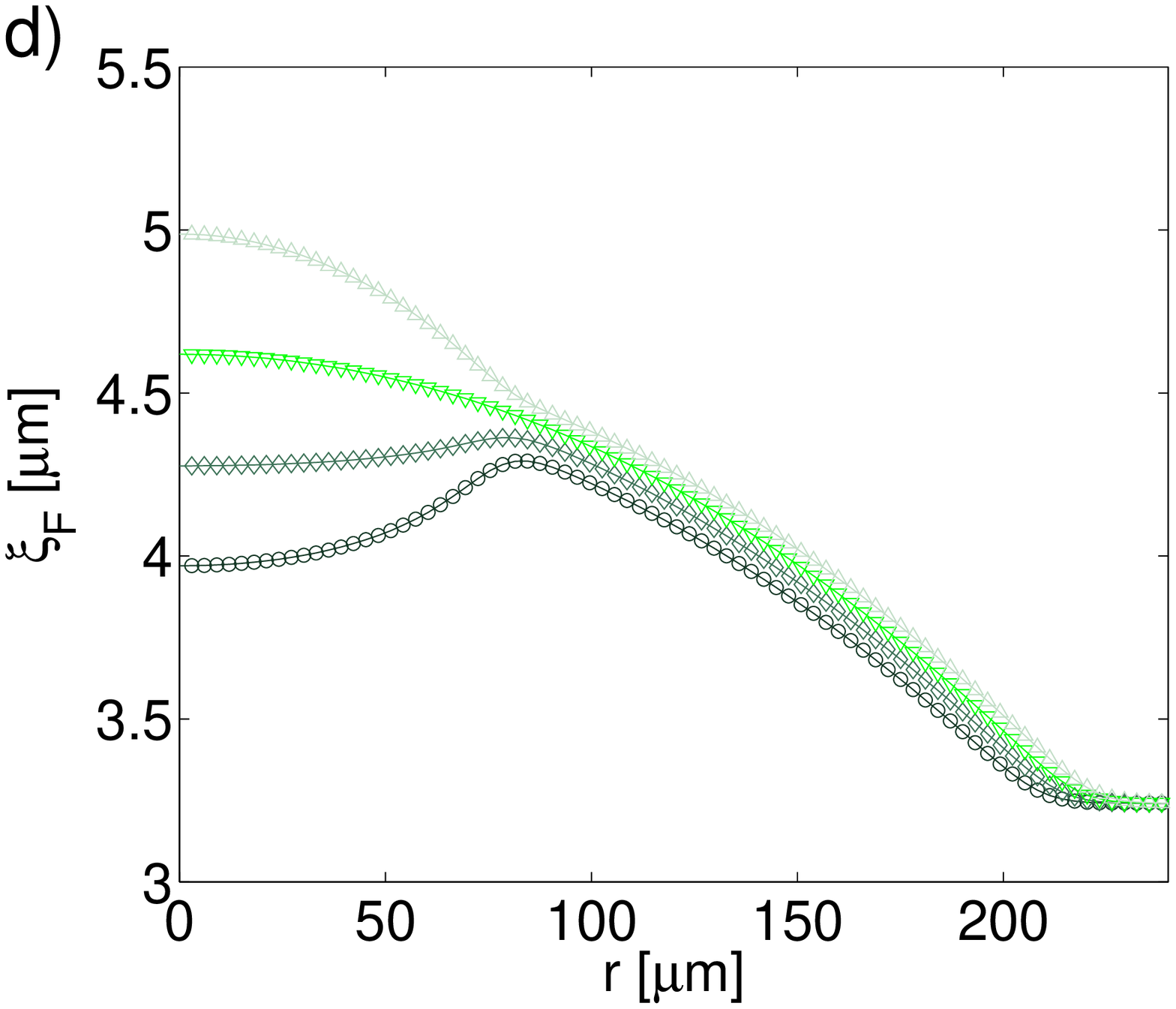}
}
\caption{(Color online) The radial profile of the 2D particle density
(for the VA and 3D simulations, see the inset in panel (b)), and
the respective width in the transverse direction ($z$) of the bosonic and
fermionic profiles, for five values of $a_{\mathrm{BF}}$ (see insets in the panels):
(a) $n_{2\mathrm{D},\mathrm{B}}$, (b) $n_{2\mathrm{D,F}}$, (c) $\protect\xi %
_{\mathrm{B}}$, and (d) $\protect\xi _{\mathrm{F}}$. The parameters are $N_{%
\mathrm{B}}=5\times 10^{4}$, $N_{\mathrm{F}}=2.5\times 10^{3}$, $a_{\mathrm{%
B/F}}=5$ nm, $\protect\omega _{z,\mathsf{B/F}}=1000$ Hz, and $\protect\omega %
_{x,\mathrm{B/F}}=\protect\omega _{y,\mathrm{B/F}}=30$ Hz. The inset in panel (a)
shows the difference between both methods, VA and 3D, where $\Delta n_{\mathrm{2D}}\equiv \rho_{\mathrm{2D}}-n_{\mathrm{2D}}$.}
\label{fig:7}
\end{figure}

In the 2D setting, we studied the GS of the BFM, fixing the number of bosons
to $N_{\mathrm{B}}=5\times 10^{4}$ (similar to the value used in the 1D
case), which was and much greater than the number of fermions. Fig.~\ref%
{fig:7} shows the profile of the 2D bosonic and fermionic densities, $n_{2%
\mathrm{D},\mathrm{B}/\mathrm{F}}$ (top), as obtained from the
VA and 3D simulations, and the widths
of the mixture in the transverse direction, $\xi _{\mathrm{B}/\mathrm{F}}$
(bottom), with respect to the radial coordinate. To obtain
the 2D profile from the 3D simulations (results of solving
Eqs.~\ref{E-5} and ~\ref{E-6}), we integrate the 3D
density along the $z$ so that  $\rho_{\mathrm{2D,B/F}}=\int \left\vert%
\Psi_{\mathrm{2D,B/F}}\right\vert dz$. The panels for the bosonic and fermionic
components are displayed on the left and right, respectively. The results
are very similar to what was observed in the 1D case in Fig.~\ref{fig:3},
for the same value of the physical parameters. Therefore, the following
conclusions are also similar to what inferred in the 1D setting: the
repulsive mixture concentrates the bosons at the center, while the
attractive mixture concentrates both species at the center. Panels
\ref{fig:7}(c) and \ref{fig:7}(d) show that only the width of the
fermionic density profile varies significantly with the change of the
scattering length of the inter-species interaction, which is consequence
of a greater number of bosons than fermions. It is clearly seen that
fermions are stronger confined when the interaction is attractive,
and their spatial distribution significantly expands when the interaction
is repulsive. Similar results have been observed in \cite{Adhikari08,%
Maruyama09,Salasnich07a}.

Now, to compare the results obtained
from the VA with those produced by the 3D simulations,
we note that both profiles are practically identical, except for the repulsive
case in which  a discrepancy is observed (smaller than in the 1D case). The inset in
panel~\ref{fig:7}(a) shows that the difference between the two results
has a magnitude of nearly three orders of magnitude lower than the
density itself. The error is lower than in the 1D case because the 2D reduction
is closer to the full 3D model.

\begin{figure}[tbp]
\centering
\resizebox{0.5\textwidth}{!}{
\includegraphics{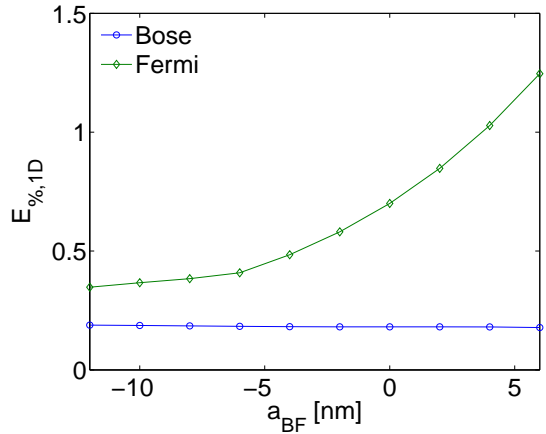}
}
\caption{(Color online) 2D overall percentage error for both species (the definition is given
in the text). Parameters are the same as in Fig.~\ref{fig:7}.}
\label{fig:13-n}
\end{figure}

Now, similar to the 1D case, we define the overall percentage error of
the VA as $E_{\%,\mathrm{2D}}=\int\int \left\vert \rho_%
{\mathrm{2D}}-n_{\mathrm{2D}}\right\vert dxdy$ (for both species).
Figure~\ref{fig:13-n} shows the error for both species as a function of
interspecies scattering parameter ($a_{\mathrm{BF}}$). For bosons it is around
of $0.2\%$ (one order of magnitude lower than in the 1D case),
and does not change much (as shown in the inset to Fig.~\ref{fig:7}(a)). For
fermions the error is greater than for bosons throughout the observed
range, but it is quite small for the attractive mixture. Like in the 1D
case, the error increases for the   repulsive mixture, but remains lower
than that calculated for bosons in the 1D case. In summary, the 2D approximation
is very accurate, implying, as shown below, that
the description of the dynamics is also more accurate in this case.

\subsubsection{Spatial correlations of the two-dimensional density}
\label{sec-3-2-2}

\begin{figure}[tbp]
\centering
\resizebox{0.5\textwidth}{!}{\includegraphics{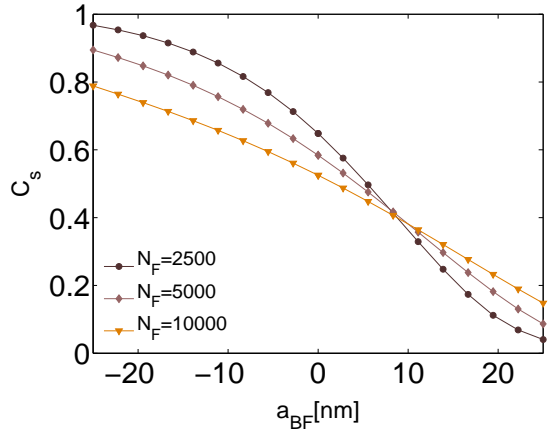}
}
\caption{(Color online) Spatial correlation $C_{s}$ of the ground state of
the 2D mixture as a a function of $a_{\mathrm{BF}}$, for three values of the
number of fermions. The other parameters are: $N_{\mathrm{B}}=5\times 10^{4}$%
, $a_{\mathrm{B/F}}=5$ nm, $\protect\omega _{x,\mathrm{B/F}}=\protect\omega %
_{y,\mathrm{B/F}}=30$ Hz and $\protect\omega _{z,\mathrm{B/F}}=1000$ Hz.
Fermions are in the fully polarized state.}
\label{fig:8}
\end{figure}

Here we again apply the formalism to the calculation of the spatial
correlation $C_s$. The Fig.~\ref{fig:8} shows the dependence of the spatial
correlation $C_{s}$,
defined as per Eq. (\ref{E-34}), on the scattering length $a_{\mathrm{BF}}$
for three values of $N_{\mathrm{F}}$ and fixed $N_{\mathrm{B}}=5\times 10^{4}
$, in the regime of the fully polarized fermionic component. The case with a
smaller number of fermions ($N=2500$, the same as in Fig.~\ref{fig:7}) gives
rise to the greatest contrast in the values of the correlation: from a value
close to $1$ for $g_{\mathrm{BF}}=-25$ nm to a value near zero for $g_{%
\mathrm{BF}}=25$ nm. This transition is more drastic than that observed for
the same parameters in the 1D case. This is mainly due to different factors
multiplying the interaction terms in the 2D equations (\ref{E-29}), ( \ref%
{E-30}), in comparison with their 1D counterparts (\ref{E-15}), (\ref{E-16}%
). The contrast between the mixing and demixing decreases as the number of
fermions increases, due to the strengthening of the self-interaction of the
fermions, hence the correlation gets less sensitive to the inter-species
scattering length, $a_{\mathrm{BF}}$.

\begin{figure}[tbp]
\centering
\resizebox{0.5\textwidth}{!}{\includegraphics{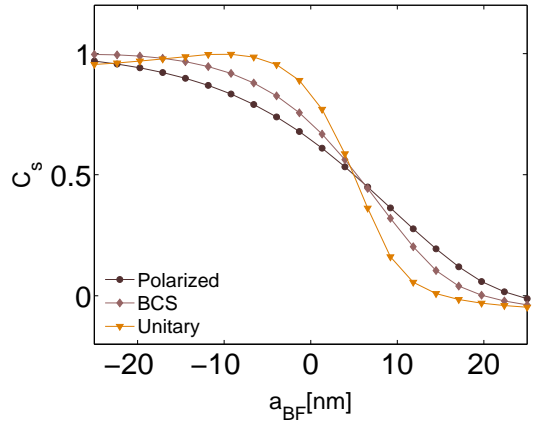}
}
\caption{(Color online) Spatial correlation $C_{s}$ of the ground state of
the 2D mixture as a a function of $a_{\mathrm{BF}}$, for three fermionic
regimes: polarized, BCS, and unitarity. The fixed parameters are: $N_{%
\mathrm{B}}=5\times 10^{4}$, $N_{\mathrm{F}}=2.5\times 10^{3}$, $a_{\mathrm{B/F}}=5$ nm, $%
\protect\omega _{x,\mathrm{B/F}}=\protect\omega _{y,\mathrm{B/F}}=30$ Hz and
$\protect\omega _{z,\mathrm{B/F}}=1000$ Hz.}
\label{fig:9}
\end{figure}

The Fig.~\ref{fig:9} shows the spatial correlation, defined as per Eq. (\ref%
{E-34}), for the three fermionic regimes: polarized, BCS and unitarity. The
figure is a counterpart of Fig.~\ref{fig:6} drawn for the 1D case, but only
for the spatial correlation in the GS ($C_{s}$). One may conclude that these
fermionic regimes show the behavior similar ti that observed in the 1D case,
a difference being that the three curves demonstrate stronger demixing when $%
a_{\mathrm{BF}}$ changes from positive to negative values. In the unitarity
regime it is again observed that the correlation reaches a maximum close to $%
1$ at $a_{\mathrm{BF}}\approx -10$ nm, dropping to negative values when the
mixture is strongly repulsive.

\subsubsection{Accuracy of the variational method for the spatiotemporal dynamic}
\label{sec-3-2-3}

\begin{figure}[tbp]
\centering
\resizebox{\textwidth}{!}{
\includegraphics{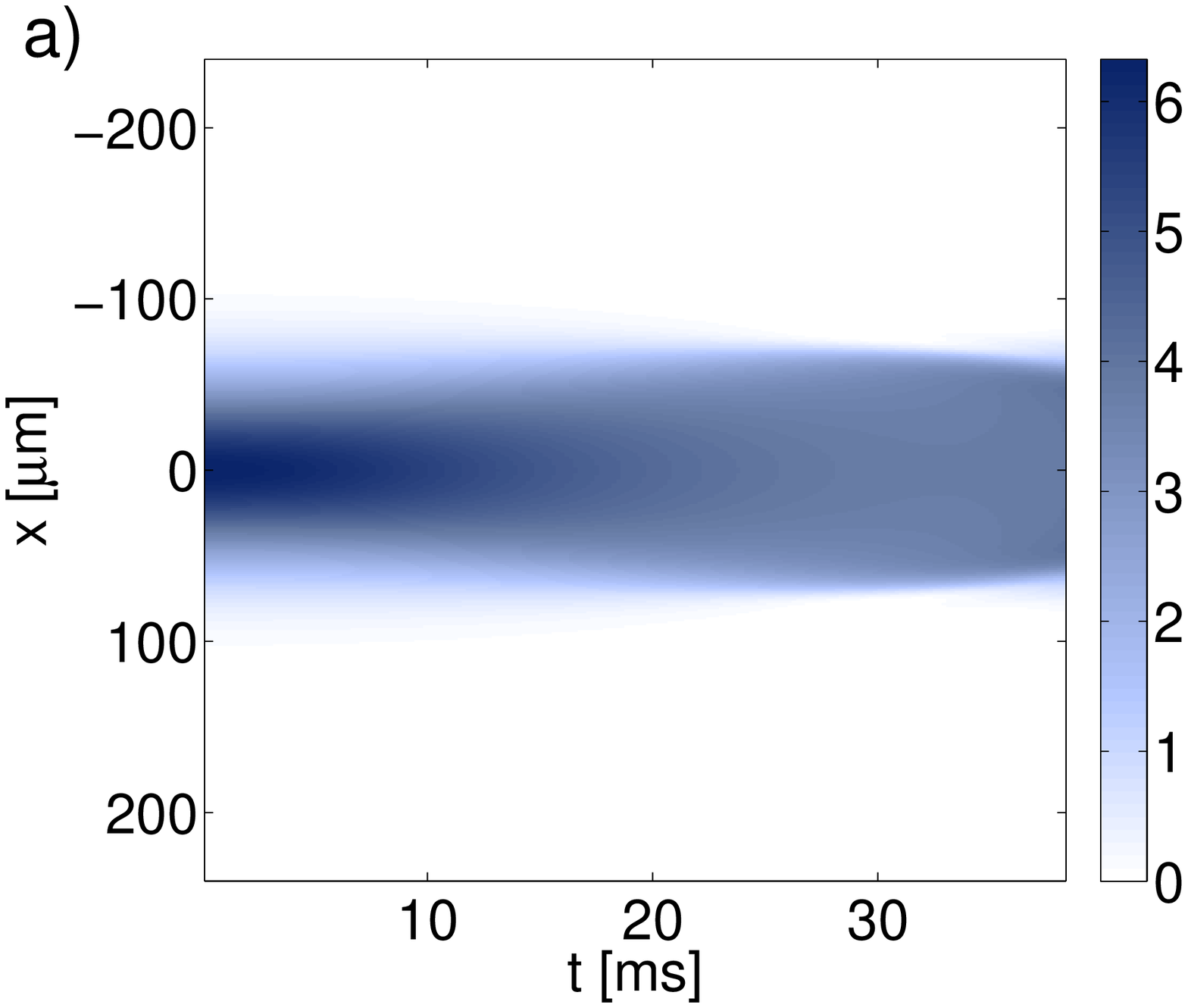}
\includegraphics{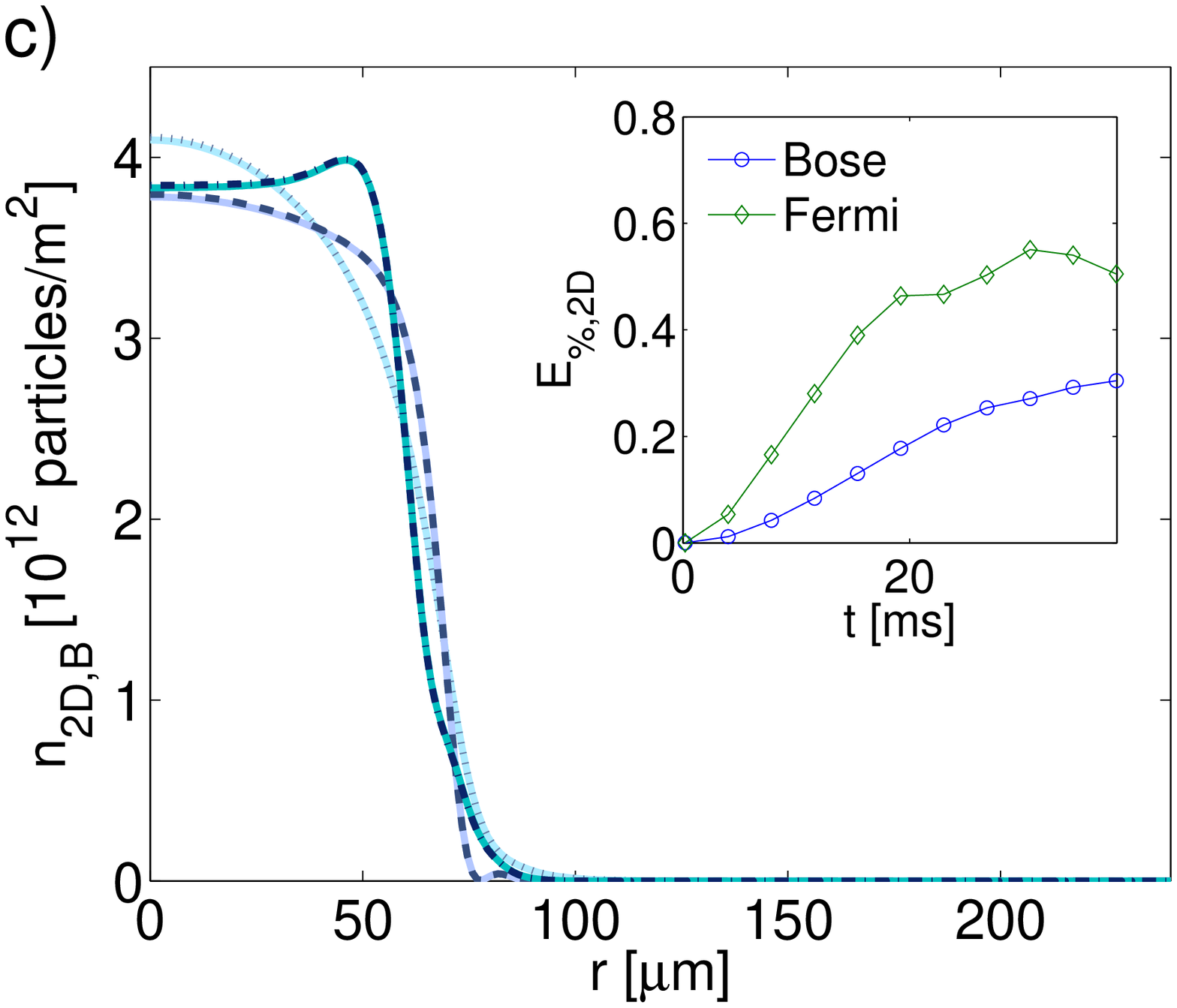}
\includegraphics{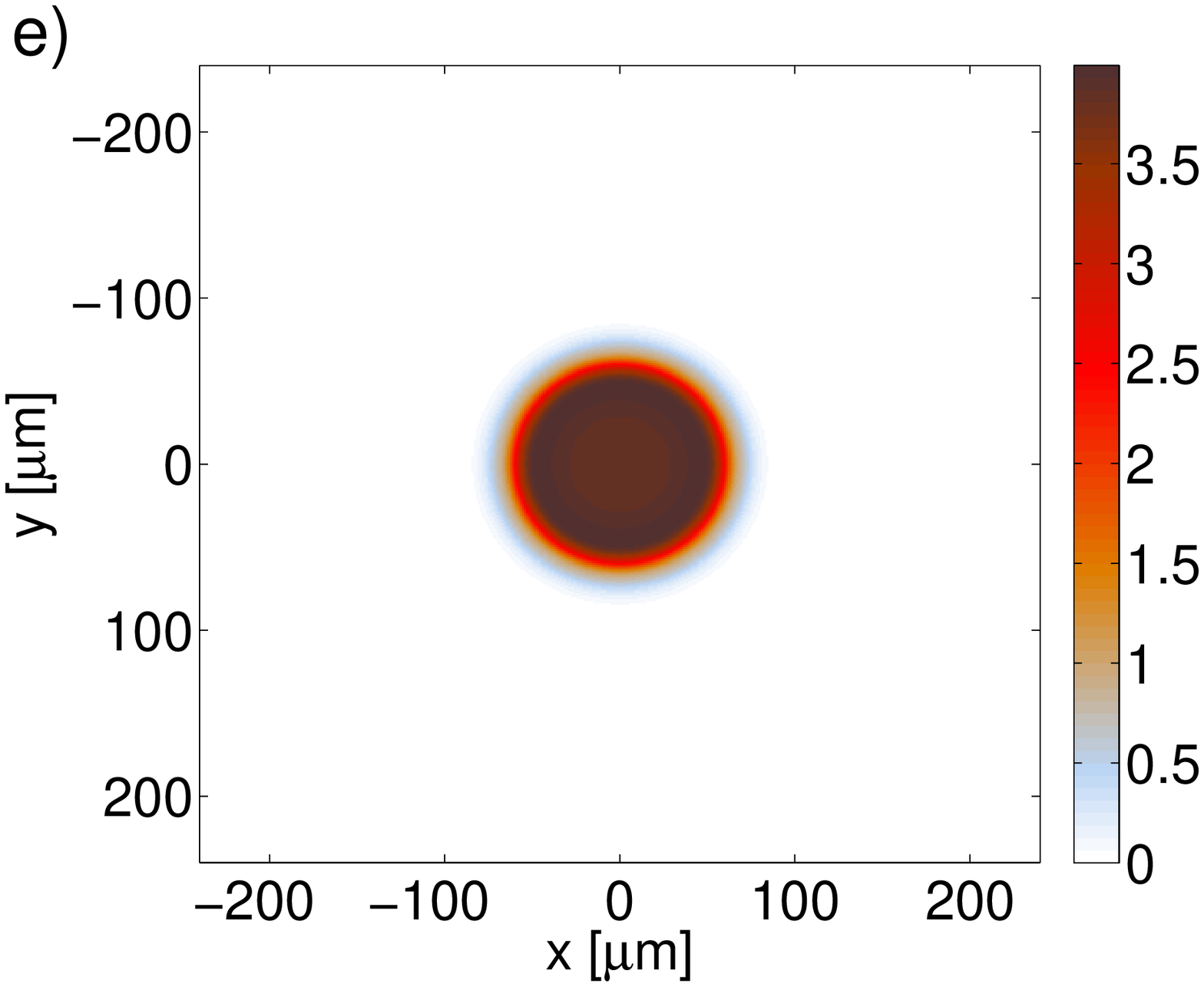}
}
\resizebox{\textwidth}{!}{
\includegraphics{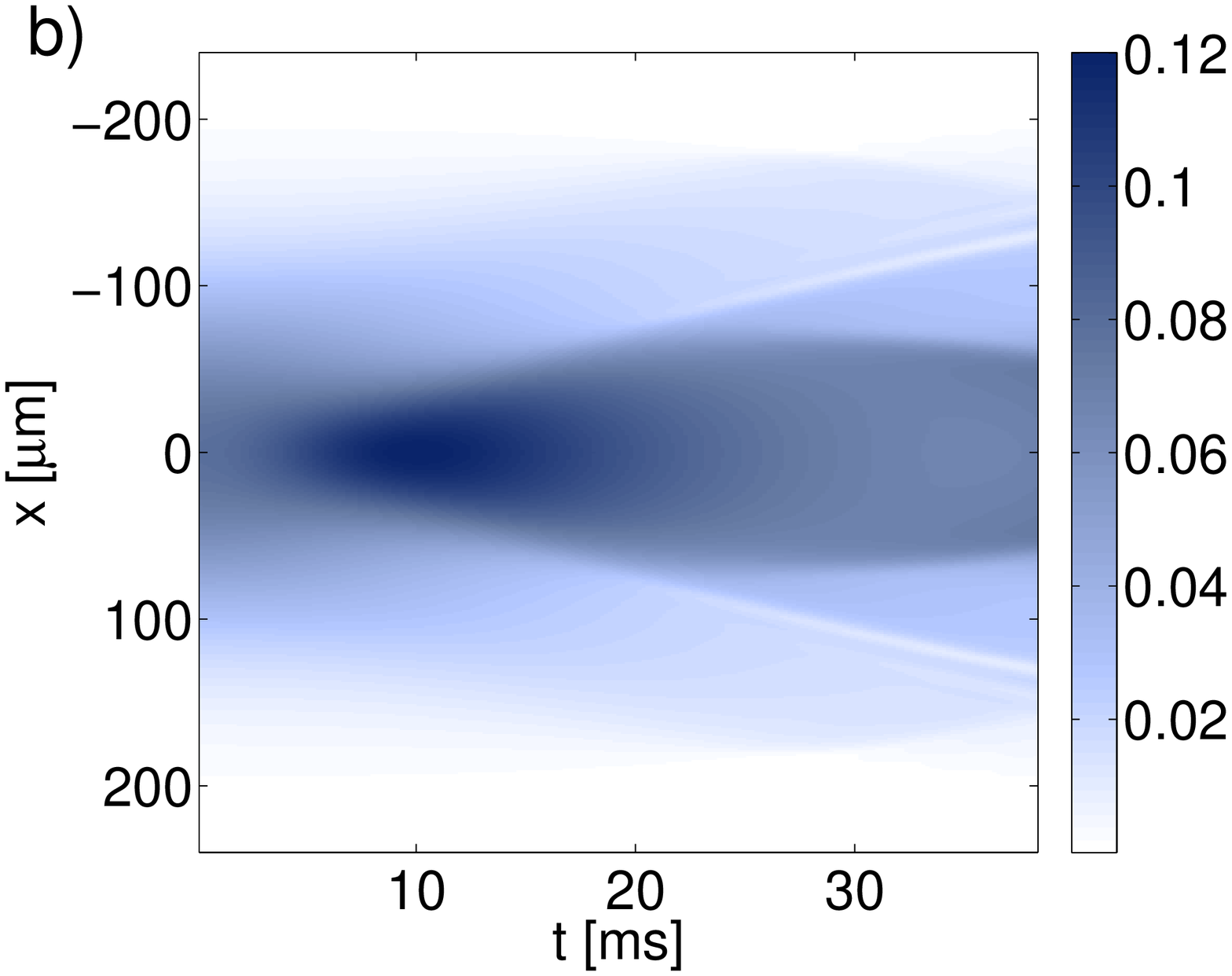}
\includegraphics{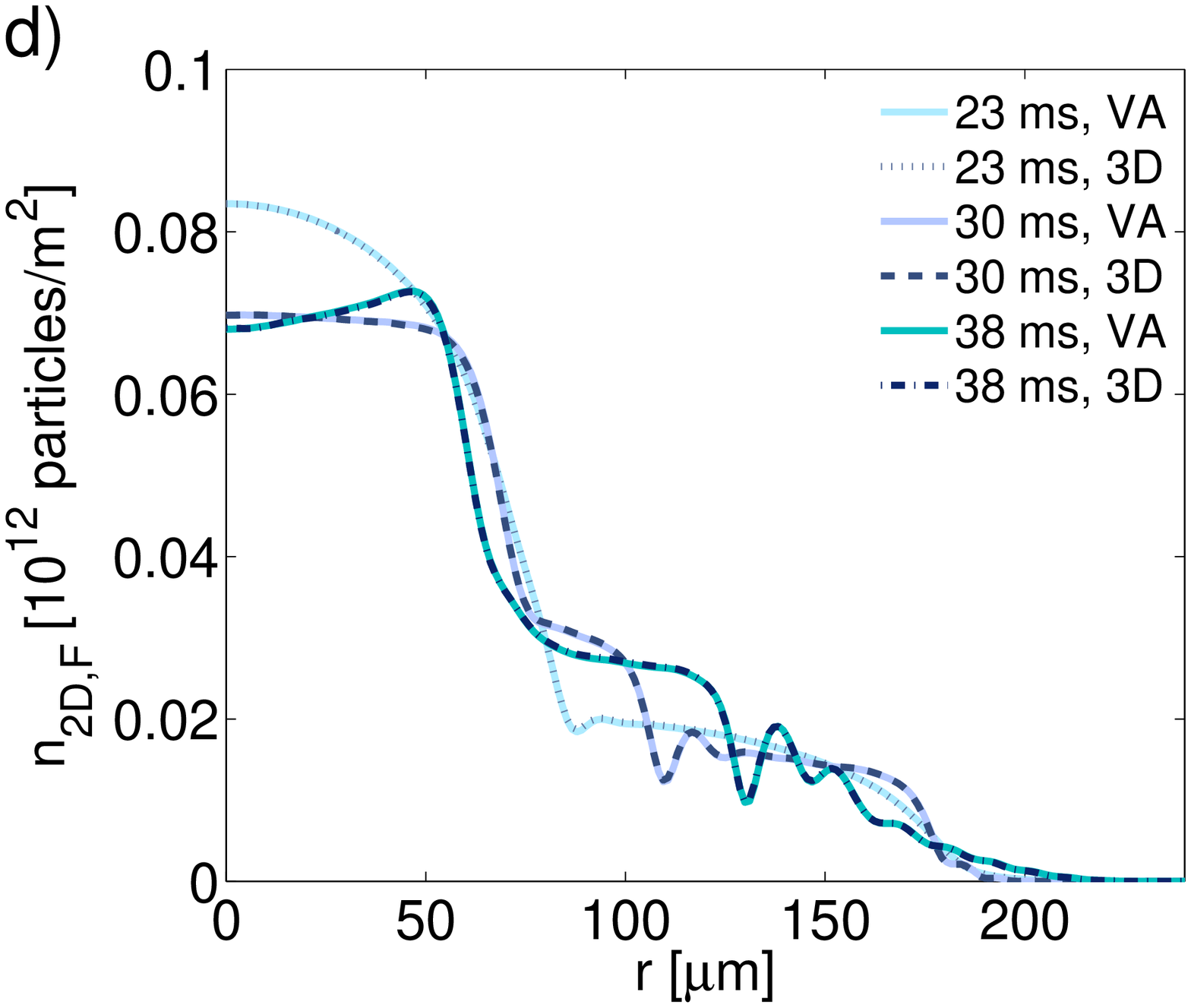}
\includegraphics{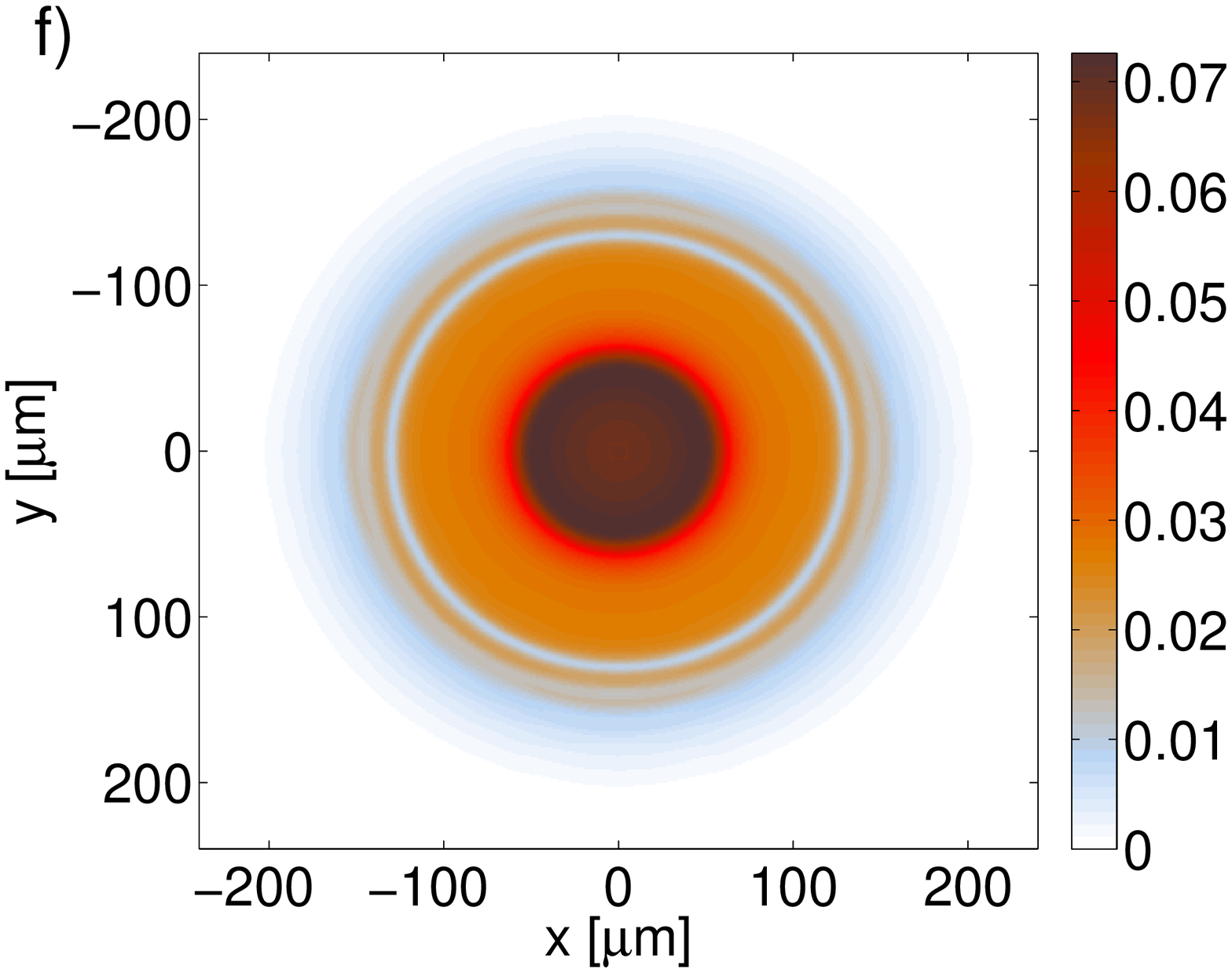}
}
\caption{(Color online) 2D dynamics, as produced by the 2D VA, and by the 3D simulations for the attractive mixture of bosons (above) and fermions (below). Spatiotemporal diagrams (a) and (b), produced by the 3D simulations, show how the profiles of the densities of particles in the $x$ direction evolve in time (it is a cut of the 2D density with cylindrical symmetry). The curves in panels (c) and (d) shows the radial profiles for three different instants of time (see the inset in panel (d)), as produced by both the VA and 3D simulations. Diagrams (e) and (f) show 2D densities in the $xy$-plane for the last moments of time in panels (a) and (b). The inset in panel (c) shows the evolution of 2D error ($E_{\%,\mathrm{2D}}$) in time. Initial conditions are described in the text, and parameters are the same as in Fig.~\ref{fig:7} with $a_{\mathrm{BF}}=-12 nm$.}
\label{fig:16-n}
\end{figure}

Figure~\ref{fig:16-n} shows, for the attractive mixture ($g_{\mathrm{BF}}=-12nm$), the comparison between the results obtained by the t3D integration (Eqs.~\ref{E-5} and \ref{E-6}) with the results obtained by the 2D integration of the variational equations (Eqs.~\ref{E-28} and~\ref{E-29}). The approach is similar to that developed in Sec.~\ref{sec-3-1-4} for the 1D case. It starts from arbitrary initial conditions in 2D, {\it viz.},  2S fields with Gaussian shapes in the radial direction, a standard deviation of $50 \mu m$ for bosons and $100 \mu m$ for fermions being assumed. The initial conditions for the 3D dynamics are obtained with the ansatz given by Eq.~\ref{E-21} and assuming a Gaussian in the $xy$-plane. Panels~\ref{fig:16-n}(a) and~\ref{fig:16-n}(b) display two spatiotemporal diagrams, obtained from the 3D integration, showing a cut along the $x$ axis of the 2D density; since the dynamics, for the time range observed, obeys the cylindrical symmetry, this cut is representative. It is seen that the boson profile clearly affects the fermion profile. As the initial conditions do not correspond to the ground state, the fermions are attracted to the center by the bosonic pattern and are subject to the confinement imposed by the harmonic potential, generating strong interaction with the boson core. A gray soliton ring is generated (in panel \ref{fig:16-n}(b) it is seen as a gray soliton). In panels~\ref{fig:16-n}(c) and~\ref{fig:16-n}(d),  the radial profile obtained by the 3D integration and by VA for three instants of time are compared, showing that it is not possible to distinguish them. The inset in panel \ref{fig:16-n}(c) shows that, in the observed range, the 2D overall error is less than $1\%$ (at initial time, it is zero because the 3D initial conditions are obtained from the 2D configuration using the ansatz, Eq.~\ref{E-21}). The approach delivers very good results. In addition, in panel~\ref{fig:16-n}(d) it was observed  that the rings are gray (the minimum value of the fermion density is not zero). Panels~\ref{fig:16-n}(e) and~\ref{fig:16-n}(f) show diagrams of the 2D density in the $xy$-plane for the last moment of time, highlighting the gray ring soliton observed in the fermionic component, which expands wherein the boson density is zero. We note that, in this case, the approximation is very good and allows one to considerably reduce the computing time needed to observe the 2D dynamics. The dynamics of the mixture thus produces a vast variety of localized nonlinear structures.

\section{Conclusions}

\label{sec-4}

The incorporation of the width of the confined species as a variational
variable shows that the term accounting for the interaction between the
bosonic and fermionic species in the lower-dimensional equations is
inversely proportional to the width in the confined direction(s). The
derived systems of equations, which include the algebraic equations for the
widths as functions of the 1D or 2D bosonic and fermionic densities.
These approximations yield solutions which agree well
with the full 3D theory. It is worthy to note that the ensuing
relation between the widths and densities
significantly varies with the strength of the inter-species interaction,
which can be tuned by means of the Feshbach resonance. For very low
densities, our equations are similar to the simplest models which postulate
constant confinement width in the transverse direction(s). As an application
of the variational method, we have studied the spatial correlation between
the species, in which a strong dependence on the interaction parameter was
observed. This sensitivity becomes stronger for the repulsive interaction,
which causes demixing of the species.

It was observed too that, in 1D and 2D alike, the spatial correlation
between the species is also strongly affected by the inter-species
interaction strength. In addition, it was concluded that the greatest
contrasts in the correlation, changing from the large correlation in the
highly attractive mixtures to a weak correlation in the strongly repulsive
ones, occur in the unitarity fermionic regime. Further, in the 1D setting
the dynamics near the ground state demonstrates that the spatiotemporal
correlation follows the same trends as the spatial correlation, the largest
(but still small) difference being observed for strongly repulsive mixtures

Finally, it is relevant to mention that the mixture can also give rise to
stable dark solitons in the Fermi field, in both the 1D and  in 2D cases.
However, as the hydrodynamic conditions was assumed, the solitons may occur
in some form but are unconfirmed and should not be believed in too strongly.
A detailed characterization of these states and the validity of the hydrodynamic
approximation for them will be presented elsewhere.

\section*{Acknowledgements}
P.D. acknowledges financial support from DIUFRO project under grant $DI14-0067$ and the Center for Modelling and Scientific Computing (CMCC)
of the Universidad de La Frontera. D. L. acknowledges partial financial support from EPSRC under grant
$EP/L002922/1$, FONDECYT
1120764, Millennium Scientific Initiative, $P10-061-F$, Basal Program Center
for Development of Nanoscience and Nanotechnology (CEDENNA) and UTA-project $
8750-12$. B.A.M. appreciates partial support from the German-Israel
Foundation (grant no. I-1024-2.7/2009) and Binational (US-Israel) Science
Foundation (grant no. 2010239).

\end{document}